\documentstyle[aas2pp4]{article}
\newcommand{\mc}{\multicolumn}
\newcommand{\ph}{\phantom}

\newcommand{\xx}{\phantom{xx}}
\newcommand{\gtsim}{\mbox{{\raisebox{-0.4ex}{$\stackrel{>}{{\scriptstyle\sim}}$}}}}
\newcommand{\cosone}{\mathversion{bold}${\Omega_{\rm M} = 0\ \&\ \Omega_{\Lambda} = 0} $}
\newcommand{\costwo}{\mathversion{bold}${\Omega_{\rm M} = 1\ \&\ \Omega_{\Lambda} = 0} $}
\newcommand{\costhree}{\mathversion{bold}$\Omega_{\rm M} = 0.1\ \&\ \Omega_{\Lambda} = 0.9 $}
\renewcommand{\tabcolsep}{0.5mm}

\slugcomment{}

\lefthead{Blundell, Rawlings \& Willott}
\righthead{The Nature and Evolution of Classical Double Radio Sources}
\input epsf

\begin{document}
\title{The Nature and Evolution of Classical Double Radio Sources \\
from Complete Samples}

\author{Katherine M.\ Blundell, Steve Rawlings and Chris J.\ Willott}

\affil{University of Oxford, Astrophysics, Keble Road, Oxford, OX1 3RH, U.K. }

\centerline{\small To appear in AJ}

\begin{abstract}
We present a study of the trends in luminosity, linear size, spectral
index, and redshift of classical double radio sources from three complete
samples selected at successively fainter low radio-frequency
flux-limits. We have been able to decouple the effects of the tight
correlation between redshift and luminosity (inherent in any {\em single}
flux-limited sample) which have hitherto hindered interpretation of the
relationships between these four source properties.  The major trends
found are that {\em (i)} spectral indices increase with linear size, {\em
(ii)} rest-frame spectral indices have a stronger dependence on luminosity
than on redshift except at high (GHz) frequencies, and that {\em (iii)}
the linear sizes are smaller at higher redshifts. We reproduce the
observed dependences in a model for radio sources (born throughout cosmic
time according to a radio-source birth function) whose lobes are fed with
a synchrotron-emitting population from compact hotspots, and which suffer
inverse Compton, synchrotron and adiabatic expansion losses.  The magnetic
energy density within each hotspot is proportional to the jet-power, and
synchrotron losses suffered in the hotspot mean that the energy spectrum
of the emitting particles fed to the lobes is governed by the jet-power.
The axial ratios of radio sources in our model increase as the sources
age, and axial ratios are higher in sources with higher jet-power.  In
simulating the basic observed dependences, we find that there is no need
to invoke {\em any} systematic change in the environments of these objects
with redshift if the consequences of imposing a survey flux-limit on our
simulated datasets are properly included in the model. It is also
necessary to include appropriate energy loss mechanisms (such as the
effects of the cosmic microwave background and feeding the lobes from a
compact hotspot) which cause decreasing luminosity through the life of a
source.  Although our study has broken the luminosity--redshift
degeneracy, we present evidence that for such studies there is an
unavoidable ``youth--redshift degeneracy'', even though radio sources are
short-lived relative to the age of the Universe; it is imperative to take
this into account in studies which seemingly reveal correlations of source
properties with redshift such as the ``alignment effect''.
\end{abstract}


\section{Introduction and scope}

In order to explore the physical mechanisms by which classical double
radio sources are governed, observations of {\em samples} of such
objects are needed to discern the salient relationships between
different source properties, such as spectral index or luminosity.

Even observations of a complete sample will inevitably be convolved
with the cosmology of the Universe as well as with the consequences of
the sample selection process. We first describe (in
\S\ref{sec:selection}) how to avoid compounding the latter problem and
giving rise to spurious, or hard to interpret, results.

In \S\ref{sec:samples} we briefly describe the samples we have used in
this new analysis, including the new 7C Redshift Survey, which allow us to
avoid problems inherent in the sample selection processes. In
\S\ref{sec:measure} we describe the principles of how we use these data,
the types of objects included in the analyses, and how these principles
were put into practice. We then examine in \S\ref{sec:results} the
dominant trends to emerge in the various source properties.  In
\S\ref{sec:param} we present a parameterisation of our results for ease of
comparison with previous results.  In \S\ref{sec:compare} we then review
and briefly discuss results from the past which have hinted at the results
we find.  In order to interpret these results, we develop our model of
radio sources in \S\ref{sec:model}; within this section, we summarise in
\S\ref{sec:consensus} basic properties of radio sources which are widely
believed to be correct and discuss what may be gleaned about the
environments in which radio sources reside in \S\ref{sec:env}. We then
consider the expansion of the source and observational constraints on
expansion speeds in \S\ref{sec:sourcesize}. A model for the luminosity
evolution for a given source is considered in \S\ref{sec:pmod}; in this
section we present a model for the r\^{o}le played by the hotspot in
governing the luminosity of radio source lobes.  After considering in
\S\ref{sec:indiv} the development of individual radio sources we describe
in \S\ref{sec:zmod} how to correctly sample radio sources born at
successive cosmic times in order to reproduce simulated versions of the
data from our complete samples.  In \S\ref{sec:pararesults} we present
simulated versions of our complete samples using the modelling described
in previous sections.  In \S\ref{sec:remarks} we discuss what has been
learned about the physical mechanisms by which radio sources are governed
from our data and simulations, and briefly consider some implications for
cosmological studies. We summarise the salient conclusions of our study in
\S\ref{sec:conc}.

We assume that $H_{\circ}~=~50~{\mathrm km~s^{-1}~Mpc^{-1}}$ and,
except where specifically stated otherwise, that $\Omega_{\rm M} = 1$
and $\Omega_\Lambda = 0$. We use the convention for spectral index
($\alpha$) that $S_{\nu} \propto \nu^{-\alpha}$, where $S_{\nu}$ is
the flux density at frequency $\nu$.

\section{Sampling and selection}
\label{sec:selection}
\subsection{The ineluctable biases in complete samples}

\begin{figure}[!h]
\begin{picture}(50,200)(0,0)
\put(-58,-220){\includegraphics{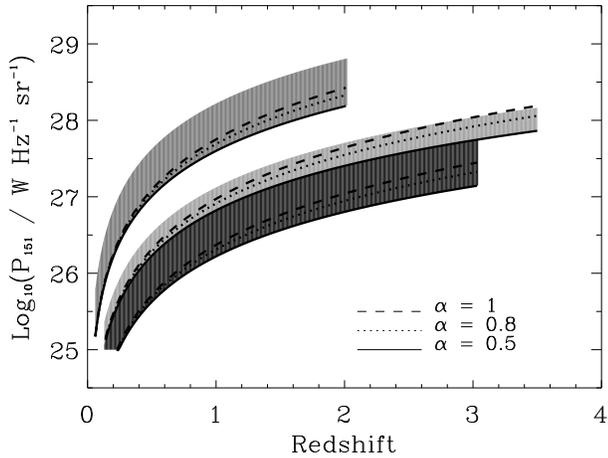}}
\end{picture}
{\caption[junk]{\label{fig:pzcartoon} A schematic illustration of the
coverage of the luminosity (at 151 MHz rest-frame) versus redshift
($P$--$z$) plane from a single sample selecting all sources brighter than
12 Jy in a certain sky area (coloured mid-grey), a single sample of
sources with flux densities between 2 and 4 Jy (coloured light-grey) and a
single sample which has selected sources brighter than 0.5 Jy in a much
smaller area of sky (coloured dark-grey).  Each coloured region contains a
similar number of sources.  The solid black lines in each case correspond
to the minimum luminosity which a source at that redshift may have, and
yet still be above the minimum flux-limit of the sample assuming source
spectral indices are 0.5; other linestyles indicate the influence of other
spectral indices on the shape of the flux-limit on the $P$--$z$ plane. The
upper edge of the mid-grey region arises because of a maximum observed
luminosity in radio sources. The upper-edge of the dark-grey region arises
because in its {\em small} sky area the probability of finding more
luminous sources is low.}}
\end{figure}

The simplest sample-type, referred to here and in companion papers as a
`complete sample' has as its selection criteria only a single flux-limit
(determined ultimately by the finite sensitivity of the survey telescope)
in a chosen observing band and a chosen sky area.  However, for any {\em
single} flux-limited sample chosen in this way there will be an inevitable
and tight correlation between luminosity and redshift. This
arises\footnote{For further discussion of the details and subtleties which
contribute to this close relationship see \S\ref{sec:pztight}.}  because
for a fixed flux-limit, only powerful sources can be detected out to great
distances.  It is therefore impossible to determine for a single
flux-limited sample whether the dependence of a given source property,
such as linear size, is primarily on redshift or on luminosity.
Figure~\ref{fig:pzcartoon} shows the coverage of the luminosity--redshift
($P$--$z$) plane by particular complete samples selected at different flux
limits.

It is clear that just one such sample neither achieves a good range in
luminosity out to high redshift nor a good range in redshift at low
luminosity.  The use of only one sample severely undermines any study
pertaining to the evolution of radio sources.  Extensive coverage of the
luminosity--redshift plane is essential if we are to decouple the
luminosity/redshift dependence of a given source property and this is only
achievable, given restricted telescope time, using a {\em number} of
complete samples selected at increasingly fainter flux-limits.

\subsection{The choice of selection frequency}
\label{sec:freq}

Samples selected at high frequency (\gtsim\ 1 GHz) come with
complications: at these frequencies the cores of radio sources can be a
significant contribution to the total flux density. Such cores are thought
to be increasingly Doppler boosted as the angle between the jet axis and
the line-of-sight is decreased (see e.g.\ Vermeulen \& Cohen 1994). This
means that objects at a small viewing angle are preferentially included in
the sample, since their total flux measurement is promoted over the sample
flux-limit because of their orientation, rather than their intrinsic
power.  In addition to having lower power in extended components, these
objects will have shorter projected angular sizes.  Moreover, the GHz part
of the spectrum of radio sources is reduced due to synchrotron losses as
the source radiates, adiabatic expansion losses as the source grows and
expands and, with increasing importance with increasing redshift, due to
inverse Compton losses from the equivalent magnetic field of the cosmic
microwave background radiation. This is further compounded because of the
increasing difference between emitted frequency and observed frequency as
redshift increases. Thus observing at a high frequency (\gtsim\ 1 GHz)
will result in a vast loss in the number of high redshift sources which
make it above the sample limit, since inverse Compton losses increase
dramatically with redshift [the equivalent magnetic field due to the
cosmic microwave background (CMB) is given by $0.318(1 + z)^2$
nT]. Inverse Compton losses have been invoked to explain the X-ray flux
from the lobes of the nearby radio galaxy Fornax A (Kaneda et al.\ 1995)
and also as an explanation of the high X-ray luminosity of at least some
distant powerful radio galaxies (Brunetti, Setti and Comastri
1997). Sources with large linear sizes at intermediate and high redshifts
are most unlikely to be found in high frequency surveys since larger
sources tend to have steeper spectral indices (see
\S\ref{sec:alphaD}). [Thus high frequency surveys are a poor probe of
evolution in the high redshift universe, except for the Doppler boosted
population.]

Surveys carried out at lower frequencies (a decade lower $\sim$ 0.1 GHz)
detect emission largely from that part of the spectrum which is least
subject to synchrotron/inverse Compton losses and orientation biases. They
thus inform on the intrinsic energy spectrum injected into the radio lobes
in a way which is less affected by the {\em history} of that particular
source. There is thus a strong argument for using complete samples
detected at such frequencies. However, in making a radio survey at
arbitrarily low frequency one encounters different problems: there are
three mechanisms by which a survey conducted at $\sim 10$ MHz would fail
to detect sources which surveys at higher frequency would detect.  One of
these mechanisms is synchrotron self-absorption; this arises when the
intensity of synchrotron radiation within a source becomes sufficiently
high that re-absorption of the radiation by the synchrotron emitting
particles is important.  Another mechanism is free--free absorption:
absorption of the synchrotron radiation by thermal material intervening
along the line-of-sight, possibly in the immediate vicinity of the
synchrotron source --- e.g.\ by its own cluster environment; an example of
this mechanism in 3C\,295 is discussed by Taylor \& Perley (1992).

The third mechanism arises because of a low-energy cut-off to the
relativistic particles which give rise to the synchrotron emission.
Arguments that there is a departure from a power-law in the
electron-energy distribution at low energies were advanced by Leahy,
Muxlow \& Stephens (1989) who made observations of 3C radio sources at 151
MHz with MERLIN and at 1.5 GHz with the VLA and found spectral flattening
in the hotspots of three of the powerful sources at low frequency.

An extreme example of sources which are rarely detected in
low-frequency surveys are the Giga Hertz Peaked Spectrum (GPS)
population [O'Dea, Baum \& Stanghellini (1991)]. It is the turnover in
the spectrum of these sources which gives rise to the absence of these
sources in low-frequency surveys, and their presence in high-frequency
surveys.

The lower the frequency of the detection survey, the greater is the chance
of losing small sources via one of these mechanisms.  However, for a given
turnover frequency, this chance decreases with increasing redshift. A
sample selected at a fairly low frequency (such as tens of MHz) might miss
small low redshift sources which would be detected at higher redshift
because the observing frequency is likely to be less than the turnover
frequency. In order to quantify the significance of this effect in samples
selected at a frequency of 38 MHz, we considered that complete sample
which is a subset of the 3C sample (described in \S\ref{sec:samples3c})
defined by declination, $\delta > 60^{\circ}$. We then considered which of
the sources in this sample of 29 objects would {\em not} have been
included in a survey selected in the same sky area at the lower frequency
of 38 MHz, due to spectral turnover.

The limiting flux density of the 8C survey carried out at 38 MHz
[\cite{Ree90}; catalogue revised by \cite{Hal95}] is 1 Jy over most
of the survey area. We set a threshold of 25 Jy for our 38-MHz
selected sample since this is the flux density which an object on the
3C flux-limit at 178 MHz would have at 38 MHz if it had a spectral
index of 0.6 --- i.e.\ a spectral index which can plausibly arise
without any recourse to one of the spectral turnover mechanisms
discussed above.

Three bona fide sources from a sample of 29 objects would be lost because
of spectral turnover; relevant properties of these objects are shown in
Table~\ref{tab:rees}. Thus the deficit of small objects (none of which are
at high redshift) missing from a sample selected at 38 MHz rather than a
somewhat higher frequency e.g.\ 178 MHz is at the $\sim 10$\% level. We
therefore note that stronger linear size evolution might in principle be
found from surveys at 38 MHz because of the absence of small (low
redshift) sources.

\placetable{tab:rees}

Observing at a higher frequency will tend to detect a higher number of
small low-redshift sources which have a spectral turnover, since a
higher observing frequency is more likely to be above typical turnover
frequencies. Even with this, we suggest that a high frequency like 408
MHz might measure spuriously high linear size evolution, [since a
survey at 408 MHz selects objects at $z \sim 2$ at rest-frame 1 GHz],
since large and small low redshift sources will still make it into the
sample but the only high redshift sources which make it into the
sample are very short ones which are too young to have suffered
catastrophic inverse Compton losses to their lobes.

To investigate the effects of increasing the survey frequency on the
selection process, we considered the differences between the 6C sample
selected at 151 MHz (described in \S\ref{sec:samples6c}) and a sample
selected over the same region of sky but at 408 MHz. We were able to do
this using the sample of Allington-Smith (1982) which is roughly the same
region of sky as the 6C sample but selected at 408 MHz to be brighter than
1 Jy as measured in the Bologna (B2) survey.  Those sources which are only
selected at 408 MHz in this way are listed in Table~\ref{tab:AS}.  We have
redshift information for all but one (B2\,0927+35) of these objects hence
we could overlay the points from this sample on our various plots of the
four different source properties (luminosity, spectral index, redshift and
linear size; an example is presented in Figure~\ref{fig:da}). We found
that while selection of this sample at the higher frequency does result in
the inclusion of more objects which tend to have fairly small projected
linear sizes, these are over a range in redshift and we discern no
redshift systematic bias implied by the properties of the sources included
in the sample selected at 408 MHz rather than 151 MHz, at these
flux-limits.

\placetable{tab:AS}

\subsection{The use of one selection frequency}
\cite{Nee95} pointed out that comparing samples with very different
selection frequencies will find a spurious evolution in linear size,
if a faint high frequency (e.g.\ $\sim 1$ GHz) sample with the biases
described in \S\ref{sec:freq} were to be compared with a bright low
frequency (e.g.\ 0.1 GHz) sample without these biases. A study of the
evolution of radio sources based on samples selected at different
frequencies is not helpful because it amounts to a comparison between
the same types of objects at different parts of their radio spectra.

\subsection{Source type}

It should be clear that samples which are incomplete or inhomogeneous
should not be used for a study pertaining to evolutionary properties
because of the subtle and unsubtle biases which will be present. Even
``well-defined'' samples are not without their problems. For example, past
studies [e.g., \cite{Bar88}] have used samples comprised only of steep
spectrum quasars. The motivation for such an approach is understandable:
the spectroscopy is relatively easy hence good coverage out to high
redshift may be achieved. However, if we believe the unification by
orientation model [\cite{Sch87}; \cite{Bar89}] at some level (and thus
that for quasars the jet axis is at a small angle to the line of sight),
then the projected angular sizes which we measure will be heavily
foreshortened. One cannot begin to deproject the sizes unless one has very
large numbers of sources and a good idea what the mean and spread in the
transition angle are.  This problem is potentially compounded if quasars
are used which are not from complete samples but have optical selection
criteria as well, since this might preferentially bias towards boosted
objects which are pointing towards us (see Willott et al.\ 1998a).

In principle therefore, studies which have used only radio galaxies, which
are presumed to lie closer to the plane of the sky and therefore will have
a smaller difference between the projected size we measure and the
deprojected size we would wish to measure, might be a more reliable
indicator of the true dependences. In fact we chose to use all radio
galaxies and quasars in our samples which fulfilled the criteria discussed
in \S\ref{sec:measure}. Such a selection criterion should ensure a random
source orientation, allowing statistical deprojection of the linear size
information.

\subsection{Spectroscopy}

Many previous studies of samples of radio sources have required the use of
redshift estimators for a significant fraction of the sample (e.g. Dunlop
\& Peacock 1990). The most common estimator employs the $K$--band (2.2
$\mu$m) magnitude.  Use of this estimator is motivated by the small spread
in the near-infrared Hubble Diagram ($K$--$z$ relation) for radio galaxies
in the 3C sample (Lilly \& Longair 1984; Best, Longair and R\"{o}ttgering
1998). It is now known that simple extrapolation of the 3C relation to
higher redshifts and/or to fainter radio samples (which contain fewer
radio luminous sources at any given redshift) leads to both large
uncertainties in the estimated redshifts, and to systematic biases. At
redshifts of $z > 2$ the $K-$magnitude becomes contaminated by luminous
emission lines, and other sources of non-stellar radiation [Eales \&
Rawlings (1993, 1996); Simpson, Rawlings \& Lacy (1998)]. This is at least
partially responsible for a scatter of 3 magnitudes in the $K$--$z$
diagram at $z > 2$, and the little, if any, correlation between the
variables (Eales et al.\ 1993). This large scatter may also be influenced
by a genuine increase in the spread of stellar luminosities at early
cosmic epochs. Comparison of the $K$--$z$ relations of 3C and 6C samples
(Eales et al.\ 1997) reveals a significant difference in the sense that 3C
radio galaxies are about one magnitude brighter than the 6C galaxies at $z
\approx 1$.  Further studies (Best et al.\ 1998; Roche, Eales \& Rawlings
1998) suggest that this effect is caused by a (positive) correlation
between radio luminosity and the stellar luminosity of the host galaxy
which is not yet well understood.  To summarise, although it may now be
possible to construct more reliable single-colour redshift estimators
using $K$--band imaging of the combined 3C/6C/7C dataset, the use of
almost exclusively spectroscopic redshifts has been an essential
ingredient of the work described in this paper. In the very few cases
where we have been unable to secure spectroscopic redshifts
(\S\ref{sec:samples6c} and \S\ref{sec:samples7c}), we have estimated
redshifts by obtaining multi-colour (optical and near-infrared)
information and fitting spectral energy distributions (SEDs) with template
galaxy spectra (e.g. Dunlop \& Peacock 1993). This method is likely to be
robust whenever non-stellar contributions to the SED can be safely
neglected.  Evidence that this is the case comes from the lack of any
detectable emission lines in both optical and near-infrared spectroscopy
of the relevant objects.

\section{Samples used}
\label{sec:samples}
\subsection{3C}
\label{sec:samples3c}
The complete sample is that of the revised 3CR catalogue of Laing, Riley
\& Longair (1983), which contains 173 objects in an area of sky of 4.239
sr with $\delta \ge 10^{\circ}$ and $\vert b \vert \ge 10^{\circ}$ each
with a flux density\footnote{For the analysis described in this paper,
flux densities quoted by Laing et al.\ (1983) for all objects in this
sample were multiplied by 1.09 to put them on the same flux scale as
\cite{Baa77}.} greater than 10.9 Jy at 178 MHz\footnote{This roughly
corresponds to a flux-limit of 12 Jy at 151 MHz.}.  We excluded 3C\,231
(M\,82) since its radio emission is due to star-formation (Condon et al.\
1990) rather than an active galactic nucleus. We also excluded another two
objects, 3C\,345 and 3C\,454.3, because their emission at 178 MHz only
exceeds the flux-limit because of Doppler boosting of their core emission.
The final redshift of this sample, for 4C\,13.66, was obtained by Rawlings
et al.\ (1996b); the penultimate redshift, for 4C\,74.16, is 0.568 (R.A.\
Laing, {\em priv.\ comm.}). The spectroscopic redshift completeness for
this sample is thus 100\%.

\subsection{6C} 
\label{sec:samples6c}
The complete sample we used is based on Eales' (1985) sample, but
re-selected from the final versions of parts II and VI of the 6C survey at
151 MHz [Hales et al.\ (1988) and Hales et al.\ (1993) respectively].
This revised sample is for objects with $2.0 \le S_{151} < 3.93$ Jy (where
$S_{151}$ is the flux density in Jy at 151 MHz), with 08 20 30 $<$ RA
{\rm(B1950)} $<$ 13 01 30 and also with +34 01 00 $< \delta $ {\rm(B1950)}
$<$ +40 00 00.  The sky coverage of this sample is thus 0.102 sr.  Details
of the revised sample are in Blundell et al.\ (1998b) and redshifts are
in Rawlings et al.\ {\em (in prep.)}.  Structural classifications and
angular sizes were re-measured from maps by Law-Green et al.\ (1995),
Naundorf et al.\ (1992), Naundorf (1992) and Tansley (1997). The revised
complete sample would consist of 59 objects but one (6C1036+3616) is
formally excluded for a reason which is not statistically biassed, namely
because a bright star lies too close to its optical identification to
measure the redshift of the radio source.  Only one redshift for this
final sample of 58 sources is not spectroscopic giving 98\% spectroscopic
redshift completeness for the 6C sample.

\subsection{7C} 
\label{sec:samples7c}
The 7C survey was carried out with the Cambridge Low Frequency Synthesis
Telescope (CLFST) at 151 MHz [\cite{McG90}] at a resolution of 70 $\times$
70 cosec({\em Dec}) arcsec$^2$.  In this paper we use the 7C$-$I and
7C$-$II parts of the 7C Redshift Survey, which were chosen by Rossitter
(1987) to overlap with the 5C6 and 5C7 fields of Pearson \& Kus (1978),
respectively. The 7C 151-MHz data were subsequently re-analysed (and in
the case of the 5C6 field, re-observed) by Julia Riley. VLA data, where
available, were retrieved from the archive (by kind permission of Barry
Clark), or were recently obtained by us: full details of all these radio
observations and the sample selection are described by \cite{Blu98b}. The
7C$-$I region covers an area of 0.0061 sr and contains 37 sources with
$S_{151} \ge 0.51$ Jy. The 7C$-$II region covers an area of 0.0069 sr and
contains 40 sources with $S_{151} \ge 0.48$ Jy. All the sources in the
final sample have now been mapped with the VLA giving high-resolution
radio maps. We used these maps to deduce the structural information and
angular size of each object in the 7C sample.  Spectroscopic redshifts
have been obtained for 71 members of the 7C complete sample, thus
spectroscopic redshift completeness is 92\% (Willott et al.\ {\em in
prep.}).  For the remaining objects, their redshifts are estimated from 4
or 5 optical/near-infra red magnitudes as described in Rawlings et al.\
{\em (in prep.).} Only one source in the small sky area of the 7C sample
(3C200) is common to the 3C sample, and this source is just included as a
member of 3C for the purposes of the analysis presented in this paper.

We excluded one object, 5C7.230 (also known as 7C\,B082134.3+244829), from
the sample because its emission at 151 MHz only exceeds the flux-limit
because of Doppler boosting of its core emission (Blundell et al.\ 1998b).

\section{Classifications \& measurement techniques}
\label{sec:measure}

\subsection{Source structure\ldots}
\label{sec:structure}
Neeser et al.\ (1995) demonstrated the perils of trying to analyse the
changes in maximum angular extent of the `edge-darkened' FR\,I-class
[\cite{FR74}] radio sources with increasing redshift and attributed the
inclusion of such objects to the extremely strong linear size evolution
with redshift which has been reported in the past (e.g.\ \cite{OKW87}).
To avoid spurious correlations and to focus our analysis on the evolution
of one physically distinct class of object out to very high redshift,
i.e., to study the same physical phenomenon at different redshifts in
order to compare like with like, we investigated the radio spectra and
structure of the objects in our samples.  One analysis which is part of
this paper --- the linear size evolution of classical double radio sources
--- amounts to considering how the separation of working surfaces at the
ends of oppositely directed radio jets depends on luminosity and/or
redshift.  For this reason we excluded from our analysis radio sources of
the FR\,I structural type. We regard objects of this {\em morphological}
type as the inevitable manifestation of low {\em jet-power} objects, where
the jet-power is insufficient to form a supersonic shock, given its
environment, where the jet impinges on the inter-galactic medium [see
e.g.\ Falle (1991)]. The 6C sample and 7C sample selected to be a factor
of 6 and a factor of nearly 25 respectively fainter than 3C, in fact
contain relatively few objects of FR\,I structure. This arises because in
the relatively small sky areas of 6C and 7C few nearby bright but
low-luminosity objects are found. Note that luminosity is not a
fundamental quantity in these objects in the way that the power
transported by the jet is. Distant low-luminosity objects which are found
will not be the exact counterparts of low-redshift objects of the same
luminosity: they must have higher jet power to compensate for the fact
that for an object of redshift $z$, inverse Compton losses increase as $(1
+ z)^{4}$. Put another way, an object with a given jet power observed in
an environment with a given density profile at high redshift will have a
much lower luminosity at, say, 80\% of the way through its life than a
lower redshift object with otherwise the same parameters at the same
point.  [Note that this is {\em not} an observational error: the selection
is made at a sufficiently low frequency and resolution that there is no
question of missing sources because of undersampling smooth extended
flux.]

For this reason we do not base the criteria of which objects should be
included in our analysis on their {\em luminosity}, but on their {\em
structure}.

How then to identify whether a given source is type FR\,I or FR\,II?  We
deemed that an object was a classical double (FR\,II) if at least one
hotspot was at the leading edge of a working surface. In practice this
meant a bright, compact feature with a sharp discontinuity in surface
brightness was at the leading edge of emission expanding outwards from the
core or location of the optical identification. We did not require that
this was the case on {\em both} sides of the core [see e.g., 5C6.62 also
known as 7C\,B021029.7+325402 (Blundell et al.\ 1998b)] to allow for the
fact that although a working surface can exist, it can have a slightly
weird appearance due to e.g., Doppler suppression, but the presence of one
such hotspot is a good indicator that in the environment in which a given
source is located, the jet-power is sufficiently high that a classical
double radio source will be manifested. 

In some objects there was clear evidence of emission beyond a possible
working surface, which had decreasing surface brightness with increasing
distance from the core. We considered these to be a somewhat different
physical phenomenon from the `working surface + backflow towards the core'
type object, and called them FR\,Is and excluded them from our analyses.
For those intermediate cases where we could not be totally sure that an
object should be assigned to the FR\,II category, we considered by
examination of their integrated radio spectra (see \S\ref{sec:spectra})
whether there was any evidence of trailing outflow with increasing
distance from the core, as might be evidenced by a greater flux from a
survey measurement (e.g., from VLA D-array or single dish measurements)
than from our higher resolution (VLA A- or B-array) measurements. Thus,
objects which do not apparently contain any clearly defined working
surface, but in which we {\em believe} all the flux is contained within a
well-defined measureable extent, are classed as having defined angular
sizes (DAs). We believe that most, if not all, the angular sizes measured
for these objects do correspond to the separation of working surfaces at
the ends of oppositely directed radio jets, though we cannot eliminate the
possibility that some might be FR\,Is.

Our angular size measurements were our best attempt at summing the
separations of the two working surfaces from the core; in the absence
of a radio core we took our best estimate of the separations of the
working surfaces from one another.

\subsection{\ldots and Spectra}
\label{sec:spectra}
Fitting of the radio spectra for the members of all three samples was made
possible by consideration of data from many radio surveys\footnote{In
addition to the original survey data at 151 or 178 MHz, we obtained from
the literature flux densities at 365 MHz, 408 MHz, 1.4 GHz and 4.86 GHz,
and in the case of 3C we also obtained flux densities at 10, 22, 38 and 85
MHz, chiefly from Laing \& Peacock (1980).}, and in the case of the 7C
sample in combination with data from our own VLA maps as tabulated in
\cite{Blu98b}.

We fitted the flux density data using a Bayesian polynomial regression
analysis which assessed the posterior probability density function ({\sc
pdf}) for the required order of a polynomial fit (e.g.\ Sivia 1996); we
chose $x = \log_{10} (\nu/\rm MHz)$ as the independent variable and
expanded the dependent variable $y = \log_{10} S_{\nu} = \sum_{r=0}^{N}
a_{r} x^{r}$.  In a number of cases the peak in the {\sc pdf} led us to
prefer a first-order polynomial, and the gradient of the fit ($a_{1}$) was
multiplied by $-1$ to give the spectral index.  In many cases the {\sc
pdf} led us to prefer a second-order fit, implying significant curvature
in the radio spectrum: in these cases we calculated the curvature $\beta =
- 2 a_{2}$, and hence the spectral index at (rest frame) 151 MHz which
equals $-a_{1} + \beta\log_{10}(151/(1. + z)).$ In no case was a higher
order fit preferred.

With a fitted functional form for the spectrum of each object, we
interpolated (in the case of 3C) or extrapolated (in the case of 6C and
7C) to derive the flux densities at rest-frame 151 MHz and together with
the redshifts we derived the luminosities in ${\rm W}\,{\rm Hz}^{-1}\,{\rm
sr}^{-1}$ in a number of different assumed cosmologies, according to the
formulae presented in Carroll, Press \& Turner (1992).

\section{Results}
\label{sec:results}
We derived all plots in the following three assumed cosmologies: (1)
$\Omega_{\rm M} = 0$ and $\Omega_\Lambda = 0$, (2) $\Omega_{\rm M} = 1$
and $\Omega_\Lambda = 0$ and (3) $\Omega_{\rm M} = 0.1$ and
$\Omega_\Lambda = 0.9$; similar dependences were seen in each case.  We
also considered separately just the class of certain FR\,IIs and also 
the combined set of FR\,IIs and DAs (see \S\ref{sec:structure}). No
significant difference was found. The plots we present here are for the
combined set of FR\,IIs and DAs and, except for Figure~\ref{fig:dz3}, an
assumed cosmology with $\Omega_{\rm M} = 1$ and $\Omega_\Lambda = 0$ but
the statistical analysis presents the numbers for all three cosmologies we
considered (see \S\ref{sec:stats}).

\subsection{Plots of source dependences}
\label{sec:depend}

\begin{figure}[!t]
\begin{picture}(50,170)(0,0)
\put(-56,-230){\includegraphics{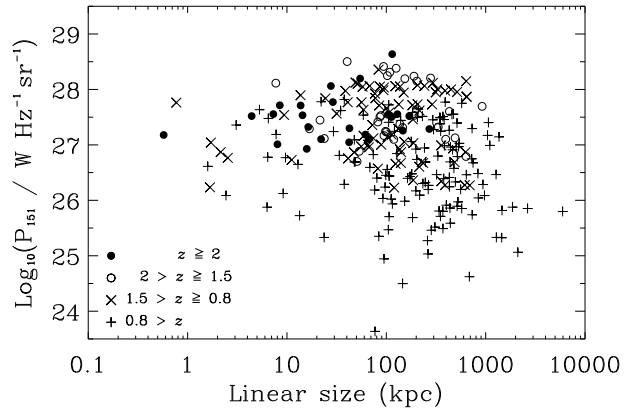}}
\end{picture}
{\caption[junk] {\label{fig:pd} This graph shows for all members of the
complete samples included in our analysis their luminosities plotted
against their projected linear sizes.  These properties are calculated
assuming $\Omega_{\rm M} = 1$ and $\Omega_\Lambda = 0$. Four different
redshift bins are indicated by symbol type.}}
\end{figure}

Figure~\ref{fig:pd} shows the luminosities {\em vs} projected linear sizes
(the $P$--$D$ plane) of the radio sources. No striking dependence of the
luminosity of an object on its linear size is seen, though there is a
deficit of small, low-power objects in the lower-left quadrant of the
$P$--$D$ plane, and also of large, high-power objects in the upper-right
quadrant of the $P$--$D$ plane.  It can be seen that the region most
densely occupied by the solid circles (corresponding to objects with $z$
\gtsim\ 2) is displaced to the left [smaller projected linear sizes] of
the region occupied by the main density of plus signs (corresponding to
objects with $z < 0.8$), indicating a distinct anti-correlation of
linear-size with redshift.  

\begin{figure}[!t]
\begin{picture}(50,475)(0,0)
\put(-47,105){\includegraphics{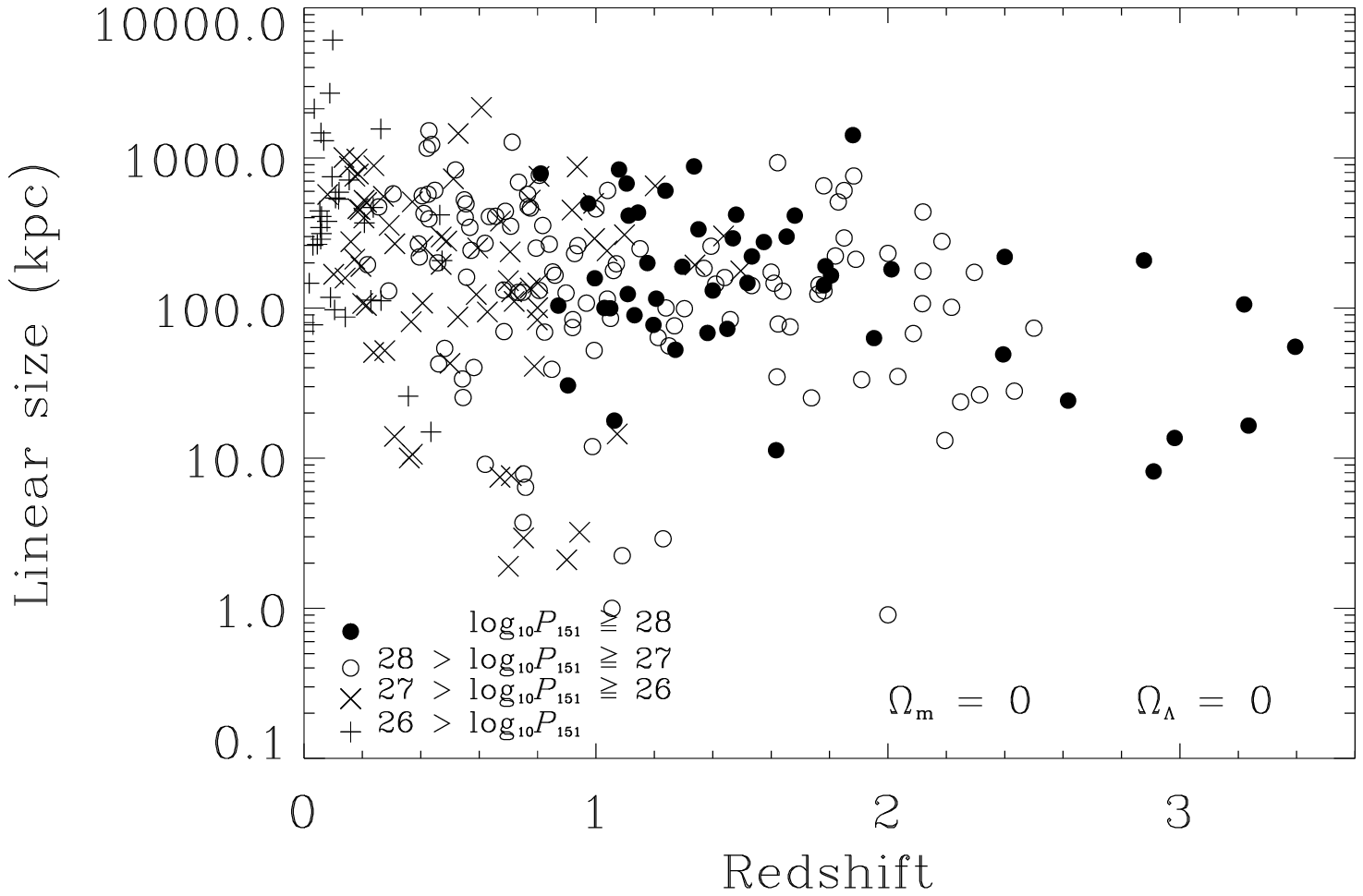}}
\put(-47,-52.5){\includegraphics{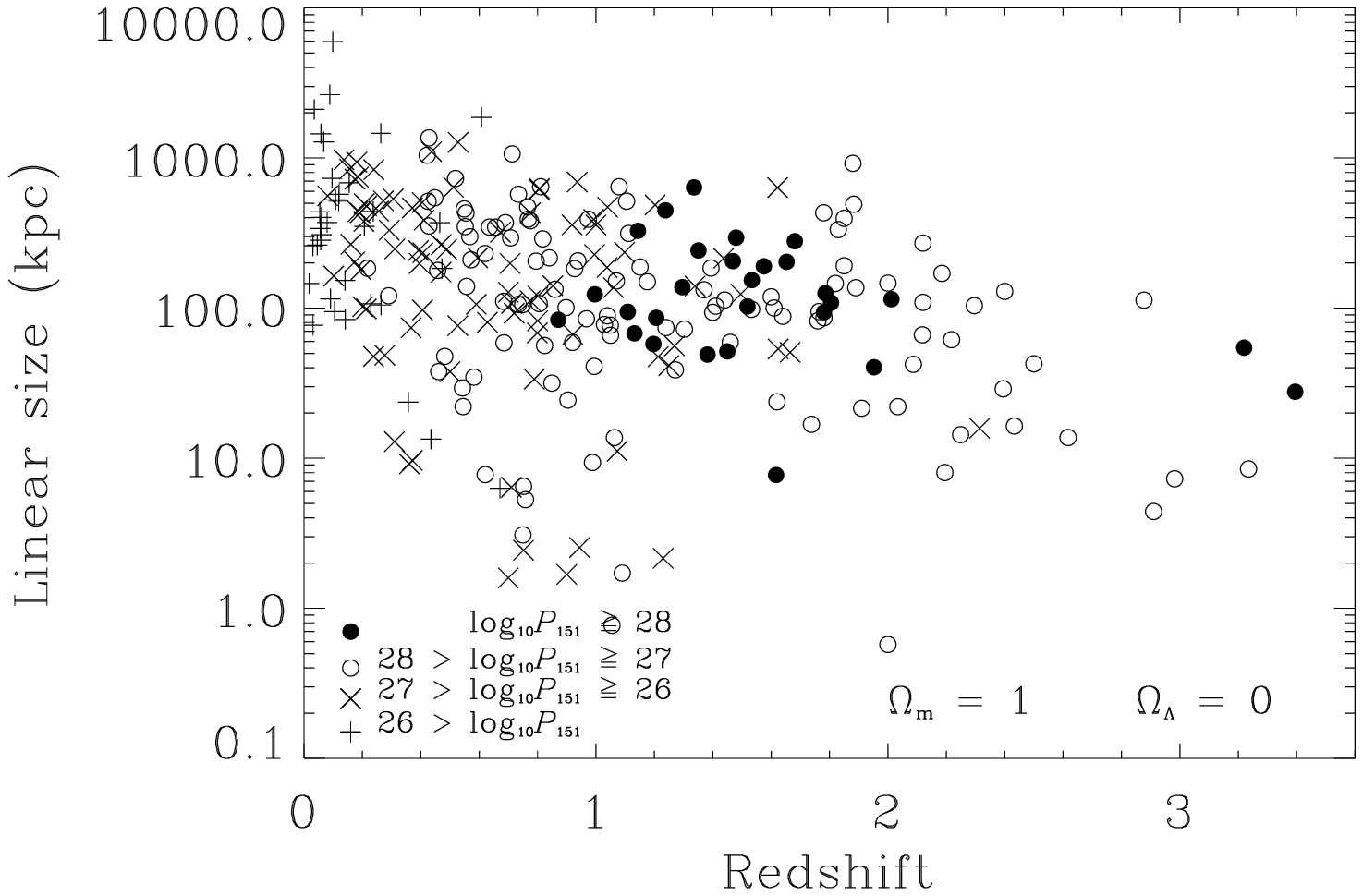}}
\put(-47,-210){\includegraphics{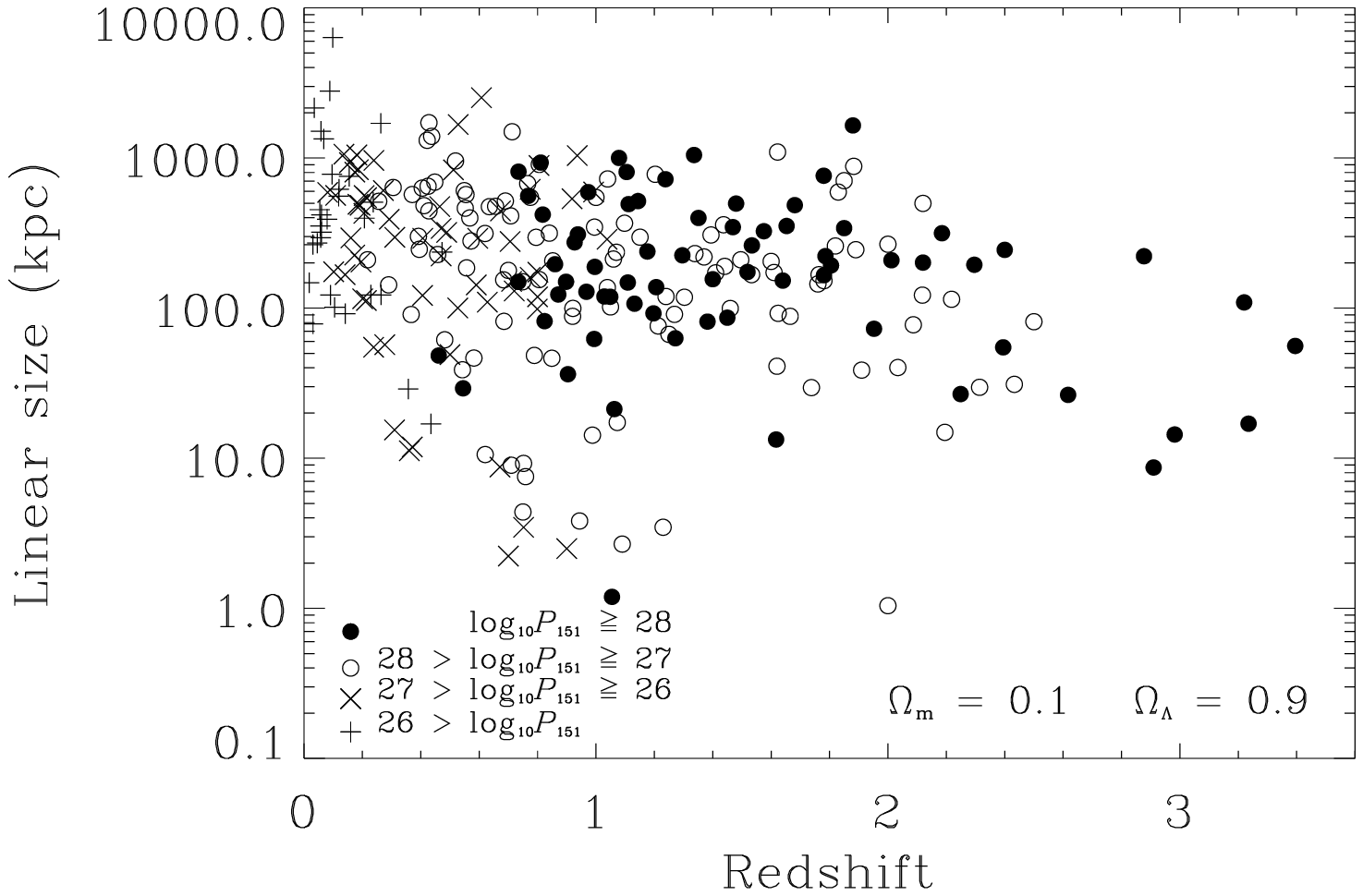}}
\end{picture}
{\caption[junk] {\label{fig:dz3} These graphs show for all members of the
complete samples included in our analysis their projected linear sizes
against their redshifts, with these properties calculated for three
different assumed cosmologies. The luminosity bin an object falls into is
denoted by symbol, with the same threshold values in each case. }}
\end{figure}

Figure~\ref{fig:dz3} shows more directly that sources at higher
redshift tend to have shorter projected linear sizes. This trend is
seen throughout the redshift range, and is seen in each of our three
considered cosmologies. The symbol type in this figure delimits the
luminosity bin an object is in. In the middle plot where the assumed
cosmology is $\Omega_{\rm M} = 1$ and $\Omega_\Lambda = 0$, the most
powerful sources are not as powerful as in the cosmologies with low
$\Omega_{\rm M}$. This plot demonstrates that in breaking the
degeneracy between luminosity and redshift, above a redshift of 2 the
tendency for the sources to be smaller seems to be independent of the
fact that the majority of these sources are not in the highest power
bin. 

\begin{figure}[!t]
\begin{picture}(50,300)(0,0)
\put(-40,-50){\includegraphics{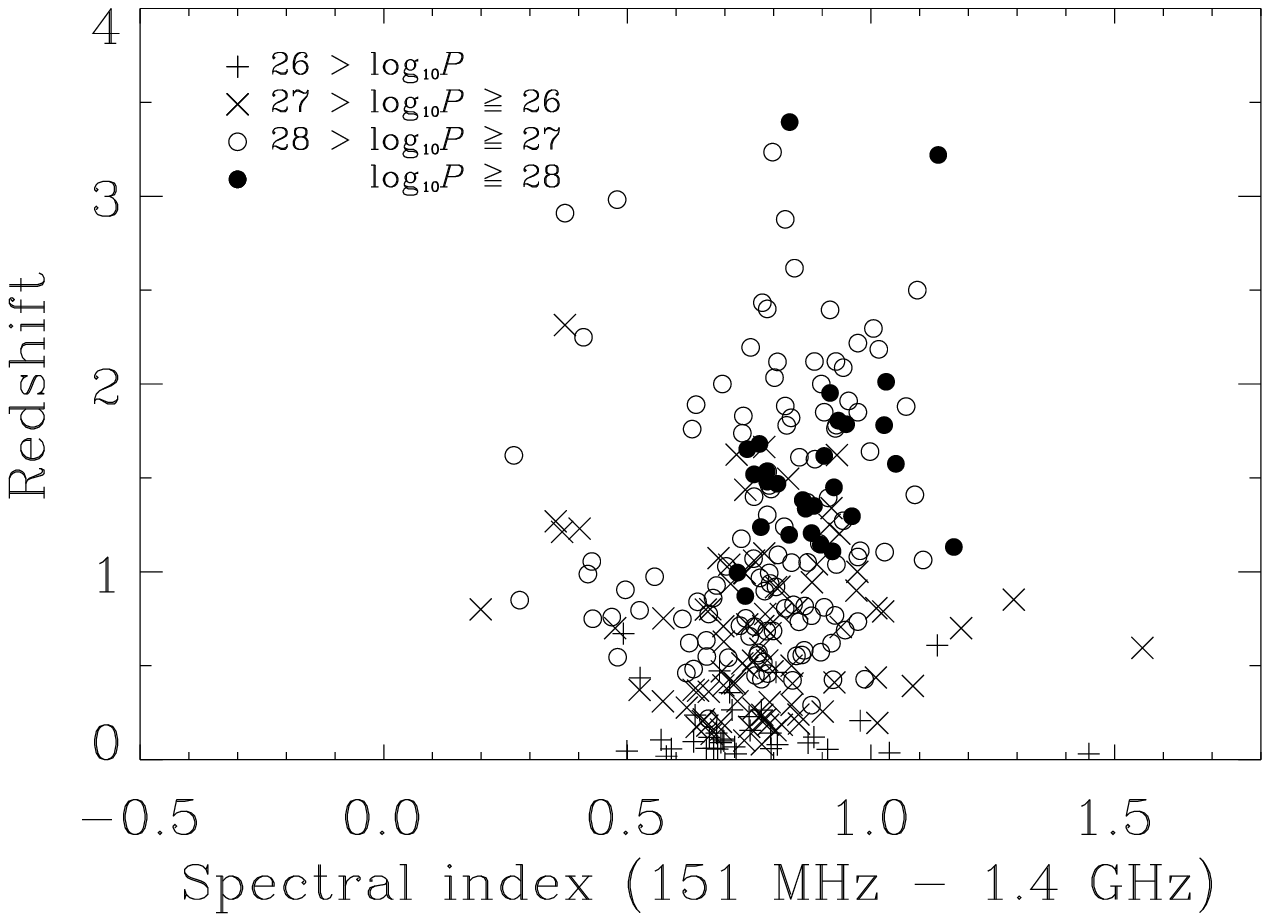}}
\put(-40,-195){\includegraphics{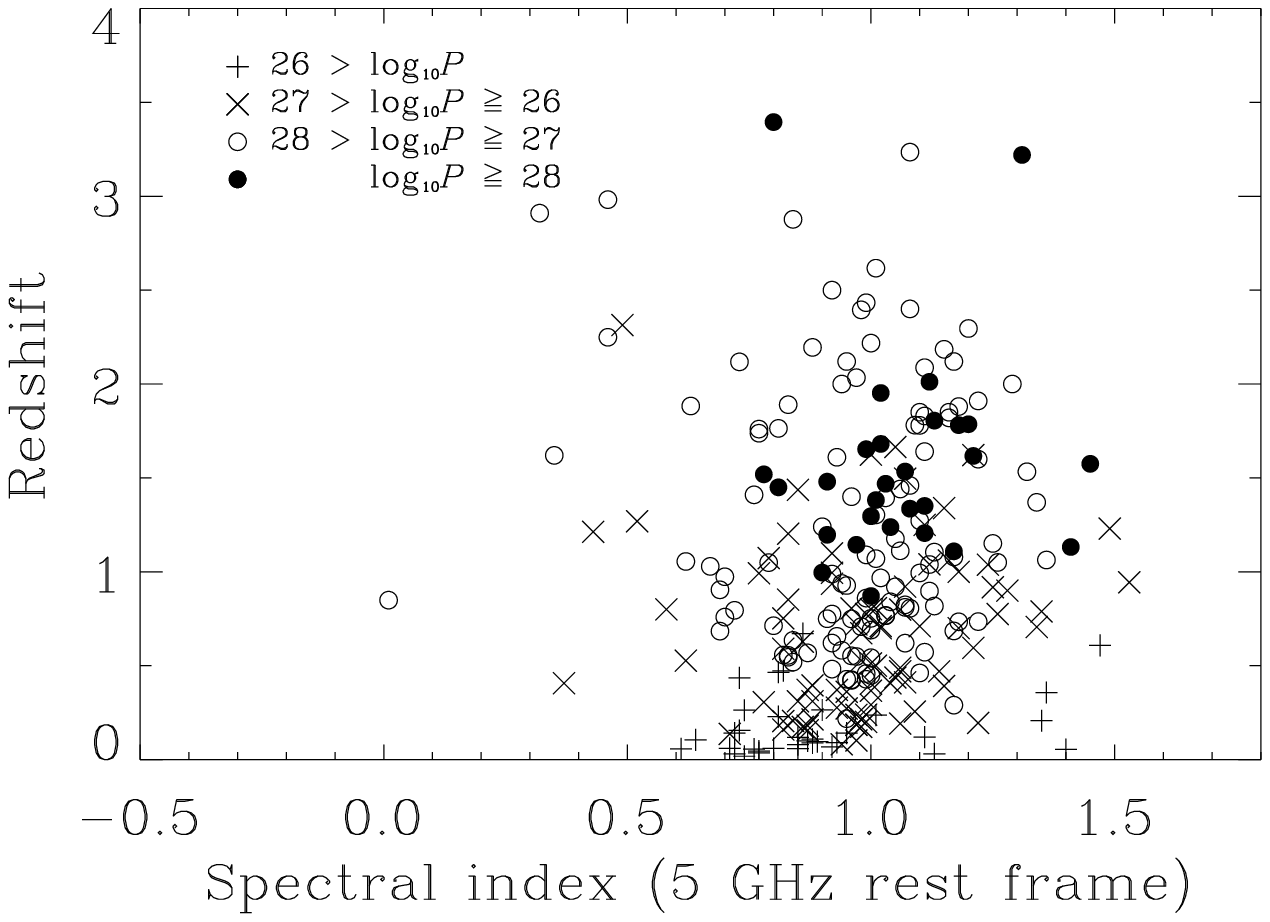}}
\end{picture}
{\caption[junk] {\label{fig:za1} These graphs show for all members of
the complete samples included in our analysis redshift against
spectral index measured in two different ways. The top panel shows the
spectral indices before fitting for spectral curvature and
$K$--correction and assuming a single power-law between 151 MHz and
1.4 GHz while the lower panel shows spectral indices evaluated at 5
GHz in the rest frame of each source. Source luminosities are
indicated by symbol-type. }}
\end{figure}

\begin{figure}[!t]
\begin{picture}(50,300)(0,0)
\put(-40,-50){\includegraphics{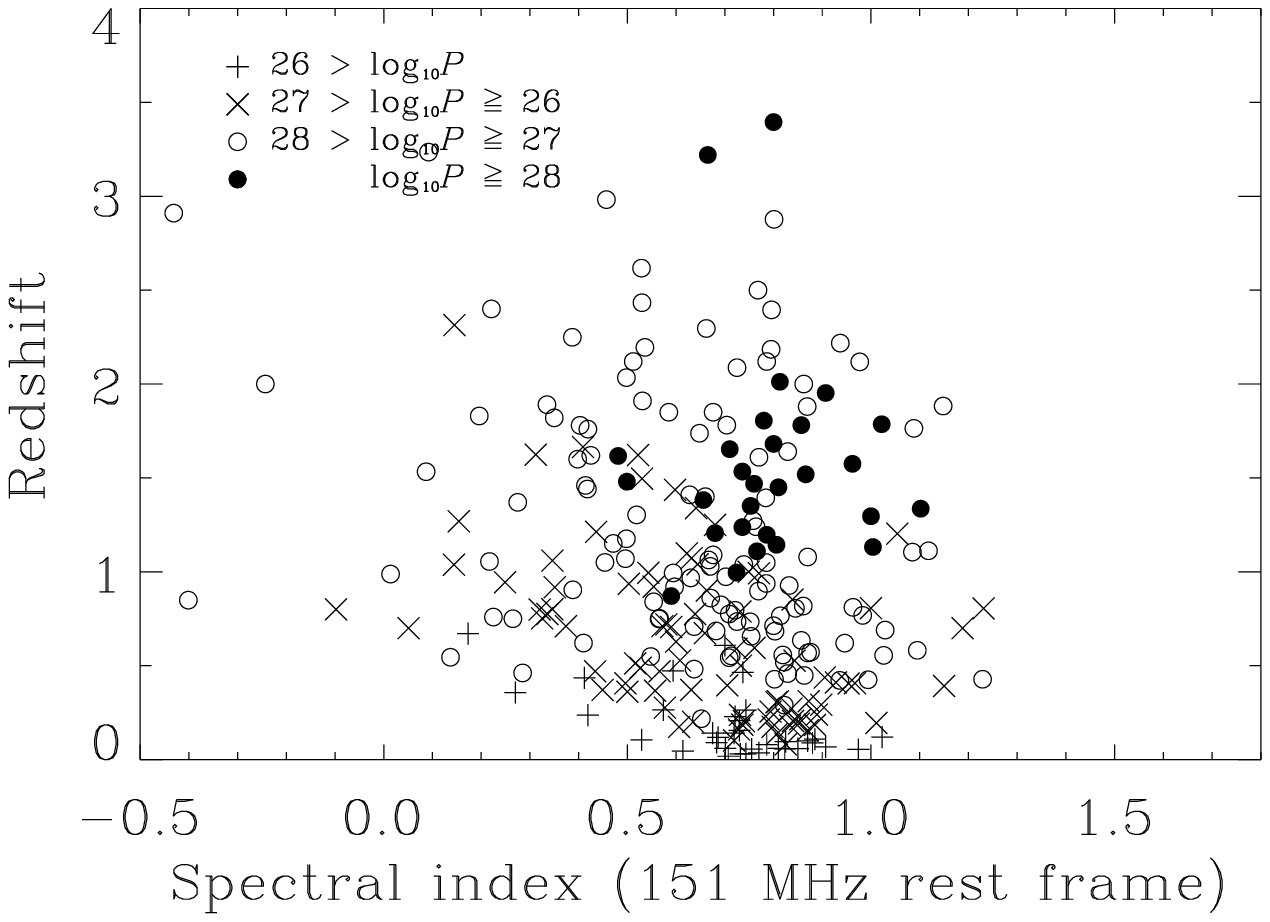}}
\put(-40,-195){\includegraphics{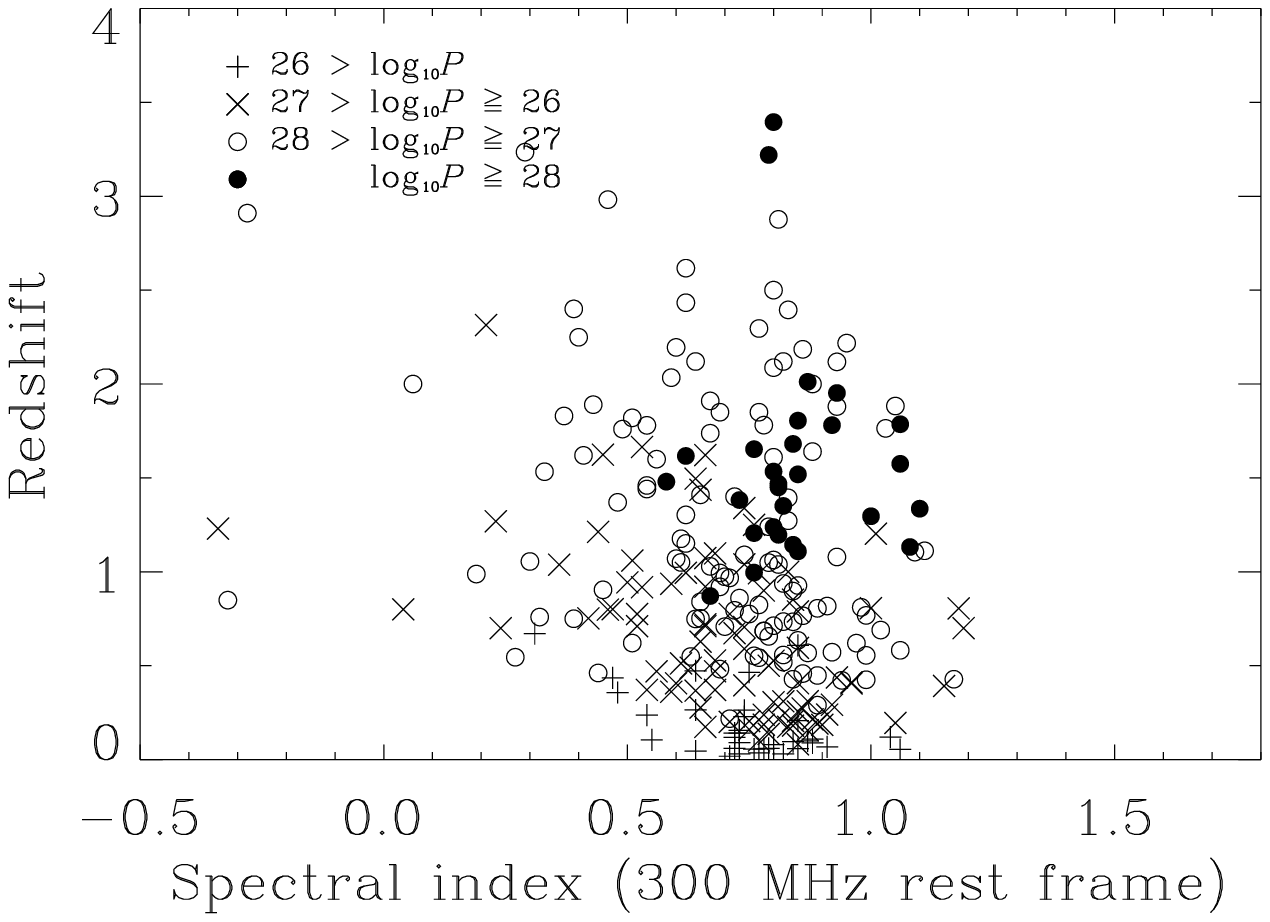}}
\end{picture}
{\caption[junk] {\label{fig:za2} These graphs show for all members of
the complete samples included in our analysis their redshifts against
spectral indices measured in two different ways, in both cases fitting
for spectral curvature and $K$--correction. The upper panel has the
spectral indices evaluated at 151 MHz in the rest frame while the
lower panel shows the spectral indices evaluated at 300 MHz in the
rest frame of each source. Source luminosities are indicated by
symbol-type. }}
\end{figure}

The operational definition of spectral index used in the past has been to
approximate the radio spectrum of a source by a single power law and to
derive $\alpha$ as $-\partial \log S_\nu/\partial\log \nu$, invariably in
the observed frame, using flux density measurements at two frequencies.
When this usage of spectral index is employed, there is a significant
spectral index dependence on redshift (see Figure~\ref{fig:za1}) which has
been exploited (Chambers, Miley \& van Breugel 1990; Blundell et al 1998a)
to find the most distant radio galaxies. This precise technique was used
to find the most distant radio galaxy currently known (Rawlings et al.\
1996a).  Once the curvature and redshifting of the spectra
(``$K$-correction'') are taken account of, this dependence of spectral
index on redshift is dramatically different, in a way which depends on the
frequency at which the spectral index is evaluated.  When the spectral
index is evaluated at 5 GHz rest-frame, then a rather strong dependence on
redshift is seen (see Figure~\ref{fig:za1}).  This arises because the GHz
part of the spectrum is where inverse Compton losses (as well as other
loss mechanisms) are most manifest.  If spectral indices are evaluated in
the rest frame at a frequency of a few hundred MHz, this strong dependence
on redshift diminishes (see Figure~\ref{fig:za2}).

\begin{figure}[!t]
\begin{picture}(50,190)(0,0)
\put(-55,-230){\includegraphics{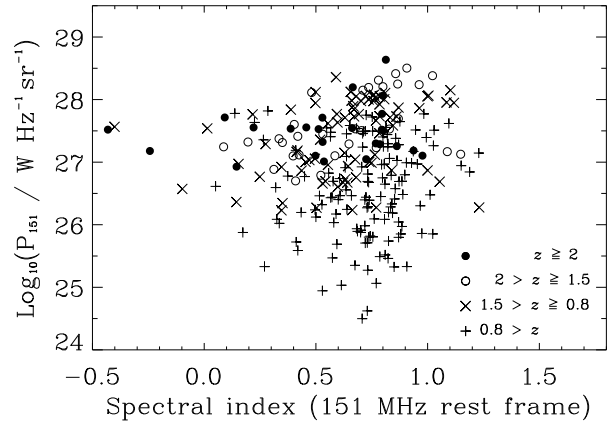}}
\end{picture}
{\caption[junk] {\label{fig:pa} This graph shows for all members of
the complete samples included in our analysis their luminosities
plotted against their spectral indices each evaluated at rest-frame
151 MHz. The redshift bins are indicated by symbol type.}}
\end{figure}

Figure~\ref{fig:pa} shows a plot of the luminosity of each object
against its spectral index, each evaluated at rest-frame 151 MHz.
There is a clear trend for the most powerful objects to have the
highest spectral indices. 

At a frequency as low as rest-frame 151 MHz, the spectral index is
informing us on the energy index of the synchrotron particles as injected
into the lobes.  We hereafter refer to this power-law exponent of the
energy distribution of particles injected into the lobes as the {\em
`injection index'}, reserving the term {\em `acceleration index'} for the
exponent of the energy distribution to which particles in the hotspot may
accelerated [and described by a model such as diffusive shock acceleration
e.g.\ Bell (1987)].  We explore in \S\ref{sec:injection} how this
injection index may depend in a simple way on the intrinsic jet-power of
an object and hence how such an injection-index--luminosity dependence can
arise.

\begin{figure}[!t]
\begin{picture}(50,370)(0,0)
\put(-50,-45){\includegraphics{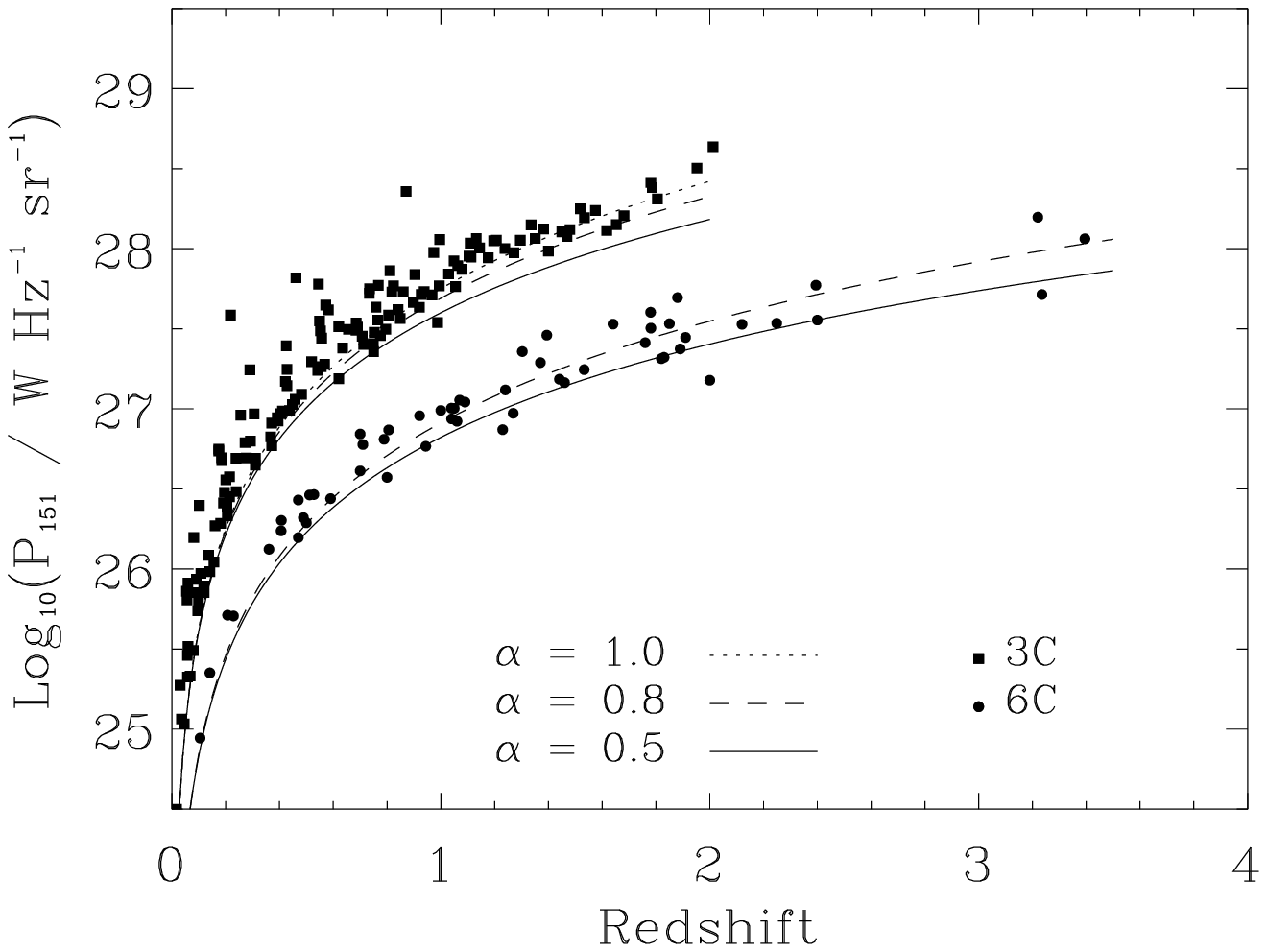}}
\put(-50,-225){\includegraphics{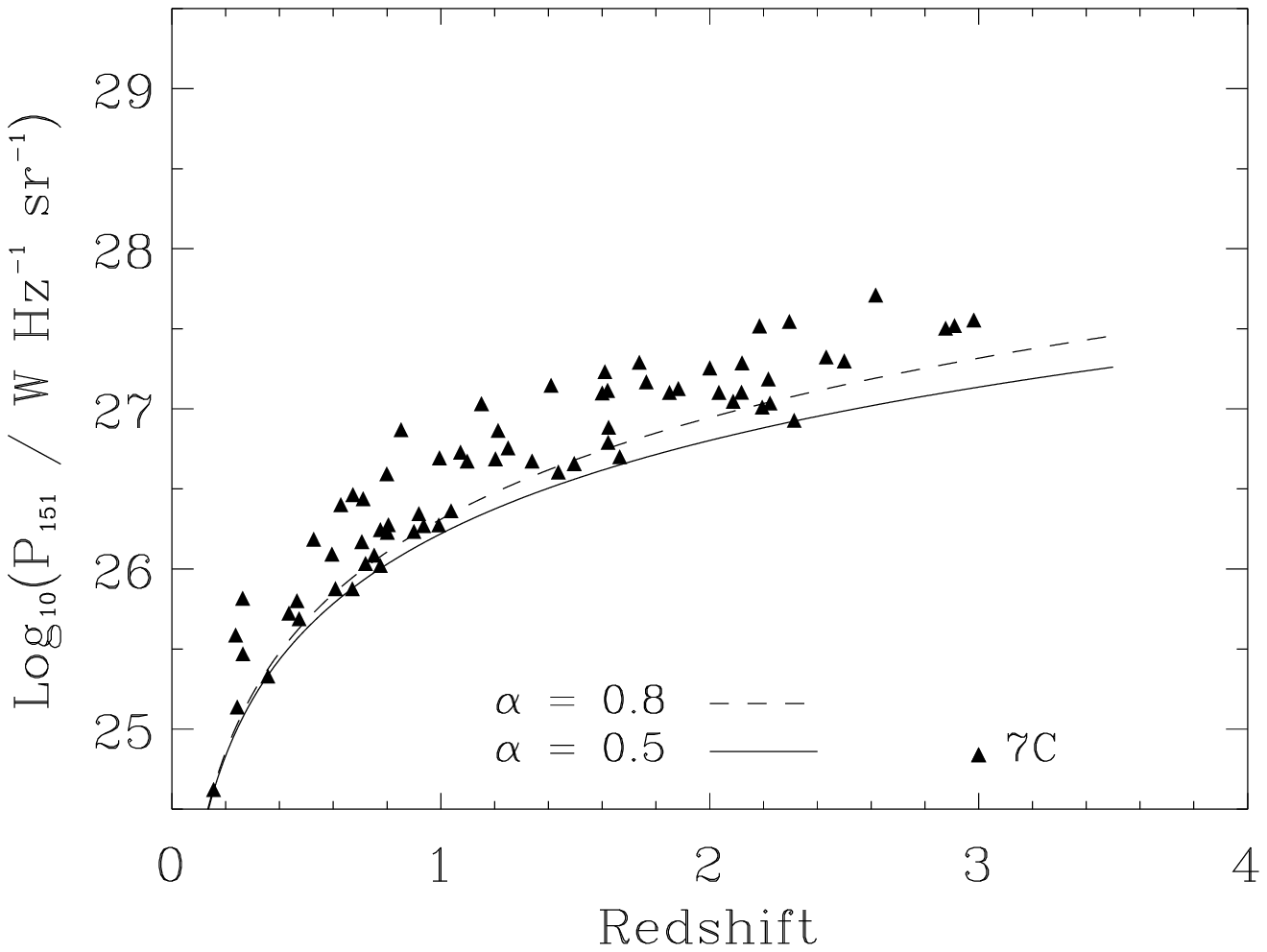}}
\end{picture}
{\caption[junk] {\label{fig:pzlim} These graphs show for all members of
the complete samples included in our analysis their luminosities at
rest-frame 151 MHz against redshifts. The top panel shows the $P$--$z$
plane for the 3C and 6C samples; the lower panel shows the $P$--$z$
plane for the 7C sample.}}
\end{figure}

A neat manifestation of the $P$--$\alpha$ correlation is seen in the
luminosity--redshift plane. Figure~\ref{fig:pzlim} shows lines which
trace the flux-limits for 3C, 6C and 7C, i.e.\ 12, 2 and 0.5 Jy
respectively. The solid black lines in each case correspond to the
minimum luminosity which a source at a given redshift may have, and
yet still be included in a given sample if its spectral index is 0.5;
other line-styles indicate the influence of other spectral indices on
the shape of the flux-limit on the $P$--$z$ plane. It can be seen that
those 3C sources with luminosities above $\sim 10^{27.5} {\rm
W\,Hz^{-1}\,sr^{-1}}$ all avoid the region between the flux-limit line
for a spectral index of 0.5 and that for a spectral index of 0.8. For
those 3C sources which have rather lower luminosities, it is
impossible to distinguish any preferred flux-limit line. In this lower
luminosity regime the fainter samples do not show any such avoidance
of the flux-limit line given by a spectral index of 0.5. 

\begin{figure}[!t]
\begin{picture}(50,170)(0,0)
\put(-48,-210){\includegraphics{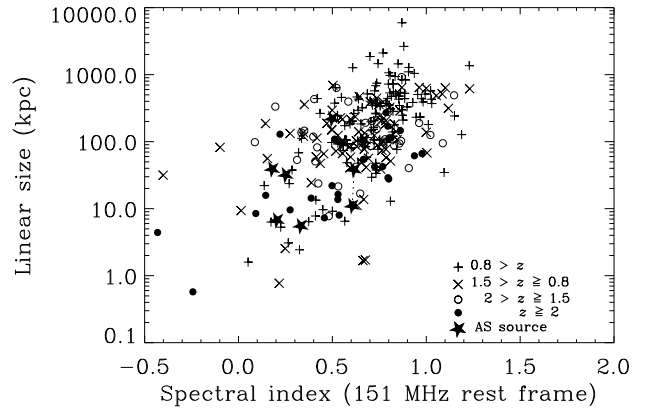}}
\end{picture}
{\caption[junk] {\label{fig:da} This is a plot of projected linear size
against spectral index evaluated at 151 MHz rest-frame for all members of
the complete samples. We also overlay those sources which would have been
members of the 6C complete sample had the selection frequency been 408 MHz
[taken from Allington-Smith (1982); see \S\ref{sec:freq}]. The redshifts
of the other sources are indicated by symbol type.}}
\end{figure}

Inspection of the $D$--$\alpha$ plane depicted in Figure~\ref{fig:da}
shows a correlation of projected linear size with spectral index, although
it also suggests that sources with even quite large linear sizes may not
necessarily have very steep spectra at rest-frame 151 MHz. It is possible
that some of these spectra are flattened by free-free absorption from
intervening material.  Taylor \& Perley (1992) present a study of 3C295
which has a (complex) turnover in the spectrum below 100 MHz; they
attribute this to thermal absorption from multi-component ISM gas at the
cluster.

\subsection{Disentangling the different dependences: statistics}
\label{sec:stats}

We display in Tables~\ref{tab:fr23stats} and \ref{tab:fr2stats} the
results of the statistical analysis using the 4-way Spearman partial-rank
correlation coefficient described by \cite{Mac82}, to indicate the true
dependences of the different source properties.  In these Tables, $r_{ab
\vert cd}$ is the correlation coefficient between $a$ and $b$ when $c$ and
$d$ are held constant.  We repeat this analysis for the same three assumed
cosmologies, {\em (i)} $\Omega_{\rm M} = 0$ and $\Omega_\Lambda = 0$, {\em
(ii)} $\Omega_{\rm M} = 1$ and $\Omega_\Lambda = 0$ and {\em (iii)}
$\Omega_{\rm M} = 0.1$ and $\Omega_\Lambda = 0.9$.  The dependences
largely persist in all the cosmologies; we quantify and discuss the
meaning of the dependence on cosmology of the strength of the
anti-correlation of linear size with redshift in a forthcoming paper.

\placetable{tab:fr23stats}

\placetable{tab:fr2stats}

\subsection{A cautionary tale: using just one flux-limited sample}

To demonstrate how the tight correlation between luminosity and
redshift in a single flux-limited sample can obfuscate the true
dependences between these source properties and linear size and
spectral index, we show in Table~\ref{tab:3cstats} the results of
the statistical analysis carried out above on only the 3C sample
(\S\ref{sec:samples3c}).  Although the relationship between linear
size and spectral index is still discernible, (indicating that it is a
real effect independent of luminosity or redshift) the extremely
strong correlation between luminosity and redshift swamps the other
dependences which do emerge when a number of complete samples with
different flux-limits are taken together.

\placetable{tab:3cstats}

\section{Parameterisation of the linear sizes}
\label{sec:param}
Although we later demonstrate that there is no direct single physical
mechanism by which linear size depends on redshift, for comparison of
our data with work previously done by others we calculate the
parametric dependence of the linear size on redshift and on luminosity
with the form:

\begin{equation}
\label{eq:nm}
D \propto (1 + z)^{-n}P^{m}.
\end{equation}

Following Neeser et al.\ (1995), values of the exponent $n$ are found by
multiplying individual source sizes by $(1 + z)^{n}$ until the
correlation coefficient (hence its associated significance) goes to
zero. A formal error-bar is derived by considering the values of $n$ at
which the significance of the correlation coefficient equals unity. The
exponent $m$ was found in a similar manner.  In Table~\ref{tab:exps} we
list the various exponents thus derived. The notation we use is that
$n[r_{{ab}\vert{cd}}]$ (or $m[r_{{ab}\vert{cd}}]$) is the exponent
required to obtain a correlation coefficient of zero between quantities
$a$ and $b$ when $c$ and $d$ are held constant. For comparison with
previous studies, which have not included the spectral index dependence,
we also present $n[r_{{ab}\vert{c}}]$ and $m[r_{{ab}\vert{c}}]$ which are
the exponents required in equation~\ref{eq:nm} to make the 3-way
correlation coefficient between $a$ and $b$ equal zero, when $c$ is held
constant.

\placetable{tab:exps}

\section{Comparison with previous work}
\label{sec:compare}
\subsection{The variation of linear size}
It has never been in dispute that the linear sizes of classical double
radio sources are larger in the local Universe, although this has often
come from studies restricted to very low redshifts. Indeed, this together
with the fact that the angular sizes of classical double radio sources are
relatively easy to measure led some to hope that it might be possible to
use these objects as `standard rods' to constrain cosmic geometry (see
e.g.\ \cite{Uba95}), if it could be demonstrated that the linear size of
an object is not influenced by its redshift.  However, past reports of the
true dependence of the linear sizes of radio sources have been equivocal
as a brief survey of the literature indicates:

\begin{itemize}
\item \cite{Kap87} found a strong evolution in linear sizes of the form $D
\propto (1 + z)^{-3 \pm 0.5}$ for a cosmology with $\Omega_{\rm M} = 1$
and $\Omega_\Lambda = 0$, but at constant redshift he found that $D
\propto P^{0.3 \pm 0.1}$. This was for a set of FR II radio galaxies from
four samples: {\em (i)} The BDFL sample (Bridle et al.\ 1972) selected at
1.4 GHz; these galaxies had almost complete spectroscopic information and
have $0.08 < z < 0.2$. {\em (ii)} The MC$-$I and MC$-$II samples
(Machalski \& Condon 1983; Machalski \& Condon 1985) selected at 1.4
GHz. Out of these two samples only 4 galaxies had spectroscopic redshifts;
redshift estimates from $R$-band magnitudes were used and redshift limits
inferred for the two samples were 0.4 and 0.6 respectively. {\em (iii)}
The LBDS sample which was selected at 1.4 GHz by \cite{Win84}. For the
analysis Kapahi presented, all sources in this sample were deemed to have
a redshift of 1.25.  {\em (iv)} Galaxies from the revised 3CR sample which
was selected at 178 MHz by Laing et al.\ (1983).

\item \cite{Bar88} found a weaker, and more explicitly uncertain,
dependence of linear size on redshift and essentially no dependence on
luminosity ($D \propto (1 + z)^{-1.5 \pm 1.4}P^{-0.03 \pm 0.3}$) from
a sample of 134 steep spectrum radio quasars from \cite{Hew80} known
to exhibit radio emission. In the context of finding for their sample
that distant quasars have a more bent and distorted appearance than
nearby ones, they postulated that the ambient medium, interacting with
the radio jets, played an important r\^{o}le in governing linear size
evolution.

\item \cite{Sin93} found for radio galaxies in his chosen sample that
there was a strong anti-correlation of linear size with redshift but a
correlation with luminosity [$D \propto (1 + z)^{-3.0}P^{+0.35}$], while
for quasars an anti-correlation of linear size and luminosity was found as
was marginal evidence for a dependence of linear size on redshift. His
sample consisted of 789 radio sources, taken from 3CRR selected at 178 MHz
by Laing et al.\ (1983), the 1-Jy sample selected at 408 MHz
(Allington-Smith et al.\ 1988; Lilly 1989), and the Molonglo (McCarthy et
al.\ 1991) and B3 samples (Vigotti et al.\ 1989; McCarthy 1991) which were
selected at 408 MHz.

\item \cite{OKW87} found that the dependence of linear size on luminosity
and on redshift, assuming that $\Omega_{\rm M} = 0$, was described by: $D
\propto (1 + z)^{-3.3 \pm 0.5}P^{0.3 \pm 0.05}$.  Their analysis was based
on 163 radio galaxies from a number of different samples which included
FR\,I type radio galaxies.

\item \cite{Nil93} found from a sample of 540 FR\,II double sources
taken from the literature that for (high power) radio galaxies and
quasars linear size and power are anti-correlated; they did not find
any evidence for linear size evolution.

\item The first study to use only complete samples, and using almost
entirely spectroscopic redshifts, was that by \cite{Nee95}. Using the
revised 3CR and 6C samples [Laing et al.\ (1983); \cite{Eal85}],
described in \S\ref{sec:samples3c} and \S\ref{sec:samples6c}
respectively, they found $D \propto (1 + z)^{-1.2 \pm 0.5}$ for
$\Omega_{\rm M} = 0$ and $\Omega_\Lambda = 0$ and $D \propto (1 +
z)^{-1.7 \pm 0.5}$ for $\Omega_{\rm M} = 1$ and $\Omega_\Lambda =
0$. They found no intrinsic correlation between size and radio
luminosity. \cite{Nee95} suggested the presence of dense line-emitting
gas as being responsible for the observed linear size evolution. They
also pointed out that such an effect could be achieved if radio jets
were switched on for shorter periods of time at high redshift than at
low redshift.
 
\item A recent study (Buchalter et al.\ 1998) was based on the 1.4 GHz
FIRST survey; this study focussed on a sample of 103 quasars from the
quasar catalogue of \cite{Hew80} which had radio counterparts in the FIRST
survey. In complete contrast with \cite{Nee95}, they found no evidence for
a dependence of linear size on redshift, other than that which arose from
an anti-correlation of linear size with luminosity.
\end{itemize}

It is clear that sample inhomogeneity and redshift estimates rather
than measurements, as examined in detail by \cite{Nee95}, have played
a significant r\^{o}le in giving rise to this plethora of
contradictory results.

\subsection{Spectral index dependences}

Almost 40 years ago, evidence for a relationship between spectral
index and luminosity was suggested by \cite{Hee60}. Since that time
other studies have confirmed his initial findings, e.g.\ \cite{Blu79}
studied 3C and 4C sources, though they were unable to identify whether
the dominant dependence was on redshift or on radio power.

\cite{Lai80} reported that for sources with hotspots, spectral index and
the `degree of spectral curvature' were correlated with luminosity. Their
study was based on two complete samples of radio sources: one was 3CR
(Bennett 1962) selected at 178 MHz and the other was selected at 2.7 GHz
[later published by Peacock and Wall (1985)].  They were similarly unable
to say whether the dependence of spectral index was predominantly with
luminosity or with redshift. \cite{Kro91} discussed a number of causes of
these results: {\em (i)} since a typical radio galaxy spectrum steepens
to higher frequency, in part a simple redshifting effect will give rise to
the observed correlation between redshift and spectral index. They also
pointed out {\em (ii)} the importance of inverse Compton losses in higher
redshift sources which will give rise to steeper spectrum sources and {\em
(iii)} the fact that an increased magnetic field which will increase the
synchrotron emission (and hence increase the likelihood of a source being
above a survey flux-limit) will also increase the rate at which spectral
steepening occurs.  No $K$-corrections were made to Laing \& Peacock's
(1980) spectra so the correlation they found probably arose from some
combination of all three of the causes discussed by Krolik \& Chen (1991).
Employing (a large proportion of) redshift estimates for a faint
(8C-selected) sample, Lacy et al.\ (1993) showed that high-frequency (2
GHz) spectral index correlated more closely with redshift than with
luminosity pointing towards the dominant importance of inverse Compton
losses on high-frequency spectra.

\section{Modelling the complete samples}
\label{sec:model}
In order to gain insight into the distribution of objects in the six
planes [$P$--$D$, $P$--$z$, $P$--$\alpha$, $D$--$z$, $D$--$\alpha$ and
$z$--$\alpha$] of the [$PDz\alpha$] parameter space occupied by the radio
sources which are members of our complete samples, we performed
Monte-Carlo simulations which we describe later in this section.  We begin
in \S\ref{sec:consensus} with a pr\'{e}cis of assumptions about the basic
properties of radio sources which bear the test of observations and
numerical simulations with particular reference in \S\ref{sec:headhs} to
the distinct r\^{o}les of the hotspot and the head in a radio source.  In
\S\ref{sec:vmod} we discuss the constraints we have on the way in which
radio sources expand with time giving us the formalism for their linear
sizes. We describe in \S\ref{sec:pmod} the change of the luminosity (and
the radio spectra) with time, of classical double radio sources and
produce tracks for individual objects in \S\ref{sec:indiv}.  We then
explain in \S\ref{sec:zmod} our formalism for obtaining the redshift
distribution, and then how the model properly takes into account the
sample selection process.  We discussed in \S\ref{sec:selection} how it
was imperative that a {\em combination} of complete samples should be used
to infer the true dependences of radio source properties discussed in
\S\ref{sec:results}. A necessary condition for a satisfactory model of
radio sources to explain the dependences from the combined samples is that
it can reproduce the [$PDz\alpha$] parameter space for each single
flux-limited sample individually. We therefore conclude this section with
a presentation of the six planes from the [$PDz\alpha$] parameter space
for each of a simulated and real dataset for the 3C and 7C complete
samples.

\subsection{Assumptions about radio sources}
\label{sec:consensus}

\subsubsection{Basic formalism for synchrotron radiation}
The synchrotron luminosity of an emitting region with volume $V$, threaded
by magnetic field $B$ is given, in units of ${\rm W\,Hz^{-1}\,sr^{-1}}$,
after averaging over pitch angles by:
\begin{equation}
P_{\nu} = \frac{1}{6\pi}\sigma_{\rm e}c\frac{B^2}{2\mu_{\circ}}
          \frac{\gamma^3}{\nu} n(\gamma) V,
\label{eq:sync}
\end{equation}
where $\gamma$ is the Lorentz factor of the electrons radiating at
frequency $\nu$, $n(\gamma)$ is defined in terms of $n(\gamma) d\gamma$ as
the number of electrons per unit volume per unit Lorentz factor between
$\gamma$ and $\gamma+ d\gamma$, $\sigma_{\rm e}$ is the Thomson scattering
cross-section, $\mu_{\circ}$ is the permeability of free-space and $c$ is
the speed of light. Adopting a delta-function approximation for the
spectral emission from a given relativistic electron, a given frequency of
emission $\nu$ is related to the Lorentz factor of the radiating electron
by:
\begin{equation}
\label{eq:lorentz}
\gamma = \biggl(\displaystyle\frac{m_{\rm e}}{eB} 2\pi \nu \biggr)^{
          \frac{1}{2}},
\end{equation}
where $m_{\rm e}$ is the rest mass of an electron and $e$ is the charge on
an electron.  Although it is the case that a delta function profile is not
a good representation of the power spectrum radiated by a
synchrotron-emitting particle (particularly in the high--$\gamma$ regime),
as long as the underlying energy distribution of the radiating particles
is smooth, then the energy emitted by the population at a particular
frequency can be reasonably calculated using this approximation. The
approximation will be particularly poor close to energy cutoffs.

\subsubsection{Jet behaviour}
\label{sec:jet}
We assume throughout a model in which a jet emanating from the central
engine of an Active Galactic Nucleus (AGN), which transports a total power
$Q_{\rm o}$, remains relativistic all the way from the central engine to
the jet-shock (see Wardle \& Aaron 1997) and that the mechanism for
confining the jet is via pressure balance with the lobe. We further assume
that the jet has a density much lower than the external density
[supporting evidence for this assumption comes from the simulations of
Lind et al.\ (1989) and Clarke, Norman, \& Burns (1989)]. In the case of a
low density jet, simple balancing of the momentum flux enabled Begelman,
Blandford \& Rees (1984) to argue that the jet-shock advances at much less
than the speed of light and consequently a large fraction of the bulk
kinetic energy must be converted to thermal energy in the post-jet
shock. We assume as indicated by the simulations of Lind et al.\ (1989)
that the jet thrust is applied (even instantaneously) over an area
significantly larger than the cross-sectional area of the jet itself and
we identify this region with the hotspot, as discussed in the next
section.

\subsubsection{Heads and hotspots}
\label{sec:headhs}
\begin{figure}[!h]
\begin{picture}(50,100)(0,0)
\put(-35,-310){\includegraphics{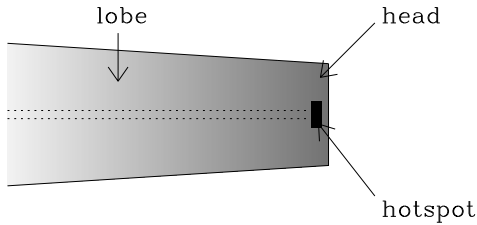}}
\end{picture}
{\caption[junk]{\label{fig:toy} A schematic illustration of the basic
features of a radio source which we wish to model: one half of a radio
source is shown. Close to the outermost edge of the radio source is the
hotspot where the bulk kinetic energy of the jet is thermalised and where
particle acceleration may take place; the hotspot feeds the `head' of the
source, and it is the head pressure which governs the rate at which the
source grows into its environment. The decreasing depth of the grey colour
in the lobe represents the decrease in pressure between the head of the
source back towards the active nucleus --- the pressure gradient which
drives backflow of plasma from the head. Note that depending on the
variation of ambient density with radius, that the pressure may increase
once more as the host galaxy of the AGN is approached.  }}
\end{figure}

We now carefully describe our usage of the term hotspot: we do not use the
term hotspot merely to mean the bright emission in the outermost parts
(the `heads') of the lobes. We wish to refine the definition of hotspot
from that which arises purely empirically (see e.g., Bridle et al.\ 1994).
We use the term to refer to that region in which a distinct physical
phenomenon occurs, namely emission associated with the post-jet shock
structure (i.e.\ the extremely high magnetic field region within and just
beyond the shock structure where a very large fraction of the bulk kinetic
energy from the jet is thermalised and where particle acceleration may
occur). A cartoon illustrating this is shown in Figure~\ref{fig:toy}.

Although many high-resolution high-fidelity images of hotspots reveal
complex structure (Leahy et al.\ 1997, Hardcastle et al.\ 1997 and
Black et al.\ 1992), we find no evidence in a comparison of the
high-luminosity quasars studied by Bridle et al (1994) with the
lower-luminosity objects studied by Leahy et al.\ (1997) of any very
strong dependence of hotspot size on lobe pressure; the hotspots are
of order a few kpc in diameter in all known FR\,II sources.

We also note that whatever the complexities of the physics by which
hotspots are governed, hotspots do not appear to grow self-similarly with
the radio source: i.e.\ the hotspot sizes do not scale strongly with
source length. We note that what we perceive hotspot sizes to be will
depend on the extent to which the emission from each hotspot is Doppler
boosted. 

\subsubsection{Hotspot observables} The properties of hotspots which
we are able to measure from high resolution maps made with MERLIN or the
VLA include the flux densities and the break frequencies in hotspots at
GHz frequencies, angular sizes subtended by hotspots (which in reality are
probably lower limits because Doppler boosting of the plasma emerging
through the hotspots will tend to make hotspots appear smaller).  From
these properties we are able to estimate the magnetic field strengths and
pressures. (Estimates of hotspot magnetic fields may be made using the
assumption of equipartition in the hotspots. X-ray detections of hotspots
now give considerable confidence in the validity of the assumption of
equipartition magnetic field strengths in hotspots (values which are in
the region of tens of nT) [see Harris, Leighly \& Leahy 1998].)

We describe in \S\ref{sec:hslum} our justification for neglecting the
direct contribution of the hotspots' luminosity to the overall
luminosity measured from a radio source at a survey frequency like 151
MHz. Nonetheless the hotspot and its magnetic field play a vital role
in governing the energy spectra of particles injected into the lobe,
and hence the spectrum of the total radio luminosity we observe.

\subsubsection{The contribution of hotspots to the total luminosity}
\label{sec:hslum}

We contend that at low frequencies like 151 MHz, the luminosity from a
hotspot never dominates over the luminosity from the head and the
lobe. Na\"{\i}ve application of equation~\ref{eq:sync}, using typical
magnetic fields inferred for hotspots (e.g.\ as derived from
equation~\ref{eq:QB}), for an energy distribution which extends down to
particles with Lorentz factors of $\sim 1$, for typical hotspot volumes
($\sim$ a few kpc in diameter) predicts for the highest jet-power objects
hotspot luminosities which are higher than the observed total luminosities
of radio sources. However, for frequencies of a few GHz, the predicted
luminosities appear to be much more closely matched with the luminosities
measured for resolved hotspots in observations.  Leahy et al.\ (1989)
found for a sample of 3C sources that the highest power sources had
hotspots whose spectra became flatter between 151 MHz and 1.5 GHz. (For
the calculated magnetic field strengths the synchrotron self-absorption
turnover frequencies for emitting volumes with sizes of a few kpc are in
the region of tens of MHz.) They interpreted this as the manifestation of
a low-energy cutoff in the distribution of the radiating particles giving
a deficit of hotspot emission at 151 MHz. In the context of an
electron-proton jet where particle acceleration occurs at the hotspot such
a cutoff is likely to occur when the Lorentz factor is of the order of the
ratio of the proton mass to the electron mass, since for a particle to be
efficiently accelerated, its gyro-radius must be large compared with the
shock thickness (which is of the order of the gyro-radius of the thermal
material, i.e.\ the protons, present). Although the precise nature of the
low-energy cutoff depends on details such as the incoming jet velocity, it
is likely that the value of the minimum Lorentz factor does not change
greatly from hotspot to hotspot. However, the cutoff in frequency at which
this is manifested {\em does} vary from hotspot to hotspot since for a
given minimum Lorentz factor the minimum emitted frequency is proportional
to the magnetic field in the hotspot (equation~\ref{eq:lorentz}). For
those sources whose hotspots have higher magnetic fields (which we will
argue via equation~\ref{eq:QB} are those with higher jet-powers), the
minimum frequency at which significant radiation will be observed will be
higher than for sources with lower hotspot magnetic fields (and lower
jet-powers).  Such reasoning is consistent with the low contribution
(0.04\%) the hotspots make to the total 327-MHz luminosity in the
archetypically high-$Q_{\rm o}$ object, Cygnus A (Carilli et al.\ 1991).

\subsubsection{Inverse Compton losses arising from the AGN's own radiation
field}
\label{sec:ownrf}
We now consider whether, for simulating low radio frequency surveys of the
type described earlier in this paper, it is acceptable to neglect energy
losses due to inverse Compton scattering off the AGN's own radiation
field.  This mechanism was invoked by Blundell \& Lacy (1995) to explain
the steep spectrum in the GHz regime of the extended radio emission around
the radio-quiet quasar E\,1821+643, and by Brunetti et al.\ (1997) to
explain the soft X-ray emission detected in six 3C FRII radio galaxies. An
`equivalent magnetic field' may be derived from the energy density in the
radiation field; this equivalent magnetic field falls off as $1/r$ where
$r$ is the distance out from the AGN. Significant losses arising from this
mechanism will therefore be confined to some specific spatial extent.  If
this mechanism causes spectral steepening to be manifested at frequencies
which may be redshifted to be close to the survey frequency, then its
contribution cannot be neglected; we now investigate whether this could be
the case.

\begin{figure}[!h]
\begin{picture}(50,380)(0,0)
\put(-30,-30){\includegraphics{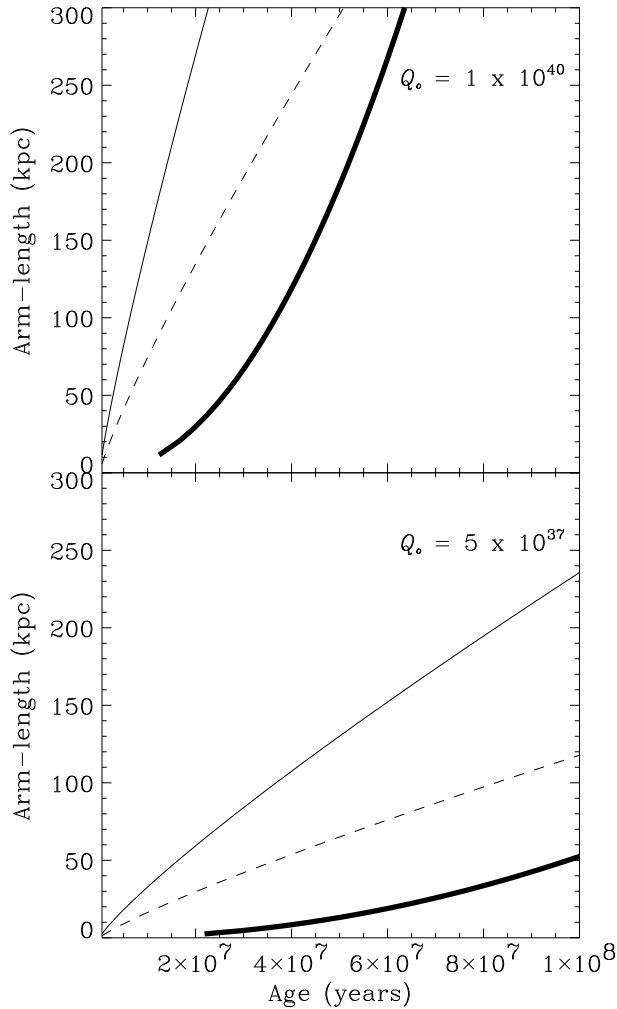}}
\end{picture}
{\caption[junk]{\label{fig:qrf} Thick solid line: the trajectory of a
point moving out from the centre of the AGN at which the break frequency
has lowered to 1~GHz; thin solid line: the trajectory of one hotspot
advancing away from the AGN; dashed line: the trajectory of the mid-point
of the arm of a source (see text for more details). {\em Upper plot:}
trajectories for a source whose jets each carry total kinetic energy
$Q_{\rm o} = 10^{40}\,{\rm W}$ and {\em lower plot:} with $Q_{\rm o} = 5
\times 10^{37}\,{\rm W}$; the value of $\phi$  was taken to be 5. }}
\end{figure}

The energy density of the AGN radiation field $U_{\rm rad}$ is given by:
$U_{\rm rad} = Q_{\rm phot}/(4\pi c r^2)$ where $Q_{\rm phot}$ is the
photo-ionising luminosity.  We assume that $Q_{\rm phot} \sim \phi Q{\rm
o}$ where $Q_{\rm o}$ is the total kinetic energy transported by one jet
and $\phi$ is a factor which Willott et al.\ (1998b) suggest might be as
high as 20. Equating the $U_{\rm rad}$ with $B_{\rm eq}^2/2\mu_{0}$ we
find,
\begin{equation}
B_{\rm eq}(r) = \left[\displaystyle\frac{\phi Q_{\rm o}}{c} \mu_0
                   \displaystyle\frac{1}{2\pi r^2} \right]^{\frac{1}{2}}.
\label{eq:bfield_qrf}
\end{equation}
We are now in a position to identify the timescale over which this
magnetic field will impact on the flux which may be detectable by a
survey, at a given distance from the AGN. We plot in Figure~\ref{fig:qrf}
a thick solid line indicating the trajectory of a point moving out from
the centre of the AGN at which the break frequency would lower to 1~GHz
(this will not directly impact on a survey at a frequency as low as 151
MHz, until the source redshifts are greater than 5) if the plasma bathed
in this radiation field had resided there since the radiation field was
switched on.  [This therefore over-estimates the rate at which the break
in the spectrum will advance out from the centre of the AGN.]  This
trajectory is calculated using equations~\ref{eq:breakfreq} and
\ref{eq:bfield_qrf} at each position $r$.  We also plot a thin solid line
which is the trajectory of one head advancing away from the AGN (as
described by halving the total linear size expressed in
equation~\ref{eq:falle}) while the dashed line is the trajectory of the
mid-point of one arm of the radio source. It is seen that the spatial
point where the break frequency becomes as low as 1 GHz is only reached
{\em after} the midpoint of the lobe has advanced beyond this.  Comparison
of the Inter-Planetary Scintillation measurements of the low-frequency
(81.5 MHz) measurements of 3C objects by Purvis et al.\ (1987) with the
flux densities of these objects at 86 MHz of Laing \& Peacock (1980)
suggest that roughly half of the low-frequency emission comes from the
head of the source. For the other half of the emission coming from the
lobe, a decreasing proportion comes from plasma with increasing proximity
to the core.  Thus, because of the specificity of the spatial extent of
the equivalent magnetic field arising from the quasar radiation field,
those regions which suffer the most from this loss mechanism are those
which barely make a significant contribution to the overall low-frequency
flux. The importance of this effect cannot be so simply analysed for
sources which are smaller in extent than e.g.\ 10 kpc. However, such
sources are known to expand very quickly [Owsianik \& Conway (1998)] and
the radiating plasma in them is continually replenished by freshly
injected synchrotron emitting particles.  Thus we conclude that the
effects of inverse Compton losses off the AGN's own radiation field should
not be significant in the selection of sources detected in surveys
conducted in these frequency regimes, although this mechanism may be
relevant in the question of the `docked tails' of radio sources --- which
we discuss in \S\ref{sec:injection}. 

\subsection{Environments}
\label{sec:env}
An important probe of the densities of the environments of radio sources
comes from studies of the depolarisation of the polarised intensity from
their synchrotron emitting lobes. Such a study by Garrington \& Conway
(1991) inferred the densities of particles responsible for the
depolarisation to be consistent with the notion that radio galaxies are in
poor group environments, with a typical value of the density given by
$\rho_{\circ} = 1.67 \times 10^{-23}\, {\rm kg\,m^{-3}}$.  We infer from
Mulchaey \& Zabludoff (1998) for density profiles of such groups which are
parameterised as follows:
\begin{equation}
\label{eq:rho}
\rho(r) = \rho_{\circ}\biggl(\frac{r}{a_{\circ}}\biggr)^{-\beta}
\end{equation}
that suitable values of $\beta$ and $a_{\circ}$ are 1.5 and 10 kpc
respectively (see also Willott et al.\ 1998b).

Many authors have used various observational data as evidence for
systematic changes in radio-source environments with redshift. Examples
include: {\em (i)} the X-ray detection of a number of high-$z$ 3C sources
interpreted as evidence that a larger proportion of these lie in rich
(X-ray emitting) clusters (Crawford \& Fabian 1996); {\em (ii)} a seeming
increase in the optical richness of group/cluster environments with
redshift (Hill \& Lilly 1991, Yates, Miller \& Peacock 1989); {\em (iii)}
the identification of 3C host galaxies at $z \sim 1$ with the brightest
cluster galaxies (Best et al.\ 1998, Eales et al.\ 1997, Roche et al.\
1998) and {\em (iv)} an increase in the radio depolarisation of radio
sources with either increasing redshift or luminosity (Garrington \&
Conway 1991).

We have attempted, and failed, to synthesise these observations into a
consensus picture for the evolution of radio-source environments. For
example, the interpretation of X-ray detections associated with distant
powerful radio sources as emission from hot cluster gas by Crawford \&
Fabian (1996) is at odds with the interpretation of Brunetti et al.\
(1997) that inverse Compton scattering from either the AGN radiation field
or from the cosmic microwave background is responsible for these X-rays.
While the finding of Garrington \& Conway (1991) --- that more distant
sources show higher depolarisation --- might appear to suggest that
environments of more distant sources are more dense, it should be borne in
mind that the more distant sources in their study, being shorter than
lower-$z$ sources, inform on the depolarisation of lobe emission at
spatial points much closer in the AGN, where inhomogenities due to the
host galaxy and the possible recent merger products which led to the
formation of the AGN will increase the measured depolarisation.  In fact
the interpretation of this very measurement is exacerbated by the
``youth--redshift'' degeneracy, a point to which we return in
\S\ref{sec:youthz}.

In the absence of any clear consensus on this question we have preferred
to adopt the simple assumption of no systematic change of radio-source
environment with redshift and to see whether, when implemented in our
model for radio sources, this necessarily presents contradictions with
observed properties of radio-source samples. One might expect information
on radio-source environments to improve rapidly with the advent of
facilities such as AXAF.

\subsection{Source sizes}
\label{sec:sourcesize}
\subsubsection{Source expansion}
\label{sec:falle}
The first essential step in modelling the time evolution of the properties
of an FR\,II radio source is to determine how the size of the source
depends on its age $t$.  Falle (1991) and Komissarov \& Falle (1998) have
shown that this dynamical problem can be solved via dimensional analysis,
and since the mass involved is dominated by the ambient material swept-up
within the bow-shock of the radio source the only characteristic
quantities required are the kinetic power transported by one jet $Q_{\rm
o}$, and $\rho_{0} \times a_{0}^{\beta}$.  This gives:
\begin{equation}
\label{eq:falle}
D(t) = 2 c_1 a_{\circ} 
       \biggl(\frac{t^3 Q_{\rm o}}{a_{\circ}^5 \rho_{\circ}}\biggr){,}^{\frac{1}{5-\beta}}
\end{equation}
where $D(t)$ is the total length of a source, i.e.\ the hotspot to hotspot
separation; see also Kaiser \& Alexander (1997).  Assumptions about the
radio source environment are therefore necessary (as summarised in
\S\ref{sec:env}), but adopting these assumptions it is possible to
constrain $c_{1}$ in three ways: (i) for those powerful FR\,II sources
with measured expansion speeds ($v = 0.5 dD/dt \approx 0.03c$; Scheuer
1995, see \S\ref{sec:vmod}) an upper limit may be obtained, similarly from
(ii) the distribution of radio sources in the $P$--$D$ plane of simulated
samples (see Figures~\ref{fig:mock3c} and \ref{fig:mock7c}) and (iii)
obtaining the axial ratios consistent with those observed. We take a value
of 1.8 for the dimensionless constant $c_{1}$.  Comparison of the
observational and simulated data presented later in this paper
(Figures~\ref{fig:real3c} \& \ref{fig:mock3c} for the 3C data and
Figures~\ref{fig:real7c} \& \ref{fig:mock7c} for the 7C data) will be used
to show that $c_{1}$ must be similar to this value (see also Willott et
al.\ 1998b).  Falle's (1991) analysis also shows that the bow-shock must
expand in a homologous (self-similar) manner.

We wish to re-emphasize the distinction between the hotspot and the head
region of the expanding radio source (see \S\ref{sec:headhs} and
Figure~\ref{fig:toy}). The hotspot, also sometimes referred to as the
`working surface', may well move around the head of the lobe as in the
``Dentist's Drill'' model of Scheuer (1982) so that it is very dangerous
to use the (instantaneous) properties of the compact hotspot to determine
how the (much larger) head region of the source expands, e.g.\ by equating
the pressure of the hotspot to the `ram pressure' of the external medium
[e.g.\ Readhead et al.\ (1996)].  It is better to consider the expansion
of the whole head region, and its associated bow shock. Similar points
have been made by Begelman \& Cioffi (1989).

From the jump conditions at a strong shock, (see e.g.\ Landau \& Lifshitz
1959) the pressure across the head ($p_{\rm head}$) is related to the
pressure in the external medium ($p_{\rm ext}$) (whose adiabatic index is
$\Gamma_{\rm x} = 5/3$) in terms of the Mach number of the bow shock
($M_{\rm b}$) by:

\begin{eqnarray}
\label{eq:mach}
p_{\rm head} &=& p_{\rm ext} \biggl(
               \frac{2 \Gamma_{\rm x} M_{\rm b}^2}{\Gamma_{\rm x} + 1} -
               \frac{\Gamma_{\rm x} - 1}{\Gamma_{\rm x} + 1}\biggr) \nonumber \\
&\approx& p_{\rm ext} \biggl(
               \frac{2 \Gamma_{\rm x} M_{\rm b}^2}{\Gamma_{\rm x} + 1}\biggr)
               \mbox{$\:\:\:\:\:$ for $M_{\rm b} \gg 1$}
\end{eqnarray}
In terms of the dimensional argument of Falle (1991) this may be
expressed as:
\begin{equation}
\label{eq:phead}
p_{\rm head} = \frac{2}{\Gamma_{\rm x} + 1}
               \rho_{\rm ext} 
               \biggl(\frac{1}{2}
               \frac{dD}{dt}
               \biggr){,}^2
\end{equation}
where $D$ is given by equation~\ref{eq:falle} and $\rho_{\rm ext}$ is the
density of the external medium (identified with $\rho(r)$ and given by
equation~\ref{eq:rho}).  This, together with equation~\ref{eq:falle},
gives an expression for the pressure in the head at time $t$:
\begin{equation}
\label{eq:head}
p_{\rm head} = \frac{18}{\Gamma_{\rm x} + 1} \frac{\rho_{\circ}}{(5-\beta)^2}
               c_1^{2-\beta} a_{\circ}^2 
               \biggl(\frac{Q_{\rm o}}{a_{\circ}^5
               \rho_{\circ}}\biggr)^{\frac{2 - \beta}{5-\beta}}
               t^{\frac{-4-\beta}{5-\beta}}.
\end{equation}

Note that this is identical to equation 12 of Kaiser \& Alexander (1997),
but that we are making a clear association between environmental ram
pressure and the average value of internal pressure across the entire head
of the source, and not just the compact hotspot. It is significant in this
regard that high resolution radio imaging (Leahy et al.\ 1997, Hardcastle
et al.\ 1997 and Black et al.\ 1992) suggests that, ignoring the possibly
ephemeral hotspot emission, there is very little pressure variation across
the head region of a given FR\,II radio source. It is also significant
that the identification of the characteristic pressure with the whole head
of a source, rather than the hotspot, avoids the `lobe drag' problem
posed, for example, by Williams (1991).

We believe that the material backflowing from the hotspot, via the head,
inflates a cocoon which we identify with the observed radio lobe [an idea
which dates back to Longair, Ryle and Scheuer (1973)] and which we assume
confines the jet.  Leahy \& Perley (1991) made images at 1.4 GHz of 23
extragalactic radio sources with the VLA. In these images they detected
the outer perimeter of the radio emission in most sources; at these outer
edges the emissivity does not fade slowly into the noise, rather it
suggests lobes which are over-pressured with respect to their local
environment and will therefore expand outwards as the source ages, and
indeed as it expands longitudinally.

Simulations show that jets with a low density ratio and high Mach numbers
must produce backflow (Norman et al.\ 1982). Observational evidence for
backflow, and high backflow speeds, comes from the spectral ageing
analysis of Liu, Pooley \& Riley (1992) which found lobe-speeds (i.e.\ the
anti-vector sum of the oppositely directed head advance speed and the
backflow speed) in some sources to exceed $0.2c$. When this is taken
together with Scheuer's (1995) analysis of arm-length asymmetries arising
from light-travel time effects which limit the head advance speed to
substantially below this, perhaps as low as $0.03c$ (see discussion of his
method in \S\ref{sec:vmod}), the lobe-speeds which Liu et al.\ derive
strongly point towards high backflow speeds out of the hotspot. This, in
turn, is strong evidence for a pressure gradient from the head of the
source back towards the central nucleus.

If one considers a slice in low-frequency surface brightness through a
radio lobe, along the jet axis, a gradual fall-off in surface brightness
is seen from the head of the source back towards the core [see e.g.\ the
slices shown in Leahy et al.\ (1989)]. Over a given volume, pressure is
related to surface-brightness under equipartition conditions by a simple
power-law scaling: $p \propto S^{4/7}$. Thus we may infer a gradual
decline in pressure from the head towards the core and that the head
pressure governs the lobe pressure.

The ratio of the pressure in the head to that in the lobe is not a
well-defined quantity, but in the interests of simplicity in the modelling
which we now describe, we will take for sources of all jet-powers a
constant ratio of six for these pressures.

\subsubsection{Source advance}
\label{sec:vmod}
A lower limit to the speed at which heads advance away from their AGN, and
hence the rate at which radio sources grow, comes from the fact that
FR\,II radio sources end in shocks which mean that the advance speed must
exceed the ambient sound speed. The sound speed in an ionized gas at
temperature $T$ K is given by $ \approx 5 \times 10^{-7} \times c
\sqrt{T}$ where $c$ is the speed of light; for $T = 10^7$K this gives an
advance speed of $0.0015c$.

Upper limits to the advance speed come from consideration of the
asymmetries in arm-lengths of radio sources. This kind of analysis was
first done by \cite{Lon79}; a subsequent and similar study was done by
\cite{Bes95}. These studies assumed that all arm-length asymmetries arose
from geometrical projection and light-travel time effects. No account was
taken of effects arising from environmental asymmetries. The former study
obtained an upper limit to the head advance speed of 0.2$c$ while the
latter study, which also took into account the effects of possible
misalignment in the directions of ejection of the jets, obtained an upper
limit of $\sim 0.4c$. These values should very much be regarded as upper
limits; tighter constraints were obtained by \cite{Sch95} who considered a
sample of quasars each of which exhibited a considerable asymmetry in the
presence of jets. The side of the brighter jet he ascribed as the nearer
to us, attributing the jet brightness asymmetry to Doppler
boosting. Arm-length asymmetries in these objects could then be more
confidently attributed to light-travel time effects since there was a
certain marker as to which was the nearer side to Earth. Scheuer's upper
limit was somewhat sample dependent hinting at a non-constant hotspot
advance speed, but gave a value of 0.03$c$ for two samples and for all of
his samples taken together a certain upper limit of 0.15$c$. This method
of considering asymmetries, gives an `instantaneous' measure\footnote{The
first direct measurements of source expansion were made by Owsianik \&
Conway (1998) from multi-epoch global VLBI observations; they measured an
advance speed of the CSO 0710+439 to be \gtsim\ 0.24$c$. In such a small
source (its linear size is 101.5 $h^{-1}$ pc) it is likely that the
expansion is strongly determined by the local density profile within the
core radius of its host galaxy and so the environmental assumptions of
\S\ref{sec:env} would be correspondingly poor.} of the speed at which the
heads were advancing out from the active nucleus at the time when the
light we observe was emitted from the radio source. The term
`instantaneous' is used somewhat liberally here, to mean `within the
light-travel time from the further hotspot to the nearer' as distinct from
the speed which would be derived by averaging over the life-time of the
source by dividing the (projected) length of the radio source by its total
age.

Speed limits in the same ball-park as Scheuer's smaller limits were
obtained by \cite{Rea96} who investigated the empirical relationships
between the external density and the size, pressure, advance speed and
flux densities of compact symmetric objects by assuming the sources to be
ram pressure confined. They argued that hotspot pressures adjusted to the
ambient density so that head advance speeds could be constant and close to
0.02$c$ (but see comments in \S\ref{sec:falle}).

De Young (1997) has pointed out that the energy density at the end of the
jet used to determine advance speeds could be the result of cumulative
effects of the jet-environment interaction over the jet length.  He
suggests a preliminary approximation for how the advance speed depends on
lossy jet propagation, namely by considering a constant entrainment rate
per unit length of jet.  He derives from this a roughly constant jet
propagation rate.

We chose to adopt the formalism for the linear size growth developed by
Falle (1991) on dimensional arguments discussed in \S\ref{sec:falle}. 

The jet-axes of the sources we simulate in our models are deemed to be
randomly oriented with respect to our line-of-sight. Each source is
allocated an angle to the line of sight which is drawn from a distribution
uniform in ($1 - \cos \theta$). The projected linear size of each
simulated object is derived by evaluating $D$, the absolute physical size
of the radio source, given in equation~\ref{eq:falle} at the time when the
source intercepts our light-cone, and multiplying this length by $\sin
\theta$.

In our model we deem that all radio sources will have their beams switched
on for some maximum length of time ($t_{\rm max-age}$, which has been
taken as $5 \times 10^8$ years).  The time at which we `make the
observation' (i.e., examine whether the source is above the flux-limit of
our simulated survey) of a given source is randomly chosen from a
distribution described in \S\ref{sec:zmod}, between 0 years and $t_{\rm
max-age}$ years.  If $t_{\rm max-age}$ were chosen to be too short, there
would be two erroneous consequences: {\em (i)} if the true value of the
coefficient of proportionality for the head advance ($c_{1}$) is known
then sources will never achieve projected linear sizes as large as those
which are observed and {\em (ii)} a distribution of points would be
generated on the $P$--$D$ plane, with an abrupt maximum size across the
whole range in power, strikingly different from that which is
observed. This would arise because the sampling of the radio sources would
occur at early stages in their lives when very few sources would have
fallen below the sample flux-limit. If the sampling of the radio sources
broadly occurs over a time when the `natural selection' process (the
catastrophic decrease in luminosity due to synchrotron, adiabatic
expansion and inverse Compton losses) governs the distribution of points
on the $P$--$D$ plane, then it much more closely resembles that arising
from real data.

If $t_{\rm max-age}$ is chosen to be too large, then provided the
simulations are allowed to run through the birth of a sufficiently large
number of radio sources there will be no consequence on the distribution
of points on the $P$--$D$ plane. This is the case because the sampling of
radio sources during the period of their lives when they are sufficiently
bright to be detected by a survey on Earth is still random throughout all
of this period. That there may be a relatively long period of time when
radio sources have fallen below a sample flux-limit during which our
light-cone randomly samples them only has consequences for the number of
sources which have to be `observed' until the simulated survey is
complete.

\subsection{Luminosity and spectral index development}
\label{sec:pmod}

We took as our basic model for the luminosity-with-time profile of a given
radio source the model developed by \cite{Kai97a} and \cite{Kai97b}; this
is the first model of extragalactic radio sources which directly
incorporates the effects of inverse Compton losses on the luminosity of a
source.  This model assumes that source expansion is described by Falle's
(1991) model, i.e. equation~\ref{eq:falle}.

\subsubsection{The r\^{o}le of hotspots in governing the lobe luminosity}
As described in \S\ref{sec:pararesults} each object in our model is assigned two
jets which each transport a bulk kinetic power of $Q_{\rm o}$, assumed to be
constant throughout the lifetime of the source.

Under the assumptions of \S\ref{sec:jet} we can assume that the thrust of
the jet is given by ($Q_{\rm o}/c$).  The stagnation pressure $p_{\rm s}$
in the post-jet shock material will then be approximately given by:
\begin{equation}
\label{eq:pQ}
p_{\rm s} = \frac{Q_{\rm o}}{c A_{\rm hs}} \approx p_{\rm hs}
\end{equation}
where $A_{\rm hs}$ is the area perpendicular to the jet over which the jet
thrust acts. We equate this stagnation pressure with $p_{\rm hs}$, the
pressure in the hotspot itself.  If we assume that the energetics of the
hotspot are described by equipartition, then the total energy density in
the hotspot is due to two equal contributions, one from the particles and
one from the magnetic field (this latter being equal to $B_{\rm
hs}^2/2\mu_{\circ}$).  We may then express the magnetic field (assuming it
to be tangled) in the hotspot ($B_{\rm hs}$) in terms of the jet power as
follows:
\begin{equation}
\frac{Q_{\rm o}}{c} = \frac{B_{\rm hs}^2}{3 \mu_{\circ}}A_{\rm hs}, 
\label{eq:QB}
\end{equation}
since for a relativistic fluid, and also for a tangled magnetic field,
pressure is one-third of the total energy density.  The break frequency in
the hotspot $\nu_{\rm bh}$, is inferred from the usual expression for the
break frequency in a synchrotron source:
\begin{equation}
\nu_{\rm bh} = \frac{(9/4)c_{7}B_{\rm hs}}
               {(B_{\rm hs}^2 + B_{\rm CMB}^2)^2 t_{\rm s}^2}
\label{eq:breakfreq}
\end{equation}
where following Leahy (1991), the coefficient of proportionality $c_{7}$
is $1.12 \times 10^{3}\, {\rm nT^3\, Myr^2\, GHz}$ thus the hotspot
magnetic field $B_{\rm hs}$ and equivalent magnetic field due to the
cosmic microwave background $B_{\rm CMB}$ are measured in nT, $t_{\rm s}$
the `synchrotron age' of the distribution is measured in Myr and the break
frequency of the hotspot $\nu_{\rm bh}$ is measured in GHz.  In the
context of our model, we assume that the synchrotron age of the radiating
particles depends on how long the particles are exposed to the magnetic
field of the hotspot before they escape into the lobe.  Some effect of
this type seems inescapable in any model for a hotspot [e.g.\ Heavens \&
Meisenheimer 1987 and Eales, Alexander \& Duncan 1989]. The magnetic
field has the dependence on $Q_{\rm o}$ given by equation~\ref{eq:QB}
giving a value for the break frequency which depends just on $Q_{\rm o}$,
if we make the assumption that hotspot sizes are constant (we assume for
all our sources that the radius and depth of each hotspot are 2.5 kpc 
and 1 kpc respectively).

\subsubsection{Lobe luminosity}
\label{sec:lobe}
Our formalism essentially follows the model of Kaiser et al.\ (1997) but
with two important differences: {\em (i)} instead of assuming a constant
injection index [given by $p = -\partial \log n(\gamma) /\partial\log
\gamma$] of $p = 2.14$ we assume that the injection index is governed by
the breaks in the energy distribution of the particles injected into the
lobe from the hotspot.\footnote{We remind the reader of the distinction we
draw between injection index and acceleration index as discussed in
\S\ref{sec:depend}.}  These are ultimately determined by the maximum and
minimum particle dwell-times in the hotspot together with the magnetic
field strength in the hotspot, which as seen in equation~\ref{eq:QB}
directly depends on $Q_{\rm o}$ itself. {\em (ii)}~We also assume that
adiabatic expansion losses out of the hotspot are governed by the hotspot
pressure as derived in equation~\ref{eq:pQ}, not by the pressure in the
head of the source which governs the increase in arm-length of the source.

Note that this process of adiabatic expansion of the particles out of the
hotspot into the lobes does not move the energy distribution of the lobes
out of equipartition, if the adiabatic indices of the particles and of the
fields are both 4/3 (as is the case for a highly relativistic fluid and for
the virtual photons of a magnetic field).  

A continuum of spectra may be regarded as injected to the lobe as the
supply of particles in the hotspot is tapped for those populations of
particles which have spent different dwell-times in the hotspot and thus
have different break frequencies.  Since the magnetic field in the hotspot
is governed by the jet power $Q_{\rm o}$, this ultimately governs the
spectrum of particles injected into the lobes [this is the origin of the
$P$--$\alpha$ correlation in our model].

We derive a break in the energy spectrum of synchrotron particles freshly
injected into the lobe by consideration of the corresponding break in
their frequency spectrum of synchrotron emission. Since spectra are
unchanged in shape, though translated to lower amplitudes and lower
frequencies, by adiabatic expansion (Scheuer \& Williams 1968) --- e.g.\
from the hotspot into the lobe --- the Lorentz factors of the break
frequencies of the particles injected into the lobes at some time $t_{\rm
inj}$ may be derived. Scheuer \& Williams showed that if the factor by
which the linear scale of the emitting population has changed after
undergoing adiabatic expansion is $f$, then the factor relating the break
frequency after expansion to that before is $f^{-4}$.  It is assumed that
at the instant that an elemental population of particles is injected from
the hotspot into the lobe, they undergo expansion which is approximately
adiabatic.  The break frequency of the synchrotron population as it is
injected into the lobe is related to its break frequency in the hotspots
by a factor of $(p_{\rm hs}/p_{\rm lobe}[t_{\rm inj}])^{-4/(3\Gamma)}$,
where $(p_{\rm hs}/p_{\rm lobe}[t_{\rm inj}])^{1/(3\Gamma)}$ is their
linear scaling factor ($\Gamma$ is the usual adiabatic index of 4/3) and
where $t_{\rm inj}$ is the time at which a population of particles is
injected into the lobe. This may be expressed in terms of the magnetic
energy densities instead of the pressures i.e., the break frequency of the
synchrotron spectrum of the particles freshly injected into the lobes is
$(B_{\rm hs}^2/(2\mu_{\circ}u_{\tiny B}[t_{\rm inj}]))^{-4/(3\Gamma)}
\times$ $\nu_{\rm bh}$ where $\nu_{\rm bh}$ was the break frequency in the
hotspots and $u_{\tiny B}[t_{\rm inj}]$ is the magnetic energy density in
the lobes at this time $t_{\rm inj}$ (which is directly proportional to
the pressure in the lobe and head given in equation~\ref{eq:head}).

The energy of such a particle, freshly injected into the lobe, which
is emitting at the break frequency $\nu_{\rm bl}$ is therefore given
by $\gamma_{\rm bl}m_{\circ}c^2$ where $\gamma_{\rm bl}$ is given by:
\begin{equation}
\label{eq:larmor}
\gamma_{\rm bl}[t_{\rm inj}] = \biggl( \frac{\nu_{\rm bl}[t_{\rm
                             inj}]} {\nu_{\rm g}}
                             \biggr)^{\frac{1}{2}} = \biggl(
                             \frac{\nu_{\rm bh}}{\nu_{\rm g}}
                             \biggr)^{\frac{1}{2}} \biggl(\frac{B_{\rm
                             hs}^2} {2\mu_{\circ}u_{\tiny B}[t_{\rm
                             inj}]}\biggr)^{ -\frac{2}{3\Gamma}}
\end{equation}
where ${\nu_{\rm g}}$ is the non-relativistic gyro-frequency, given by
$eB[t_{\rm inj}]/(2\pi m_{\rm e})$.

\begin{figure}[h]
\begin{picture}(50,200)(0,0)
\put(-45,250){\includegraphics{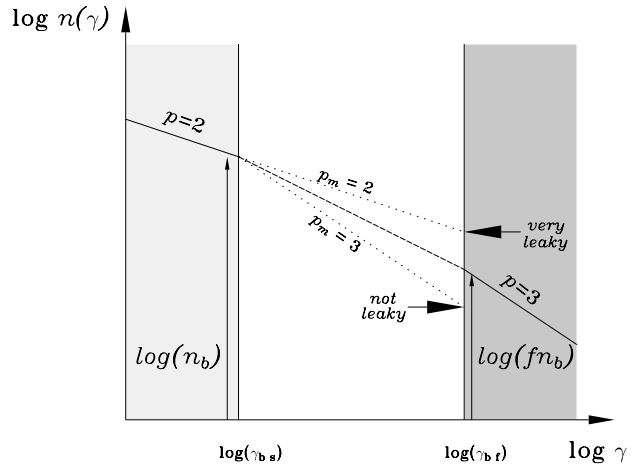}}
\end{picture}
{\caption[junk]{\label{fig:leaky} This schematic illustration shows the
different energy regimes within a hotspot. The low energy spectrum is
described by a region with index $p = 2$ (in light grey). The high energy
spectrum is governed by a region with index $p = 3$ (in dark grey). The
middle region is described by a single power law with index $p = p_{\rm
m}$ (in white). The precise value of this index depends on how leaky the
hotspot is deemed to be and the difference in dwell-times in the hotspot
of the particles which leak out most quickly and those which leak out most
slowly.}}
\end{figure}

The exponent of the energy spectrum in the low $\gamma$ regime ($p_{\rm
l}$) is taken to be 2 (corresponding to a frequency spectral index
$\alpha$ of 0.5) assuming the particle acceleration mechanism in the
hotspots to be via the first-order Fermi mechanism (Bell 1978).  We note
however that it is quite likely that the value of the acceleration index
might be lower than 0.5 (Heavens \& Drury 1988) although we do not
implement values lower than this in our model.

The injection spectrum is derived from the two break frequencies in the
hotspot emission at $\nu_{\rm b\,s}$ and $\nu_{\rm b\,f}$. The former is
the point in the spectrum where the population of particles which has
lingered for the longest time in the hotspot changes from a spectral index
of $\alpha = 0.5$ to 1 (or $p = 2$ to 3, since $p = 2\alpha + 1$). The
latter is the point in the spectrum where the population of particles
which escapes most quickly from the magnetic field of the hotspot changes
from a spectral index of $\alpha = 0.5$ to 1 (or $p = 2$ to 3).  

We deem the maximum $\gamma$ (i.e.\ $\gamma_{\rm b\,s}$) in the $p = 2$
regime to be the Lorentz factor corresponding to the break frequency of
the electrons which linger longest in the strong magnetic field of the
hotspot before being injected into the lobes.

The exponent of the energy spectrum in the high $\gamma$ regime ($p_{\rm
h}$) is taken to be 3 in identifying a hotspot as an example of continuous
injection (Carilli et al.\ 1991).  We deem the minimum $\gamma$ (which is
$\gamma_{\rm b\,f}$) in this regime to be the Lorentz factor corresponding
to the break frequency of the electrons which leak out most quickly from
the strong magnetic field of the hotspot into the lobe (see
Figure~\ref{fig:leaky}). The shape of the spectrum between $\gamma_{\rm
b\,s}$ and $\gamma_{\rm b\,f}$ is determined by the relative rates at
which particles with different dwell-times in the hotspot leak out into
the lobe. 

A refined model of the details of this mechanism might suggest that this
region should have quite a curved spectrum, but in the interests of
simplicity, we approximate this region by a straight line whose gradient
can take a maximum value of 3 ($\alpha = 1$) and a minimum value of 2
($\alpha = 0.5$).
\newpage
\onecolumn

The values of ${\gamma_{\rm b\,s}}$ and ${\gamma_{\rm b\,f}}$ are derived
by considering the longest and shortest times for which particles reside
under the influence of the magnetic field of the hotspot and deriving the
appropriate break frequencies using equation~\ref{eq:breakfreq}. These
dwell-times were taken to be $10^5$ years and 1 year respectively
corresponding to, in the former case a backflow speed of 0.03$c$ and, in
the latter case, to particles which do not fully cross the hotspot.  The
break frequencies from these may be converted to Lorentz factors
corresponding to breaks in the energy spectrum using
equation~\ref{eq:larmor}.  The derivation of $n_{\rm b} ( = n[\gamma_{\rm
b\,s}])$ (the number density of particles with a given Lorentz factor
$\gamma_{\rm b\,s}$ per unit Lorentz factor, see Figure~\ref{fig:leaky})
at a time $t_{\rm inj}$ is obtained by equating the energy density at time
$t_{\rm inj}$ with the integral over the whole energy range, from
$\gamma_{\rm min}$ to $\gamma_{\rm max}$ for each particle of energy
$\gamma$ appropriately weighted as follows:

\begin{eqnarray}
u_{\rm e}[t_{\rm inj}]  &=&  \int_{\gamma_{\rm b\,f}}^{\gamma_{\rm max}}
            n_{\rm h}(\gamma) \gamma m_{\rm e} c^2\,d\gamma
          +  \int_{\gamma_{\rm b\,s}}^{\gamma_{\rm b\,f}}
            n_{\rm m}(\gamma) \gamma m_{\rm e} c^2\,d\gamma
          + \int_{\gamma_{\rm min}}^{\gamma_{\rm b\,s}}
            n_{\rm l}(\gamma) \gamma m_{\rm e} c^2\,d\gamma
\end{eqnarray}
with the weighting factors $n_{\rm l}(\gamma)$, $n_{\rm m}(\gamma)$ and 
$n_{\rm h}(\gamma)$ from the three regimes in $\gamma$ respectively:
\begin{equation}
\label{eq:ngamma}
\begin{array}{lll}
	n_{\rm l}(\gamma) &= 
               n_{\rm b}
               \biggl(\displaystyle\frac{\gamma}{\gamma_{\rm b\,s}}\biggr)^{-2}& 
               \mbox{if $\gamma \leq \gamma_{\rm b\,s}$}, 
               \\[0.5cm]	
	n_{\rm m}(\gamma) &= 
               n_{\rm b}
               \biggl(\displaystyle\frac{\gamma}{\gamma_{\rm b\,s}}\biggr)^{-p_{\rm m}}& 
               \mbox{if $\gamma_{\rm b\,s} < \gamma \leq
               \gamma_{\rm b\,f}$}, 
               \\[0.5cm]	
	n_{\rm h}(\gamma) &=  
               f n_{\rm b}
               \biggl(\displaystyle\frac{\gamma}{\gamma_{\rm b\,f}}\biggr)^{-3}& 
               \mbox{if $\gamma > \gamma_{\rm b\,f}$}
            \end{array}
\end{equation}
where $f = (\gamma_{\rm b\,s}/\gamma_{\rm b\,f})^{p_{\rm m}}$, as shown
               in Figure~\ref{fig:leaky}.

Thus we find that the normalisation factor
$n_{\rm b}$ is given by:
\begin{eqnarray}
n_{\rm b}[t_{\rm inj}]
              &=& \frac{u_{\rm e}[t_{\rm inj}]}
              {m_{\rm e} c^2}
              \biggl(  \gamma_{\rm b\,s}^2
	      \biggl[\ln\frac{\gamma_{\rm b\,s}}{\gamma_{\rm min}}  
              \biggr]
              + 
              \frac{\gamma_{\rm b\,s}^{p_{\rm m}}}{p_{\rm m} - 2}
              \biggl[
              \frac{1}{\gamma_{\rm b\,s}^{p_{\rm m} - 2}} -
              \frac{1}{\gamma_{\rm b\,f}^{p_{\rm m} - 2}}
              \biggr]
              +
              f\gamma_{\rm b\,f}^{3}\biggl[ 
              \frac{1}{\gamma_{\rm b\,f}} - 
              \frac{1}{\gamma_{\rm max}}
              \biggr]
              \biggr)^{-1}.
\end{eqnarray}
The energy density in the lobe at time $t$, $u_{\rm e}[t]$, is directly
proportional to the pressure in the lobe and the head at time $t$, given
in equation~\ref{eq:head}, with coefficients of proportionality which are
of order unity (see Kaiser et al.\ 1997).

We take $\gamma_{\rm min}$ to be 1 and $\gamma_{\rm max}$, the maximum
value of the Lorentz factor in the hotspots, to be $10^{14}$ since emission
in the intense magnetic fields of the hotspots is observed to extend to
frequencies of several tens of GHz.

We now consider the thermodynamics of the transition of the relativistic
plasma from the jet, to the hotspot (meaning the post-jet shock region),
and then through the head to its eventual home in the lobe itself.
Following Kaiser et al.\ (1997) we equate the final energy state of an
elemental volume $\delta V_{\rm l}$ of plasma in the lobe with the energy
it had when the jet delivered it to the hotspot (a dominant fraction of
which is assumed to be in the form of bulk kinetic energy) minus the work
done in adiabatically expanding from the hotspot to the lobe. We assume
that the jet-power $Q_{\rm o}$ is completely converted into thermal energy
at the hotspot. This seems reasonable given that the flow speeds into
hotspots are close to $c$ (Wardle \& Aaron 1997) while those out of
hotspots can be up to $\sim 0.2c$ and are oppositely directed (Liu et al.\
1992).  [The energy lost to radiation mechanisms is neglected and can be
shown to be many orders of magnitude smaller than the other quantities
involved.] Hence, 
\begin{equation}
\frac{p_{\rm l} \delta V_{\rm l}}{\Gamma_{\rm l} - 1} = 
Q_{\rm o} \delta\,t_{\rm inj} + \biggl(\frac{p_{\rm l} \delta V_{\rm l}}{\Gamma_{\rm l} - 1} -
\frac{p_{\rm hs} \delta V_{\rm hs}}{\Gamma_{\rm hs} - 1} \biggr),
\end{equation}
where $\delta V_{\rm l}$ is the volume element injected into the lobe,
$\delta V_{\rm hs}$ is the volume which this lobe element occupied when it
was in the hotspot, $\delta\,t_{\rm inj}$ is the time interval at $t_{\rm
inj}$ over which this element is injected into the lobe, $p_{\rm hs}$ and
$p_{\rm l}$ are the pressure in the hotspot and lobe respectively and
$\Gamma_{\rm l}$ and $\Gamma_{\rm hs}$ are the adiabatic indices of the
lobe and hotspot respectively each taken to be 4/3.  Since the expansion
out of the hotspot is approximated to be adiabatic, we may write that:
\begin{equation}
\label{eq:workdone}
\frac{p_{\rm hs}}{p_{\rm l}}\frac{\delta V_{\rm hs}}{\delta V_{\rm l}}
= \biggl(\frac{p_{\rm hs}}{p_{\rm l}}\biggr)^{\frac{\Gamma_{\rm l} -
1}{\Gamma_{\rm l}}}
\end{equation}
and using equation~\ref{eq:pQ} we may thus write the contribution to
the volume of the lobe from the element injected in $\delta\,t_{\rm
inj}$ at $t_{\rm inj}$ as:
\begin{equation}
\label{eq:volel}
\delta V_{\rm l} = 
\biggl[\frac{(\Gamma_{\rm l} - 1)Q_{\rm o}}{p_{\rm l}}
\biggl(\frac{Q_{\rm o}}{cA_{\rm hs}p_{\rm l}}\biggr)^{\frac{1 -
\Gamma_{\rm l}}{\Gamma_{\rm l}}}
\biggr] \delta\,t_{\rm inj}.
\end{equation}
After multiplying by a term to allow for the further (adiabatic) expansion
of the element as the source evolves (namely $[t/t_{\rm inj}]^{a_1}$, where
$a_1$ is a constant which relates the expansion of volume $V$ after time
$t$ as $V \propto t^{a_{1}}$), the integral of equation~\ref{eq:volel} over
$t_{\rm inj}$ may be identified as the volume $V_{\rm l}(t)$ now occupied
at time $t$ which has been injected into the lobe since time $t_{\rm
min}$, and hence as the volume term in equation~\ref{eq:sync}; this is
given by:
\begin{equation}
\label{eq:volelint}
V_{\rm l}(t) = 
\int_{t_{\rm min}}^{t}
\biggl[\frac{(\Gamma_{\rm l} - 1)Q_{\rm o}}{p_{\rm l}[t_{\rm inj}]}
\biggl(\frac{Q_{\rm o}}{cA_{\rm hs}p_{\rm l}[t_{\rm inj}]}\biggr)^{\frac{1
- \Gamma_{\rm l}}{\Gamma_{\rm l}}} \biggr] 
\biggl(\displaystyle\frac{t}{t_{\rm inj}}\biggr)^{a_1}
dt_{\rm inj}.
\end{equation}

Note that the net result of this expansion is that only a fraction $\mu$
of the thermal energy deposited in the hotspots by the jet appears as
stored energy in the lobes, the remainder being lost as work done against
the shocked ambient medium in the bow-shock. The magnitude of $\mu$ can be
estimated from equation~\ref{eq:workdone} which is is essentially the
reciprocal of the ratio of the internal energy of a given element in the
lobe to that which it had in the hotspot.  For $\Gamma_{\rm l} =
\Gamma_{\rm hs} = 4/3$, and typical values of $p_{\rm hs}/p_{\rm lobe}
\sim$ 10 -- 100, $\mu$ varies between 0.6 -- 0.3.

\begin{figure}[!t]
\begin{picture}(50,230)(0,0)
\put(120,240){\includegraphics{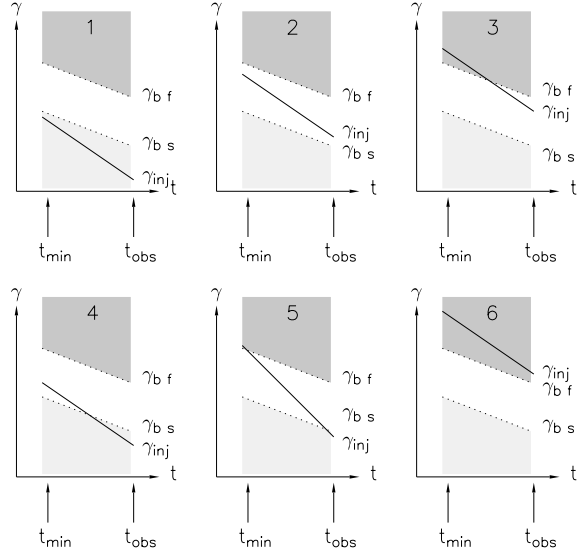}}
\end{picture}
{\caption[junk]{\label{fig:two_bf} Schematic depiction of the six possible
ways in which the Lorentz factor at injection of those particles which
ultimately contribute to the radiation we observe (emitted at $t_{\rm
obs}$) may relate to the Lorentz factors at the breaks in the injected
spectrum.  As in Figure~\ref{fig:leaky}, the light grey regions are where
$p = 2$, the dark grey regions are where $p = 3$ and the white regions are
where $p = p_{\rm m}$. }}
\end{figure}

Figure~\ref{fig:two_bf} indicates the six possible ways in which the
slope of the energy distribution of the relativistic particles
injected into the lobe (i.e.\ the gradient at $\gamma_{\rm inj}$) may
change with time, according to the relative value of $\gamma_{\rm
inj}$ compared to $\gamma_{\rm b\,s}$ and $\gamma_{\rm b\,f}$.

We now consider the relation between the Lorentz factor of a particle
injected into the lobe which will later contribute to the emission we
observe at $t_{\rm obs}$, and its Lorentz factor at the time when the
particle is injected.  At $t_{\rm obs}$ (the point when the light we
observe leaves the object, i.e.\ when the object intercepts our
light-cone) the Lorentz factor of the particles injected at that point in
time, must be given by equation~\ref{eq:lorentz} for a given emitted
frequency.  The particles which contribute to the emission leaving the
object at $t_{\rm obs}$ but which were injected into the lobe at a time
$\Delta t$ prior to $t_{\rm obs}$ must have had a Lorentz factor higher
when they were injected than they have at $t_{\rm obs}$, to compensate for
the energy depletion during $\Delta t$. The maximum value of $\Delta t$
gives $t_{\rm min}$, the time before which no injected particles
contribute to $t_{\rm obs}$. As in Kaiser et al.\ (1997) $t_{\rm min}$ is
the time before which no particles which give rise to the emission
observed (at the chosen frequency) at $t_{\rm obs}$ can be injected
because the losses they suffer, given their Lorentz factors, would be too
catastrophic. Thus for the ensemble of particles contributing to the
radiation at a given emitted frequency at $t_{\rm obs}$, those with the
largest Lorentz factors at injection are injected at $t_{\rm min}$ and
those with the lowest are those actually injected at $t_{\rm obs}$.

We now consider the relation between the Lorentz factors injected at time
$t_{\rm min}$ and $t_{\rm obs}$ and the corresponding breaks in the
injected energy spectrum (depicted in Fig.~\ref{fig:leaky}). If the
Lorentz factor of the injected electrons is always above these two or
always below or always between them then the calculation is
straightforward involving just one integral with the weighting function in
each case being just either the bottom or the top or the middle form
respectively of the full form of the weighting function listed in
equation~\ref{eq:ngamma}.  If, during the time between $t_{\rm min}$ and
$t_{\rm obs}$ the gamma-break should overlap with two distinct regimes in
the spectrum, the integral must be performed in two steps. If during the
time between $t_{\rm min}$ and $t_{\rm obs}$ the gamma-break should
overlap with all three distinct regimes in the spectrum, the integral must
be performed in three steps (see Figure~\ref{fig:two_bf}), thus:
\newpage
\begin{eqnarray}
&P_{\nu}[t_{\rm obs}]& =  \frac{1}{6\pi\nu} \sigma_{\rm e} c \times
 \nonumber \\ 
&\biggl(&             
\int_{t_{\gamma_{\rm b\,s}}}^{t_{\rm obs}}
u_{\rm B}[t_{\rm inj}]
\gamma^3[t_{\rm inj}] n_{\rm b} 
\biggl(\frac{\gamma[t_{\rm inj}]}{\gamma_{\rm b\,s}}\biggr)^{-2}
\frac{Q_{\rm o}}{p_{\rm l}}
\biggl(\frac{Q_{\rm o}}{cA_{\rm hs}p_{\rm l}}\biggr)^{\frac{1 -
 \Gamma_{\rm l}}{\Gamma_{\rm l}}}
\biggl(\frac{t_{\rm obs}}{t_{\rm inj}}\biggr)^{-a_{1}(1/3 + \Gamma_{\rm l})}
\,dt_{\rm inj}
\nonumber \\                                        
&+&
\int_{t_{\gamma_{\rm b\,f}}}^{
t_{\gamma_{\rm b\,s}}}
u_{\rm B}[t_{\rm inj}]
\gamma^3[t_{\rm inj}] n_{\rm b} 
\biggl(\frac{\gamma[t_{\rm inj}]}{\gamma_{\rm b\,s}}\biggr)^{-p_{\rm m}}
\frac{Q_{\rm o}}{p_{\rm l}}
\biggl(\frac{Q_{\rm o}}{cA_{\rm hs}p_{\rm l}}\biggr)^{\frac{1 -
 \Gamma_{\rm l}}{\Gamma_{\rm l}}}
\biggl(\frac{t_{\rm obs}}{t_{\rm inj}}\biggr)^{-a_{1}(1/3 + \Gamma_{\rm l})}
\,dt_{\rm inj}
\nonumber \\                                        
&+&
\int_{t_{\rm min}}^{t_{\gamma_{\rm b\,f}}}
u_{\rm B}[t_{\rm inj}]
\gamma^3[t_{\rm inj}] fn_{\rm b} 
\biggl(\frac{\gamma[t_{\rm inj}]}{\gamma_{\rm b\,f}}\biggr)^{-3}
\frac{Q_{\rm o}}{p_{\rm l}}
\biggl(\frac{Q_{\rm o}}{cA_{\rm hs}p_{\rm l}}\biggr)^{\frac{1 -
 \Gamma_{\rm l}}{\Gamma_{\rm l}}}
\biggl(\frac{t_{\rm obs}}{t_{\rm inj}}\biggr)^{-a_{1}(1/3 + \Gamma_{\rm l})}
\,dt_{\rm inj} \biggr)
\nonumber \\                                        
&\phantom{\biggr)}& 
\end{eqnarray}
where the volume element in equation~\ref{eq:sync} has been replaced by
the integrand on the right-hand side of equation~\ref{eq:volelint}, and
where $u_{\rm B}[t_{\rm inj}]$ represents the magnetic energy density in
the lobe at the time a given element of particles is injected.  The
pressure in the lobe $p_{\rm l}$ is taken to be a constant factor of six
lower than the pressure in the head $p_{\rm head}$ given by
equation~\ref{eq:phead}.  We use Kaiser et al.'s (1997) expression for
$\gamma[t_{\rm inj}]$ in their equation 10.
\newpage
\twocolumn
\subsection{Predictions for the evolution of individual sources}
\label{sec:indiv}
\subsubsection{Sizes and energy budgets}
\label{sec:energy}
The volume of the radio source at the time of `observation' may be found
by integration of the volume elements which have been injected over the
life of the radio source from a very early time in its life (e.g. $t_{\rm
min} = 1 \times 10^4$ years), up to $t_{\rm obs}$, i.e.\ by performing the
integral of equation~\ref{eq:volelint}, to allow for the expansion of
these elements between the time they were injected and the time the
`observation' is made.

The energy transported, $E_{\rm t}$, by the jets in a radio source, after
they have been switched on for a time $t$, is given simply by
\begin{equation}
\label{eq:et}
E_{\rm t} = 2 Q_{\rm o} t. 
\end{equation}
The energy, $E_{\rm s}$, which is stored in the lobes according to the
usual minimum energy arguments depends on the luminosity and the volume of
the lobes as follows:
\begin{equation}
\label{eq:es}
E_{\rm s} = \chi P_{\rm 151}^{4/7} \displaystyle\frac{D^{9/7}}{R_{\rm T}^{6/7}} 
\end{equation}
where $\chi$ is a constant throughout the life of a source which depends
on assumptions about the low-energy cut-off of particles and their filling
factor in the radio lobes (see Miley 1980), $P_{151}$ is the luminosity at
rest-frame 151 MHz, $D$ is the total linear size of the radio source and
where $R_{\rm T}$ is the axial ratio of the radio source, defined as the
total length divided by the width, as one would measure on a radio map.

In our model, throughout a source's life, it does not store energy at a
rate which is proportional to its age; this arises because the rate at
which work is done by the expanding source itself depends on the age of
the source.  This follows from equation~\ref{eq:workdone} and the fact
that hotspot pressure does not scale with time in the same way as lobe
pressure (unlike self-similar models).  The prediction of axial ratios by
our model comes from finding (the square root of $\pi/4 \times$) the
quotient of the cube of the linear size derived from
equation~\ref{eq:falle} and the volume obtained from the integration of
equation~\ref{eq:volelint}. We find that throughout the lifetime of an
individual source its axial ratio steadily increases thus, for an
individual source its expansion is not self-similar. For sources with
$Q_{\rm o} = 10^{38}$ W, we find that the axial ratios increase from 2 to
8, while for sources with $Q_{\rm o} = 10^{40}$ W, the axial ratios
increase from 4 to 13.  Our model thus reproduces axial ratios which are
larger in sources with higher jet-powers, consistent with the findings of
Leahy \& Williams (1984) and Leahy et al.\ (1989) who found axial ratios
between 4 and 12 (using the definition that the axial ratio is the total
projected length of the source divided by its width). We note that the
closer the jet-axis is to the line-of-sight, the smaller the measured
axial ratio of a source will be because of projection effects.  Both Leahy
\& Williams and Leahy et al.\ found evidence of a trend of increasing
axial ratio with increasing source luminosity and commented that the
scatter was greater than that expected from projection effects
alone. Leahy \& Williams' study was for a subset of the 3C sample (see
\S\ref{sec:samples3c}) with angular sizes greater than
$45^{\prime\prime}$. With the elimination of these smaller (hence, given
the sample, younger) sources it is then unsurprising in the context of our
model that they found their more powerful sources to have higher axial
ratios.

\subsubsection{$P$--$D$ tracks}
\label{sec:tracks}

\begin{figure}[!h]
\begin{picture}(50,200)(0,0)
\put(-58,-190){\includegraphics{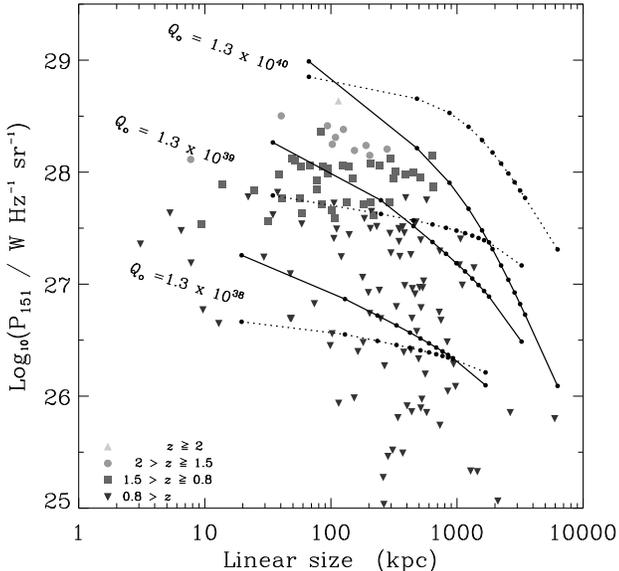}}
\end{picture}
{\caption[junk]{\label{fig:track3c} We overlay on this plot in {\em solid
lines} the tracks through the $P$--$D$ plane of three sources (upper to
lower) with $Q_{\rm o} = 1.3 \times 10^{40}$ W at $z = 2$, $Q_{\rm o} =
1.3 \times 10^{39}$ W at $z = 0.5$, $Q_{\rm o} = 1.3 \times 10^{38}$ W at
$z = 0.2$. We overlay in {\em dotted lines} the tracks from the model of
Kaiser et al.\ (1997) for sources with the same jet-powers and
redshifts. The larger symbols in this plot show the rest-frame luminosities
at 151\,MHz of members of the 3C complete sample (\S\ref{sec:samples3c})
against their projected linear sizes. Since this is a logarithmic plot,
the differences in the distributions between the projected and the true
linear sizes of sources will be slight.  Small dots lie on the tracks to
indicate the times 1, 10, 20\ldots 100, 200 Myr through the source's
life. All tracks in this figure were generated assuming an environment as
described in \S\ref{sec:env}, i.e.\ with $\beta = 1.5$ and with $c_1 =
1.8$. }}
\end{figure}

Figure~\ref{fig:track3c} shows the tracks across the $P$--$D$ plane of
sources according to our model described above with various values of
jet-power and redshift and compares them with the predictions of the model
of Kaiser et al.\ (1997). Note that in each case the tracks may be
approximately described by two power-laws, either side of the knee. In
each case on the left-hand side of the knee the gradient of our tracks are
steeper than those of Kaiser et al. This arises because in our model the
synchrotron emitting particles adiabatically expand out of a constant
pressure hotspot into a lobe whose pressure falls throughout the source
lifetime, thus the losses due to adiabatic expansion increase with
time. In the model of Kaiser et al.\ (1997), the expansion of the
synchrotron emitting population is from a head into a lobe whose pressures
are in a constant ratio throughout the lifetime of the source. Thus the
adiabatic losses in this latter model do not increase with time. The
increasing adiabatic expansion losses with time in our model ensure that
the tracks in the $P$--$D$ plane can be very steep even in an environment
whose $\beta$ exponent is not particularly large (e.g.\ 1.5 here as
described in \S\ref{sec:env}). This point plays a crucial r\^{o}le in
determining the ease with which (particularly large, powerful) sources fall
through the survey flux-limit. 

The `knee' between the two regimes of a $P$--$D$ track occurs earlier in
the lives of sources at higher redshifts. Below the knee the magnetic
field in the lobes $B_{\rm lobe}$ is stronger than the equivalent field
due to the cosmic microwave background $B_{\rm CMB}$. Spectral losses for
a particle of Lorentz factor $\gamma$ go as $\gamma^2 B_{\rm lobe}^2$, but
for a given $\gamma$ this rate of energy loss decreases with time since
$B_{\rm lobe}$ is decreasing with time.  Throughout the lifetime of the
source, $B_{\rm lobe}$ will be falling --- thus for radiation at a fixed
frequency which we might observe in our survey, the Lorentz factor
$\gamma$ will be proportional to the reciprocal of the root of the lobe
magnetic field (see equation~\ref{eq:lorentz}).  Thus as the source ages,
increasingly energetic particles are responsible for the emission observed
at some fixed frequency. Once the magnetic field of the lobe has dropped
below $B_{\rm CMB}$, the inverse Compton losses will dominate the rate at
which the high $\gamma$ particles are depleted, in this second regime in
the source's life.  This magnetic field is constant and not decreasing as
$B_{\rm lobe}$ is, so the spectral losses thereafter remain at a fixed,
not a decreasing, rate.  However, the lobe magnetic field $B_{\rm lobe}$
continues to fall requiring yet higher Lorentz factors for emission at a
given frequency, and the high Lorentz factor population is being depleted
at a steady rate due to inverse Compton scattering from the cosmic
microwave background.

\begin{figure}[!h]
\begin{picture}(50,200)(0,0)
\put(-58,-190){\includegraphics{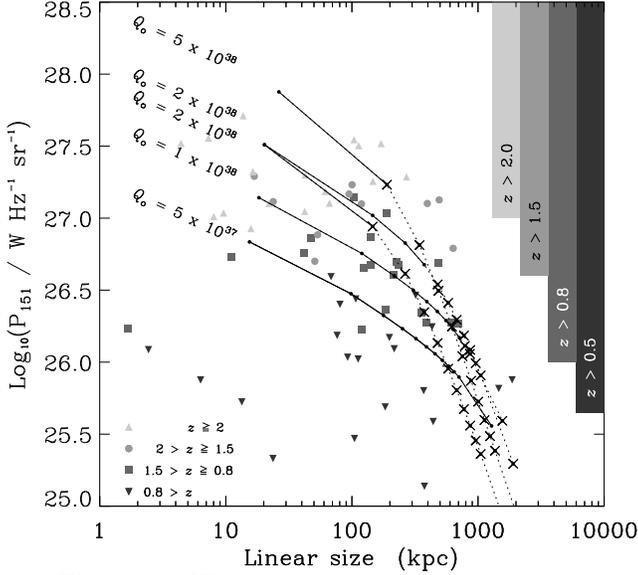}}
\end{picture}
{\caption[junk]{\label{fig:track7c} We overlay on this plot the tracks
through the $P$--$D$ plane of five sources (upper to lower) with $Q_{\rm
o} = 5 \times 10^{38}$ W at $z = 3$, $Q_{\rm o} = 2 \times 10^{38}$ W at
$z = 1.5$, $Q_{\rm o} = 2 \times 10^{38}$ W at $z = 2.5$, $Q_{\rm o} = 1
\times 10^{38}$ W at $z = 0.8$, and $Q_{\rm o} = 5 \times 10^{37}$ W at $z
= 0.5$.  The left-most light-grey bar indicates that for an object to be
brighter than the flux-limit of the 7C sample (0.5 Jy), if it is at $z >
2$ then its luminosity must be in the range indicated by the bar; the
corresponding bars going to lower luminosities are for the lower redshifts
1.5, 0.8 and 0.5 respectively. The solid lines of the tracks become dotted
when the luminosity and redshift of that source are such that they fall
below the survey flux-limit of 0.5 Jy.  The larger symbols in this plot
show the rest-frame luminosities at 151\,MHz of members of the 7C complete
sample (\S\ref{sec:samples7c}) against their projected linear sizes.
Small dots lie on the tracks to indicate the times 1, 10, 20\ldots 100,
200 Myr through the source's life; these become cross-symbols when an
object falls below the flux-limit. It can be seen that because the tracks
of the more powerful, higher-redshift sources become steeper, and because
of the consequences of the flux-limit, these sources fall below a survey
flux-limit at an earlier stage in their lives than do sources at lower
power and redshift. The same environment and value of $c_{1}$ are assumed
as for Figure 13. }}
\end{figure}

Figure~\ref{fig:track7c} shows tracks across the $P$--$D$ plane derived
from our model for various values of jet-power and redshift. The solid
lines of the tracks become dotted when the luminosity and redshift of that
source are such that they fall below the survey flux-density limit of 0.5
Jy. Comparison of the time markers along the different tracks show that
the tracks of the more powerful, higher-redshift sources become steeper,
this together with the consequences of the flux-limit, means that these
sources fall below a survey flux-limit at an earlier stage in their lives
than do sources at lower power and redshift. Here then, is an indication
of the origins of the effect we discuss in more detail in
\S\ref{sec:youthz}, the ``youth--redshift degeneracy'' which is
responsible for linear size evolution, on the assumption that radio
sources are intercepted by our light-cone in a manner which is essentially
random.  We examine in detail the correct way to calculate the
interception of the source population with our light-cone in
\S\ref{sec:zmod} and then see clearly how these two effects arise.

\subsection{Redshift distribution}
\label{sec:zmod}
In this section we consider how to model the number of radio sources born
at a sequence of cosmic times, and their subsequent interception with our
light-cone.  Successive populations of radio sources are hatched every
$10^6$ years, and are evolved, starting at some initial high redshift ($z
= 10$ is used).

We assume that the number of sources born at a given cosmic time ($t$),
per unit cosmic time, per unit co-moving volume element (defined as having
a proper volume of one cubic Gpc at redshift zero) is given by $\rho(t)\,dt
= \rho(z)\,dz$ which may be parameterised by:
\begin{equation}
\rho(z)\,dz \propto 
     \exp \displaystyle\frac{1}{2}\biggl(
          \displaystyle\frac{z - z_{1}}{\Delta z_{1}}\biggr)^{2}dz. 
\label{eq:rhoz}
\end{equation}
We further assume that at any given cosmic time, this description holds at
any spatial point in the Universe, since this is presumed to be
homogeneous and isotropic. We make the approximation that all sources are
precisely following the Hubble flow.

We wish to consider the total number of sources born at a given cosmic
time which will be intercepted by our light cone, within a certain
interval in cosmic time from the time when they were born. [This interval
is chosen to be the maximum length of time for which the beams in each
radio source are switched on. We assume that in each radio source this is
$5 \times 10^{8}$ years (see justification of this number in
\S\ref{sec:vmod}), presuming no dependence on redshift or on intrinsic
jet-power.] If a source born at a given point in cosmic time is to be
intercepted by our light-cone at some point between the time when it was
born and the time when its beam is switched off this has implications for
the distance it must be from us at the point when it is born. If the
source is intercepted by our light-cone at the time when it is born, then
its distance from us is simply given by the distance along our light-cone
to this redshift (which is hereafter referred to as $z_{\rm birth}$). If
the source is intercepted by our light-cone at the moment when the beam is
switched off (i.e.\ $5 \times 10^{8}$ years after the source was born)
then its distance from us at this later point in cosmic time must be the
distance along the light-cone to this lower redshift, but this must
correspond to a specific distance from us at the earlier cosmic time when
the source was born. With this information a thick radial shell of a
specific proper volume whose maximum and minimum radii are defined by the
maximum distance and the minimum distance a source may be from us if it is
born at a given $z_{\rm birth}$ {\em and} if it is intercepted by our
light-cone within a cosmic time of $5 \times 10^8$ years from $z_{\rm
birth}$. This region, for a given $z_{\rm birth}$, is hereafter called the
`relevant co-moving volume element'. The relevant co-moving volume element
is the maximum size, for a given solid angle, of a box which contains all
the sources which must be intercepted within an interval of $5 \times
10^{8}$ years of being born at the redshift $z_{\rm birth}$ (and no
sources born at this redshift which are not intercepted by our light-cone
within this interval).

The form of the metric which we use in this analysis is:
\begin{equation}
ds^2 = c^2 dt^2 - \frac{R^2(t)}{(1 + \frac{1}{4}kr^2)^2} [
         dr^2 + r^2(d\theta^2 + \sin^2\theta d\phi^2)],
\label{eq:metric}
\end{equation}
where $R(t)$ is the scale factor of the Universe at cosmic time $t$, $r$
is a radial co-moving co-ordinate, $\theta$ and $\phi$ are the polar and
azimuthal co-ordinates; $k$ refers to the curvature of the Universe ($k =
0$ for a spatially flat Universe, $k = 1$ for an open Universe and $k =
-1$ for a closed Universe).  Following standard practice, we introduce an
alternative radial co-moving co-ordinate $\chi$ defined in terms of $r$
by: 
\begin{equation}
\begin{array}{lll}
	\chi      & = r             & \mbox{if $k =  0$} \\
	\sin\chi  & = r/[1 + r^2/4] & \mbox{if $k =  1$} \\
	\sinh\chi & = r/[1 - r^2/4] & \mbox{if $k = -1$} \\
        \end{array}
\end{equation}
writing the space-time metric as 
\begin{equation}
ds^2 = c^2 dt^2 - dl^2,
\end{equation}
we have for $k = 0$:
\begin{equation}
dl^2 = R^2(t)[d\chi^2 + \chi^2(d\theta^2 + \sin^2\theta d\phi^2)]
\end{equation}
while for $k = 1$, we write:
\begin{equation}
dl^2 = R^2(t)[d\chi^2 + \sin\chi^2(d\theta^2 + \sin^2\theta d\phi^2)],
\end{equation}
and for $k = -1$, we write:
\begin{equation}
dl^2 = R^2(t)[d\chi^2 + \sinh\chi^2(d\theta^2 + \sin^2\theta d\phi^2)].
\end{equation}
In the three cases the proper volume $\Delta\,V$ of a spherical shell,
whose inner and outer edges are at $\chi_{1}$ and $\chi_{2}$ respectively,
(a co-moving volume element) is given by:
\begin{equation}
 \Delta\,V = \left\{ 
         \begin{array}{ll}
         2\pi R^3(t) \biggl[\frac{2}{3}\chi^3\biggr]^{\chi_{2}}_{\chi_{1}} & 
         \mbox{for $k = 0$} \\[0.6cm]
         2\pi R^3(t)\biggl[\chi - \frac{1}{2}\sin{2\chi}\biggr]^{\chi_{2}}_{\chi_{1}} & 
         \mbox{for $k = +1$} \\[0.6cm]
         2\pi R^3(t)\biggl[\frac{1}{2}\sinh{2\chi} - \chi\biggr]^{\chi_{2}}_{\chi_{1}} & 
         \mbox{for $k = -1$}. \\
         \end{array}
\right.
\label{eq:vol} 
\end{equation}

Along a light-ray, the interval $ds = 0$, and along a radial light-ray
\begin{equation}
ds^2 = c^2 dt^2 - R^2(t)d\chi^2
\end{equation}
is independent of the value of $k$ so we may write
\begin{equation}
\chi = c\int^{t_{\rm now}}_{t} \frac{1}{R(t)} dt.
\end{equation}

The limits of integration, $\chi_{2}$ and $\chi_{1}$, are respectively the
radial co-ordinates to the outer and inner surfaces of the thick radial
shell and are given by:
\begin{equation}
\chi_{2} = c\int^{t_{\rm now}}_{t_{\rm birth}} \frac{1}{R(t)} dt,
\end{equation}
and
\begin{equation}
\chi_{1} = c\int^{t_{\rm now}}_{t_{\rm birth} + 5 \times 10^8 {\rm yr}} 
                 \frac{1}{R(t)} dt,
\end{equation}
where $t_{\rm birth}$ is the cosmic time when this batch of sources is born.

In order to convert proper volumes measured in cubic Gpc at the time of
birth into units which will be cubic Gpc at redshift zero, the proper
volumes are multiplied by $(1 + z_{\rm birth})^{3}$.

\begin{figure}[!h]
\begin{picture}(50,140)(0,0)
\put(-13,-170){\includegraphics{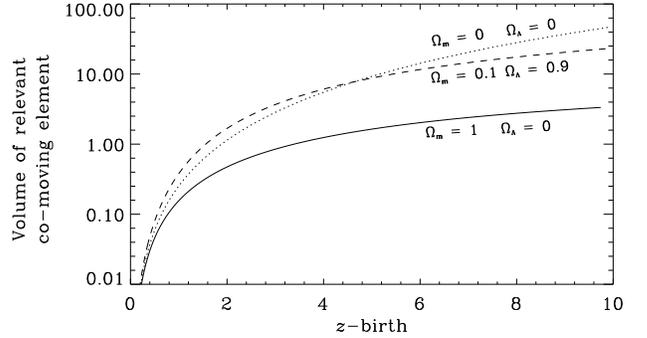}}
\end{picture}
{\caption[junk]{\label{fig:comovol.ps} This figure shows the volume of the
`relevant co-moving element' (see text for definition; the units are those
which will be cubic Gpc at redshift zero) at a given redshift, in each of
the three different cosmological models we consider in this paper.}}
\end{figure}

Once the volume of this co-moving element is calculated in a chosen
cosmological model, it is then simple to calculate the number of sources
born [per solid angle born in the chosen interval in cosmic time] at a
given $z_{\rm birth}$ which will ultimately be intercepted by our
light-cone, by multiplying this by $\rho(z)$ given in
equation~\ref{eq:rhoz}.  Precisely when each source intercepts our
light-cone is determined by its radial distance from us at the cosmic time
corresponding to $z_{\rm birth}$ within the radial shell described
above. Since we assume spatial homogeneity we deem these sources to be
randomly distributed within the volume of the box. Thus the length of time
which elapses after a given source is born before it intercepts the
light-cone, (called $t_{\rm age}$) is randomly drawn between 0 and $5
\times 10^{8}$ years, but not from a uniform distribution. The uniformity
of space requires that within equal (infinitesimal) volumes within the
thick radial shell there should be equal numbers of objects born at
$z_{\rm birth}$. Sources are randomly allocated to equal volume shells,
then their radial co-ordinate $(\chi)$ is derived by inverting the
equation~\ref{eq:vol} appropriate for the value of $k$ chosen, and solving
this using an iterative method. We see that this process means that a
source is therefore more likely to have a larger radial co-ordinate within
a given spherical shell than a smaller one, since the sources are
uniformly distributed in volume rather than in their radial co-ordinate,
and hence there is a mild bias arising from geometry for a source to be
seen at an earlier rather than a later stage in its life.  Once the
appropriate radial co-ordinate is drawn, the cosmic time ($t_{\rm obs}$)
at which the source is intercepted by our light-cone may be determined
from
\begin{equation}
\chi = c\int^{t_{\rm now}}_{\rm t_{\rm obs}} \frac{1}{R(t)} dt.
\end{equation}
By subtracting the cosmic time when the source is intercepted by the
light-cone from the cosmic time when the source was born, the time for
which a source should evolve i.e.\ $t_{\rm age}$, according to its initial
parameters, is known. Thus,
\begin{equation}
t_{\rm age}  = t_{\rm obs} - t_{\rm birth}.
\end{equation}

Therefore, at a given point in cosmic time, we know the exact number of
sources born which will intercept the light-cone thereafter within a
maximum time-period of $5 \times 10^{8}$ years, and the length of time for
which each evolves until the light ultimately observed by us is emitted.
The redshift of such an object is then derived by considering the cosmic
time at which the source was born, plus this time for which the source
lived ($t_{\rm age}$) until the light which we ultimately observe on the
Earth, left the source.

\subsection{Parameters and results}
\label{sec:pararesults}

A jet-power is allocated to each jet in a simulated object which has been
drawn randomly from a distribution of the form:
\begin{equation}
\label{eq:qpars}
 p(Q_{\rm o}) d\,Q_{\rm o} = \left\{ 
            \begin{array}{ll}
               Q_{\rm o}^{-x}d\,Q_{\rm o}  & 
               \mbox{if $Q_{\rm min} < Q_{\rm o} < Q_{\rm max}$}, 
               \\
               0 & 
               \mbox{if $Q_{\rm o} \geq Q_{\rm max}$ or $Q_{\rm o} \leq Q_{\rm min}$ }
               \\
            \end{array}
     \right. 
\end{equation}
where $x$ is a positive number.  Each source is then allowed to evolve in
luminosity (according to the prescription discussed in \S\ref{sec:pmod})
at a (rest-frame) frequency $(1 + z) \times$ the observing frequency for
the length of time $t_{\rm age}$.  If the luminosity of the source at this
frequency is such that its flux is above the deemed survey flux-limit
(e.g., 0.5 Jy for a simulated 7C sample) given its redshift, it is
regarded as being detected by the survey.  Sources which do not meet the
survey flux-limit are thereafter ignored.  Note that we do not simulate
the effect of surface brightness, as distinct from flux density, in these
simulated surveys. However, as Riley (1989) discusses the numbers of
sources omitted from real surveys due to surface brightness effects in
large angular size sources is small indeed.

For those sources which do lie within the sample flux-limits, the
following calculations are performed: the luminosities at two other
frequencies (including rest-frame 151 MHz), for the same point in the
source's lifetime, are derived as is the projected linear size (according
to the prescription discussed in \S\ref{sec:vmod}). From three
(luminosity, frequency) points now derived for each source in the sample,
a quadratic expression may be fitted from which the spectral index at
rest-frame 151 MHz is derived, together with the $a_{1}$ and $a_{2}$
coefficients of spectral shape discussed in Section~\ref{sec:spectra}.

Figures~\ref{fig:real3c} and \ref{fig:mock3c} show real and simulated 3C
data.  Figures~\ref{fig:real7c} and \ref{fig:mock7c} show real and
simulated 7C data. In these simulations the `sky-areas' (i.e.\ the solid
angle covered by each survey) differ by a factor of 430, which is of the
same order as the factor by which the true sky areas covered by 3C and 7C
differ (4.2 sr versus 0.013 sr). For the simulations presented here the
values of the $Q_{\rm o}$--related quantities included in
equation~\ref{eq:qpars} are $x = 2.6$, $Q_{\rm max} = 5 \times
10^{42}\,{\rm W}$ and $Q_{\rm min} = 5 \times 10^{37}\,{\rm W}$. The
values of the $z$--related quantities are $z_{1} = 2.2$ and $\Delta z_{1}
= 0.6$.  

\subsection{Discussion of simulations}
A comparison of the simulated and real 3C data suggests that to first
order there is agreement in the basic quantities. Sources are detected out
to approximately the same redshift, the maximum size of sources detected
is approximately correct, the dependence between linear size and redshift
shows encouraging correspondence and spectral indices increase with
luminosity rather than with redshift. These correspondences also seem to
be broadly present in the 7C datasets though there is of course
considerably less scatter seen in the spectral index distributions; note
that we have not included in our model the possibility that acceleration
indices within the hotspot might be less than 0.5 [as suggested by the
work of Heavens \& Drury (1988)] so the small angular size, $\alpha < 0.5$
objects from 3C with $\log_{10} P_{151} \sim 27.5$ will not appear in our
simulations.  Particularly in the case of the 7C sample, there is a
discrepancy between the minimum spectral indices observed and those
simulated and it is unlikely that all of the discrepant sources could be
explained by invoking synchrotron self-absorbtion at these frequencies. In
the case of the 3C sample, the simulation shown in Figure~\ref{fig:mock3c}
has too high a proportion of sources with linear sizes $< 10$ kpc. It is
likely that the simplicity of our assumptions about the environments of
radio sources at these very small separations from the AGN are responsible
for this, in particular the rapidity of source expansion in these regimes
may be underestimated (as discussed in \S\ref{sec:vmod}).  We do not
include in our simulations the analogues of those sources which we believe
to be free--free absorbed or synchrotron self-absorbed which may be
responsible for the absence of these small sources in real samples.  The
bimodality in the simulated spectral indices is likely to arise in part
because we do not model any variation in hotspot size in the lifetime of
an individual source, or in different sources.

We defer detailed statistical comparison of the differently parameterised
simulated datasets and an investigation of the constraints on the
birth-function in the different cosmological models and the purported
existence of a redshift cutoff in the radio source population, to a
forthcoming paper.

\section{Remarks on interpretation}
\label{sec:remarks}
\subsection{Linear size evolution}
The anti-correlation of linear size with redshift arises naturally as a
consequence of imposing a survey flux-limit on a population of sources
each of whose luminosity falls off with age.  For high-power sources at
high-redshift, the $P$--$D$ tracks are seen to have steepening slopes
throughout the life of the source. This arises partly because of the
effect of inverse Compton losses to the cosmic microwave background (as
found by Kaiser et al.\ 1997) and partly because of increasing adiabatic
expansion losses of successively injected particles emerging from a
constant pressure hotspot into a lobe whose pressure decreases throughout
the life of the source. In the case of the highest power sources, the
injection index itself is steeper which causes further steepening of the
luminosity of a radio source with time.  The interception of our light
cone with a particular object is random throughout its lifetime, but high
redshift sources in the later stages of their lives are increasingly less
likely to be detected by surveys because of the steep decline of their
luminosity with age.  Thus the only high redshift objects we tend to
detect are younger (more luminous) ones --- which being younger, are
shorter; hence `linear size evolution' is seen.  This means that in a
survey, there is a strong correlation between youth and redshift in the
sense that the higher the redshift of a source, the earlier in its
lifetime we are observing it.  Such a degeneracy must be taken into
account when explaining ``redshift effects'', which we explore in
\S\ref{sec:youthz}.

\subsection{Injection indices and the $P$--$\alpha$ correlation}
\label{sec:injection}

The $P$--$\alpha$ correlation may arise in the way we have suggested with
the injection spectrum of particles, and the steepness of this
distribution, governed in a way which depends on the jet-power. It is
possible however, that the acceleration index of immediate post-shock
particles is not necessarily 0.5 as implied by first-order Fermi
mechanisms, but that this spectrum itself is governed by the jet-power in
some way depending on the detailed physical mechanisms of the
hotspot. Whatever the details of the underlying mechanism, an injection
spectrum dependent on the jet-power is required to give the observed
dependence of spectral index at rest-frame 151 MHz on luminosity; if the
dominant effective break frequency were comparable with rest-frame 151 MHz
then the drastic depletion of the power reservoir of the source would have
caused it to drop below the flux-limit of a typical flux-limited survey.

Possible evidence that the $P$--$\alpha$ correlation arises in the way
modelled in this paper, namely through the steepening of the energy
distribution of particles which are ultimately injected into the lobes by
radiative losses in the enhanced magnetic fields of the hotspots of
sources with more powerful jets, comes from the non-detection of optical
synchrotron emission associated with the hotspots of very powerful radio
sources. The number of hotspots which emit optical synchrotron radiation
is small, for example there is a deficit of optical hotspots in very
powerful radio sources such as Cygnus A (R\"{o}ser, Meisenheimer \& Yates
1996), whereas optical hotspots are detected in low-power radio sources
(Meisenheimer et al.\ 1989). If it is the case, as we have modelled in
this paper, that more powerful sources have higher magnetic fields in
their hotspots than do less powerful sources, then the absence of optical
synchrotron emission from the hotspots of the former arises simply because
the spectrum of the hotspot emission is so steep towards this frequency
regime.

\onecolumn
\begin{figure}[!t]
\begin{picture}(50,500)(0,0)
\put(-40,-85){\includegraphics{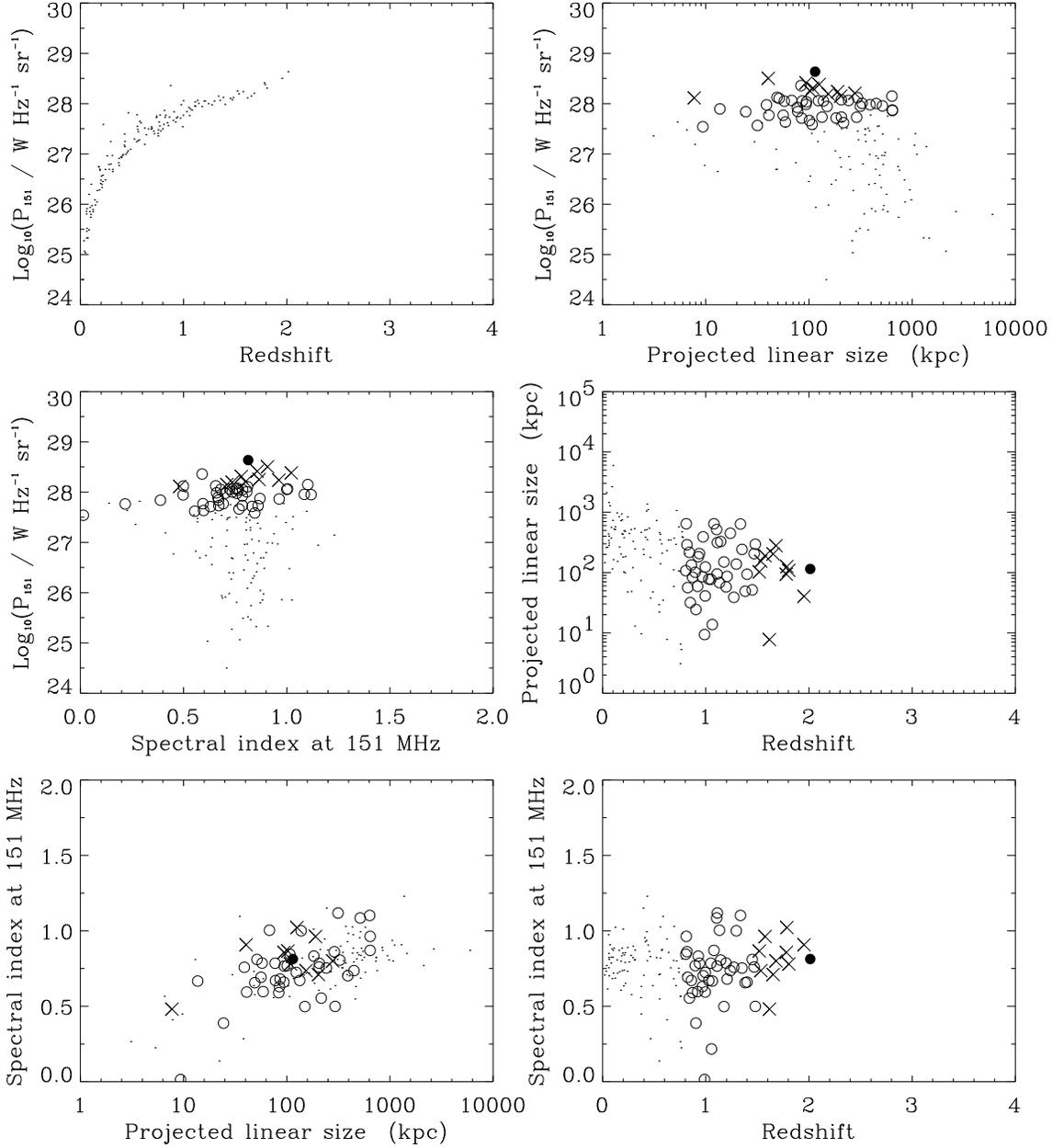}}
\end{picture}
{\caption[junk]{\label{fig:real3c} Figure showing real 3C data, for
FR\,IIs and DAs, assuming ${\Omega_{\rm M} = 1\ \&\ \Omega_{\Lambda} =
0} $, for the six planes in [$PDz\alpha$] parameter space. Solid
circles indicate objects at $z \geq 2$, crosses indicate objects
with $2 > z \geq 1.5$, open circles indicate $1.5 > z \geq 0.8$
and dots indicate sources with $z < 0.8$. }}
\end{figure}
\clearpage
\begin{figure}[!t]
\begin{picture}(50,570)(0,0)
\put(-40,-45){\includegraphics{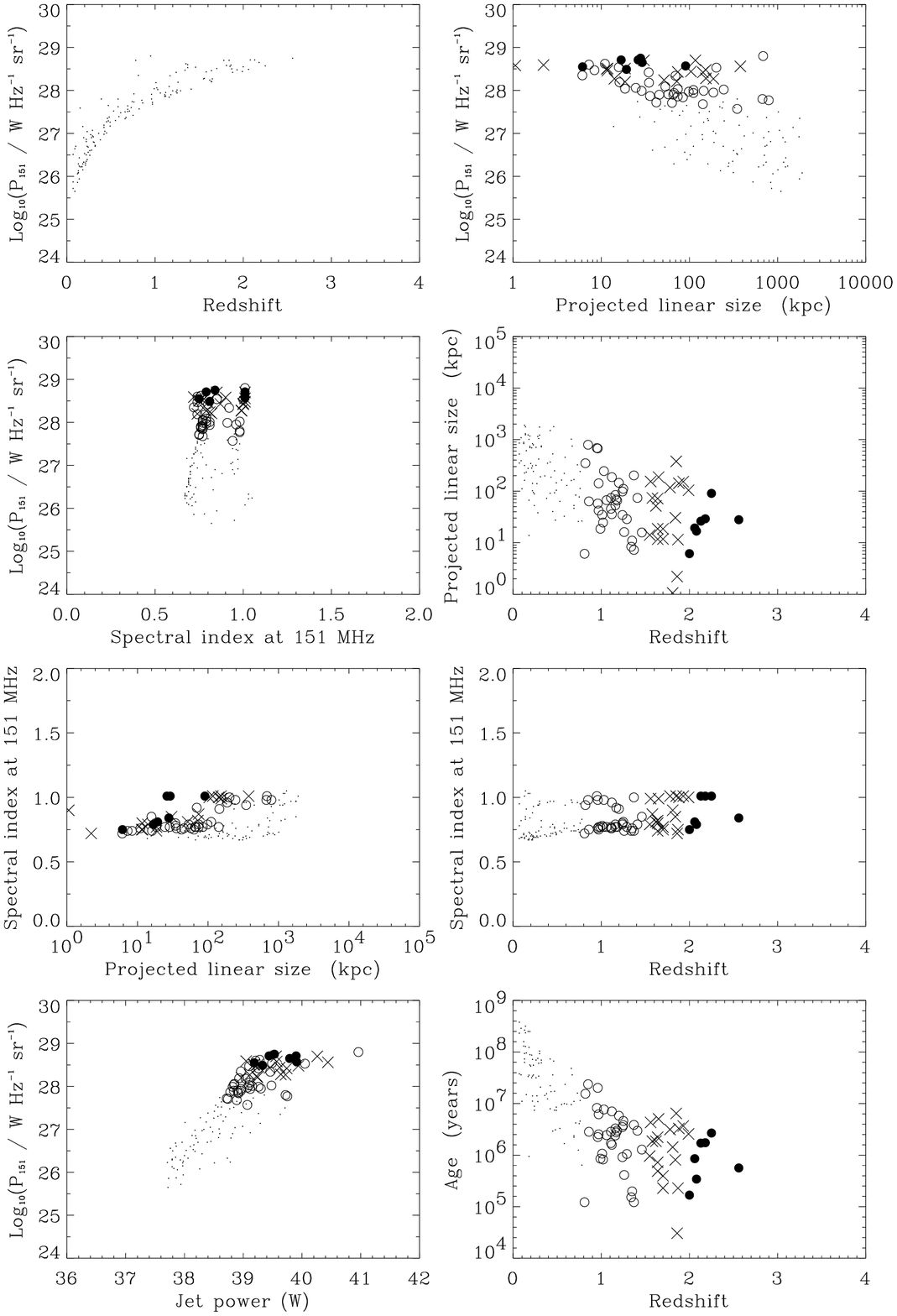}}
\end{picture}
{\caption[junk]{\label{fig:mock3c} Figure showing simulated 3C data,
assuming ${\Omega_{\rm M} = 1\ \&\ \Omega_{\Lambda} = 0} $, for the six
planes in [$PDz\alpha$] parameter space and additionally a plot of the
jet-power $Q_{\rm o}$ {\em vs} rest-frame luminosity at 151 MHz and a plot
of the age of each source at the point we observe it against redshift.
Solid circles indicate objects at $z \geq 2$, crosses indicate objects
with $2 > z \geq 1.5$, open circles indicate $1.5 > z \geq 0.8$ and dots
indicate sources with $z < 0.8$.}}
\end{figure}
\clearpage
\begin{figure}[!t]
\begin{picture}(50,500)(0,0)
\put(-40,-85){\includegraphics{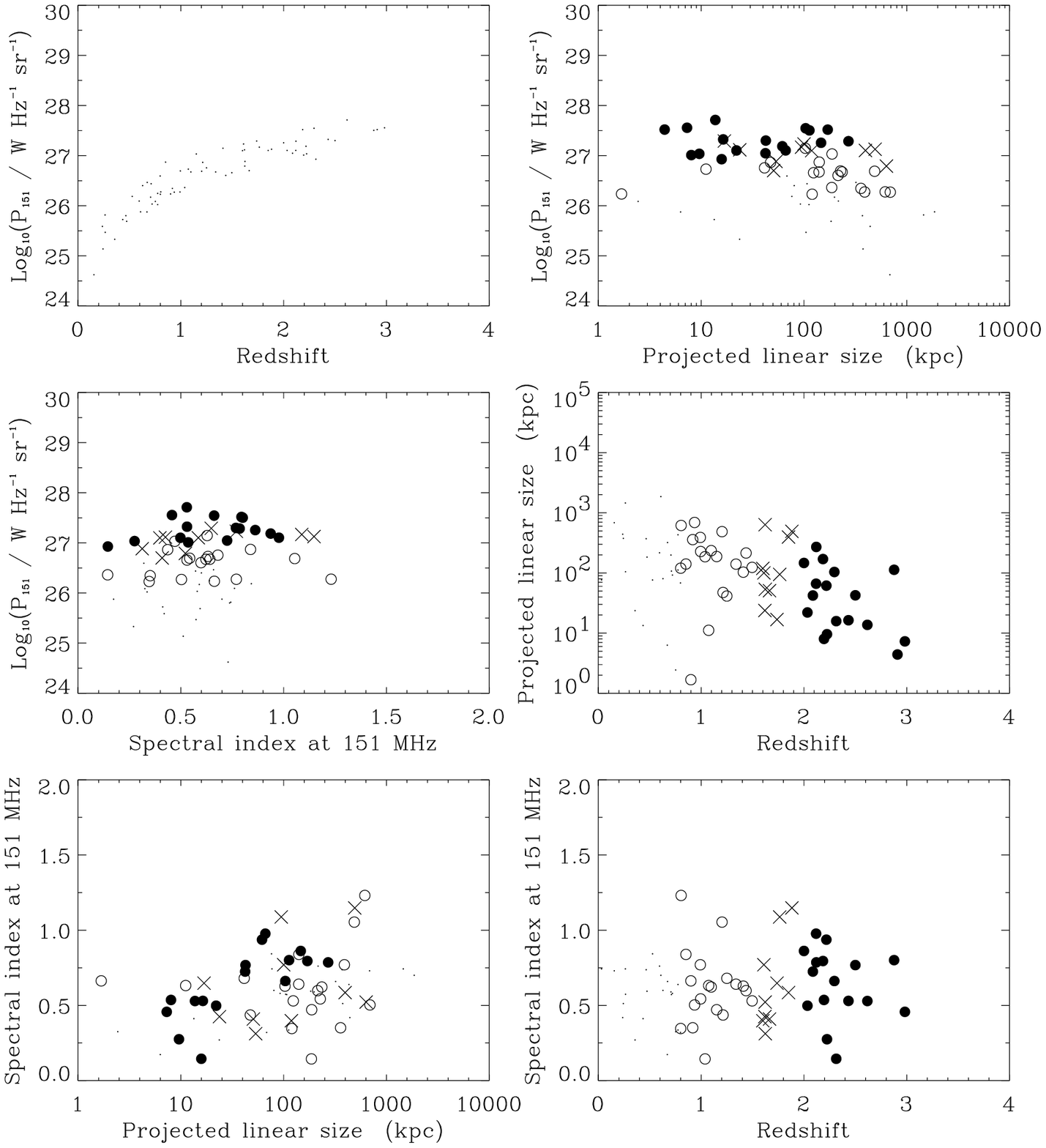}}
\end{picture}
{\caption[junk]{\label{fig:real7c} Figure showing real 7C data, for
 FR\,IIs and DAs, assuming ${\Omega_{\rm M} = 1\ \&\ \Omega_{\Lambda}
 = 0} $ for the six planes in [$PDz\alpha$] parameter space. Solid
 circles indicate objects at $z \geq  2$, crosses indicate objects
 with $2 > z \geq 1.5$, open circles indicate $1.5 > z \geq
 0.8$ and dots indicate sources with $z < 0.8$.}}
\end{figure}
\clearpage
\begin{figure}[!t]
\begin{picture}(50,570)(0,0)
\put(-40,-45){\includegraphics{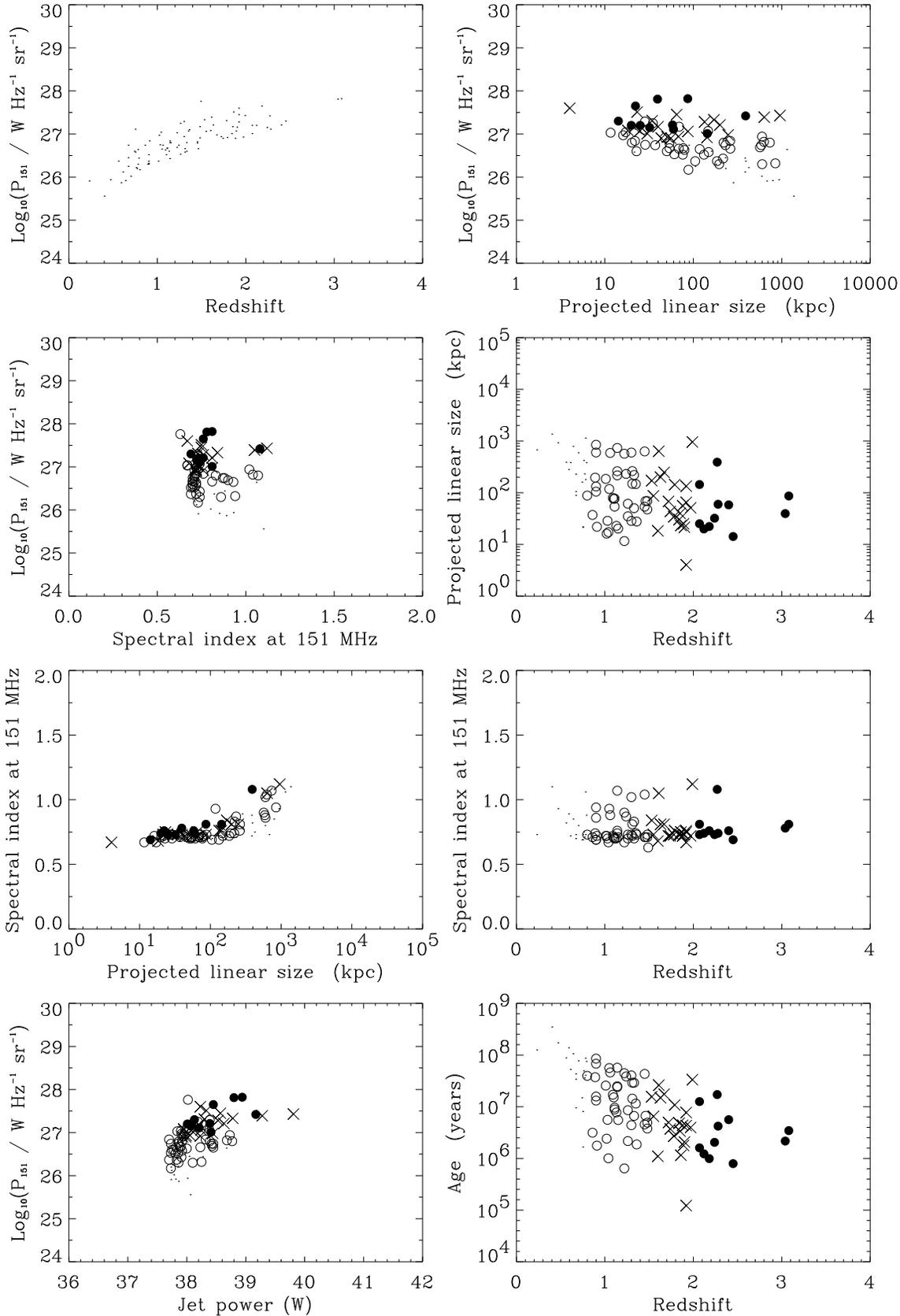}}
\end{picture}
{\caption[junk]{\label{fig:mock7c} Figure showing simulated 7C data,
assuming ${\Omega_{\rm M} = 1\ \&\ \Omega_{\Lambda} = 0} $, for the six
planes in [$PDz\alpha$] parameter space and additionally a plot of the
jet-power $Q_{\rm o}$ {\em vs} rest-frame luminosity at 151 MHz and a plot
of the age of each source at the point we observe it against
redshift. Solid circles indicate objects at $z \geq 2$, crosses indicate
objects with $2 > z \geq 1.5$, open circles indicate $1.5 > z \geq 0.8$
and dots indicate sources with $z < 0.8$.}}
\end{figure}
\clearpage
\twocolumn

If the $P$--$\alpha$ correlation arises in a way resembling that which we
have modelled here, then it follows that hotspots with more powerful
magnetic fields being fed by jets which transport higher powers, should
themselves have steeper spectra than the hotspots in sources with lower
jet-powers. Such a finding has been pointed out by Scheuer (1998), that
the pairs of hotspots in powerful radio sources have steeper spectra than
those in less powerful sources. 

Previous work [e.g.\ Baldwin 1982] has shown that simple models of radio
source evolution predict far more high $P$, large $D$ sources than are
actually observed. We are able to generate a distribution of points on the
$P$--$D$ plane which does not give rise to such a bulge. This is in part
because the higher $Q_{\rm o}$ sources have steeper spectra and so are
more easily lost below the survey flux-limits; this is in addition to the
effect whereby the sample selection process itself discriminates against
sources which have suffered large adiabatic losses.

We consider the question posed by Jenkins \& Scheuer (1976) `what docks
the tails of radio source components?'. Jenkins \& Scheuer found no
evidence that the tails (i.e.\ bridge towards the core) of radio sources
were longer at low-frequency than at high-frequency, as might be expected
if synchrotron losses play a major part in limiting the lengths of
tails. We contend that a break in the energy spectrum of particles
injected into the lobe is an increasingly relevant factor in depleting the
bridges of radio sources in increasingly powerful sources. An alternative
explanation is that the backflow speeds in these sources are too slow for
lobe emission to ever reach the core; a possible test between these two
explanations is by consideration of the polarisation patterns in the lobes 
of these sources (Miller 1985). Inverse Compton losses arising from the
AGN's own radiation field will also deplete the lobe emission close into
the core (see \S\ref{sec:ownrf}).

\subsection{The tightness of the $P$--$z$ correlation}
\label{sec:pztight}
Application of a flux-limit to a survey means that at the lower luminosity
edge there will be a tight correlation between luminosity and redshift.
Just how tight the upper edge of the correlation is depends on {\em (i)}
the steepness and/or curvature of the distribution in jet-powers (see
equation~\ref{eq:qpars}) {\em (ii)} the proximity to the upper-limit of
the distribution in jet-powers (note how the 3C $P$--$z$ distribution is
much tighter than that of the 7C distribution comparing
Figures~\ref{fig:real3c} and \ref{fig:real7c}), {\em (iii)} the energy
distribution of the particles injected into the lobes as discussed in
\S\ref{sec:lobe} and {\em (iv)} the birth function itself. The lack of
high redshift sources in 3C may arise in part because of an intrinsic
maximum jet-power (= maximum black hole mass) in these objects (as
discussed in Willott et al.\ 1998b).

\subsection{Dependence of spectral index on linear size}
\label{sec:alphaD}

The tendency for radio sources with larger linear sizes to have steeper
spectra, as seen in Figure~\ref{fig:da}, is not hard-wired into our
simulations, yet it does appear (e.g.\ Figure~\ref{fig:mock7c}).  We
interpret this tendency as being due to the decrease in magnetic field as
the lobes expand. For a fixed observing frequency, the Lorentz factor of
particles contributing most of their emission at this frequency will be
higher for a lower magnetic field (see equation~\ref{eq:lorentz}). Thus
particles with a higher Lorentz factor are required for the chosen
emitting frequency and given the power-law exponent of typical energy
distributions, the number of high Lorentz factor particles is smaller than
the number of less energetic particles. Moreover, the adiabatic expansion
process itself, while preserving the shape of the spectrum, will shift to
lower frequencies any features in the spectrum such as a break frequency.

\subsection{Ramifications of the `youth--redshift link'}
\label{sec:youthz}
Each of the Figures~\ref{fig:mock3c} and \ref{fig:mock7c} show (bottom
right hand corners) plots of the ages of sources in the given simulated
sample versus the redshifts. There is a very steep (negative) gradient
seen in the distribution of points on these plots: objects close to a
redshift of 2 have ages approaching two orders of magnitude smaller than
objects in the local Universe; the correspondence in the simulated fainter
sample (Figure~\ref{fig:mock7c}) is the same but less pronounced. We thus
conclude that in any study of the importance of redshift to a given
phenomenon, it is essential to take into account the fact that the higher
the redshift at which an object is observed, the younger (on average) it
is. Two such issues where this is an important consideration are (i) the
higher distortion in radio structure of Barthel and Miley's (1988)
high-redshift sample of steep-spectrum quasars compared with their
low-redshift sample and (ii) the alignment effect [McCarthy et al.\ 1987;
Chambers, Miley \& van Breugel 1987].  For case (i) Barthel \& Miley
suggested that at higher redshifts the environments of radio sources were
more dense and inhomogeneous than at low redshift. In case (ii) it has
been proposed that the origin of the optical and infra-red light which is
aligned with the radio axes arises either because of star-formation
stimulated by the passage of the jet through the medium [McCarthy et al.\
1987; Rees 1989] or by quasar light escaping along the radio axes
scattered into the line-of-sight by dust or electrons [Barthel 1989;
Fabian 1989].

Sources which are younger may have the interactions of their jets occuring
much closer in to the host galaxy where there is a higher density and
likely greater inhomogeneity in the ambient matter; thus it is possible
that star-formation is more easily triggered here than at distances
further out from the central engine and possible that the radio structures
seen at these times are more distorted than those seen much later in the
life of a radio source.  Although a young source is not necessarily a
short source (consider the derivative of equation~\ref{eq:falle}), in the
case of sources which are young but quite long on account of their high
jet-power $Q_{\rm o}$, the higher jet-power may well be more efficient at
triggering star formation.

We note Dey et al.'s (1997) spectropolarimetric study of 4C41.17 presents
a compelling case for jet-triggered star formation in this young, high
jet-power, radio galaxy at $z = 3.8$.

Moreover if it is the case that mergers of galaxies trigger radio galaxy
or quasar activity (see e.g.\ Sanders et al.\ 1988), then observations of
young radio sources (and thus those at high redshift) will be revealing
the emergence of a radio source through recent merger products. Such an
environment is more likely to be favourable to the triggering of
star-formation than the rather more `steady-state' environment of a radio
source which has been in existence for longer than $10^8$ years.

Consistent with this picture is the finding of Best, Longair and
R\"{o}ttgering (1996) that the smallest sources in their sample of $z \sim
1$ radio galaxies (all with very similar luminosities) are those which are
most aligned with optical emission. Best et al.\ (1996) remarked that the
sequence of changing optical aligned structure with increasing radio size
could be naturally interpreted by comparing it with different phases of
the interaction of the radio jets with the interstellar and intergalactic
medium as the radio sources age.

We have not needed to invoke any systematic change in the environments of
radio sources with redshift in order to reproduce good approximations to
the six planes in [$PDz\alpha$] space from our observations. However, we
note that Garrington \& Conway (1991) found that there was a tendency for
depolarisation to be higher in sources with a higher $P$ (and hence $z$,
since their sample mainly comprised objects in the 3C sample or similar
samples). It is possible that this is related to the above discussion:
objects in their sample with higher $P$ or $z$ which are younger may well
be in much more recently merged environments with the consequence that
inhomogeneous regions of higher ambient density will more readily
depolarise polarised intensity radiated from the lobes in these
environments.

Note that a result of this work is that we are unavoidably driven to the
conclusion that the high redshift sources we find in a radio survey are
inevitably a younger population than their low redshift counterparts.
Comparison of the simulated samples in Figures~\ref{fig:mock3c} and
\ref{fig:mock7c} suggest that the strong dependence of age on redshift is
alleviated significantly by studies using more than one complete sample 
e.g.\ the new 7C redshift survey in tandem with the 3C sample.

Even if some details of our model are not fully correct and for example,
the inclusion of a source into a survey is governed much more by its
low-frequency emission being dominated by its hotspots than we infer to be
the case on the basis of current data, then in order to avoid simulating
an excess of very powerful large linear size radio sources over that which
is seen, it would be necessary to invoke that sources at higher redshift
have their beams switched on for shorter times. The ``youth--redshift
degeneracy'' would then be truely hard-wired into the model. As it is, in
our model the ``youth--redshift degeneracy'' arises because older sources
do not survive above the survey flux-limit because the spectral losses due
to (especially) inverse Compton and also adiabatic expansion and
synchrotron losses deplete the luminosity of an object at the relevant
emitted frequency $[(1 + z) \times$ survey frequency].

The model we have presented here, which reproduces the nature of radio
sources which are included in low-frequency selected radio surveys, does
not need to invoke the `death' of radio sources. This gives us a
distribution in the {\em ages} of the sources which make it into our
samples but does not require any strong statement about their {\em
lifetimes}.  It has been suggested that radio sources which are on death
row should be seen as a population of sources which simply appear as
widely separated pairs of ``naked hotspots''. We suggest that this is
unlikely: the inclusion of such objects into low-frequency (e.g.\ 151 MHz)
surveys is most unlikely to be via their hotspots because, as discussed in
\S\ref{sec:hslum} hotspots do not radiate strongly at these low
frequencies; although such hotspots will radiate at frequencies like
1.4\,GHz [e.g.\ as in the FIRST survey of Becker et al.\ (1998)] it is
most unlikely that emission will be found to link apparently unrelated
isolated hotspots.

\newpage
\section{Summary of conclusions}
\label{sec:conc}
\begin{flushleft}
{\em New observational results} 
\end{flushleft}

The population of radio sources which are detected by a given survey {\em
cannot} be naively identified with the {\em real} population of radio
sources out there: our study of a number of complete samples of radio
sources selected at frequencies close to 151 MHz, with a coverage of the
luminosity--redshift plane which is substantially improved over previous
studies, has led us to conclude that the principal dependences between
properties of these radio sources selected in complete samples are as
follows:

\begin{enumerate}
\item The rest-frame spectral index at low-frequency depends on the luminosity
of a source ($P$--$\alpha$ correlation).
\item The rest-frame spectral index at high-frequency (GHz) depends on the
redshift of a source.
\item Observed-frame spectral indices are also governed by the redshift of
a source.
\item Sources with larger physical sizes have steeper spectra
($D$--$\alpha$ correlation).
\item Sources at higher redshift have smaller physical sizes ($D$--$z$
anti-correlation).

\begin{flushleft}
\hspace{-1.2cm}{\em New views of radio sources}
\end{flushleft}

\item We have presented an intuitive picture for the random interception of
radio sources with our light-cone, which does not necessitate the use of
assumptions concerning the dependences of the lifetimes of radio-sources.
\item The $P$--$\alpha$ correlation (result 1, listed above) 
can be understood if the hotspot is viewed as the {\em governor} of the
energy distribution injected into the lobes of radio sources, and if the
jet-power determines the magnetic field density in the hotspots.
\item The $D$--$z$ anti-correlation (result 4) arises because of
a more fundamental dependence, namely that the high-redshift sources in a
given sample will be substantially younger than their low redshift
counterparts, an effect which we call the `youth--redshift degeneracy'.
\end{enumerate}

\section{Acknowledgments}

We warmly thank collaborators past and present: Dave Rossitter, Steve
Eales, Mark Lacy and Julia Riley, for their contributions to the
acquisition of data.  It is a pleasure to thank Beatriz Bengoechea,
Stephen Blundell, Christian Kaiser, Paddy Leahy, John Miller, Lance
Miller, Prasenjit Saha, Alan Sanders, Chris Simpson, Devinder Sivia and
Graham Woan for useful discussions and/or assistance.  This research has
made use of the NASA/IPAC Extragalactic Database, which is operated by the
Jet Propulsion Laboratory, Caltech, under contract with the National
Aeronautics and Space Administration and of the Atlas of 3CRR radio
sources.  KMB thanks Balliol College, Oxford for a Research Fellowship.

\newpage

\begin{table}[!h]
{\caption[junk]{\label{tab:rees} This table lists the names, redshifts,
projected linear sizes, flux densities at 178 MHz [on the scale of Baars
et al.\ (1977)] and at 38 MHz [\cite{Hal95}] and spectral indices between
38 and 178 MHz of those sources which would not have been selected were
the sample criteria to have been $ \delta > 60^{\circ}$ and $S_{38} > 25$,
where $S_{38}$ is the flux density in Jy at 38 MHz.  }}
\begin{center}
\begin{tabular}{llcrrr}
\tableline\tableline
\ph{1234567890123} &
\ph{123456} &
\ph{123456789} &
\ph{6789} &
\ph{56789} &
\ph{3456789} \\
\mc{1}{c}{Name}&\mc{1}{c}{$z$} & \mc{1}{c}{Size} 
               & \mc{1}{c}{$S_{178}$} & \mc{1}{c}{$S_{38}$} &	$\alpha^{178}_{38}$ \\
               &                      &  (kpc)               &
\mc{1}{c}{(Jy)}  & \mc{1}{c}{(Jy)} & \\
\tableline
\ph{1234567890123} &
\ph{123456} &
\ph{123456789} &
\ph{6789} &
\ph{56789} &
\ph{3456789} \\
3C\,268.3 & 0.371 & 9           &    11.7              &  12.3               &$\ph{-}0.0$  \\
3C\,343   & 0.998 & 9           &    13.5              &   6.3               &	  $-0.5$   \\
3C\,343.1 & 0.75  & 3           &    12.5              &   9.4
&	  $-0.2$    \\[0.1cm]
\tableline
\tableline
\end{tabular}
\end{center}
\end{table}

\onecolumn
\newpage

\begin{table}[!h]
{\caption[junk]{\label{tab:AS} Table of radio sources from the sample of
Allington-Smith (1982) selected at 408\,MHz in a region of sky exactly
overlapping with part of the 151-MHz 6C sample (\S\ref{sec:samples6c}) but
not common to the 6C sample. [Objects which would be too bright at 151 MHz
to lie within the flux-limits of the 6C sample are not listed here.]  We
list their names, redshifts, projected linear sizes in kpc, their flux
densities at 151 MHz and 408 MHz and their spectral indices between 151
MHz and 408 MHz.  Those sources below the horizontal rule are those whose
flux densities at 408 MHz fall below 1.1 Jy; this flux-limit corresponds
to those sources with a flux density of 2 Jy at 151 MHz and a spectral
index of 0.6.  The angular sizes of B3\,0822+39, B2\,0927+35, B3\,1056+39,
B3\,1106+37, B2\,1129+37, B3\,1207+38 and B2\,1225+36 were taken from
Law-Green et al.\ (1995); the angular size of B3\,1101+38 was taken from
the NVSS survey (Condon et al.\ 1998) and all the remainder were taken
from the FIRST survey (Becker et al.\ 1998).  The redshifts indicated by a
\ddag\ are quoted from Allington-Smith et al.\ (1988). The redshifts
marked by *\dag\ are quoted from Allington-Smith, Lilly \& Longair (1985).
The redshifts for sources marked with a single asterisk have been measured
by S.\ Rawlings. The redshifts marked with two asterisks are quoted from
Eales \& Rawlings (1996); note that the redshift of B2\,1132+37 is highly
uncertain. The redshift of B3\,1101+38 is quoted from Weiler \& Johnson
(1980).  The redshift of B2\,1204+34 is quoted from Allington-Smith
(1984). The redshift of B2\,1225+36 is quoted from Xu et al.\ (1994). The
redshift of B2\,0927+35 was not available to us and so we considered two
limiting values for this. The 151 MHz flux densities of all these objects
were taken from the 6C survey of Hales et al.\ (1988) and Hales et al.\
(1993) which are complete to \gtsim\ 0.2 Jy. For those sources whose names
begin with B2, their flux densities at 408 MHz are taken from the B2.3
catalogue (Colla et al.\ 1973) and for those sources whose names begin
with B3, their flux densities at 408 MHz are taken from the B3 catalogue
(Ficarra et al.\ 1985).}}
\begin{center}
\begin{tabular}{llrrrr}
\tableline
\tableline
\ph{1234567890123} &
\ph{123456} &
\ph{123456789} &
\ph{6789} &
\ph{56789} &
\ph{3456789} \\
\mc{1}{c}{Name}&\mc{1}{c}{$z$}    &\mc{1}{c}{Size}  &\mc{1}{c}{$S_{151}$}& $S_{408}$&\mc{1}{c}{$\alpha^{408}_{151}$}\\
               &                  & \mc{1}{c}{(kpc)}&
\mc{1}{c}{(Jy)}  &  (Jy)    &       \\ 
               &                  &                 &
&          &       \\
\tableline
               &                  &                 &
&          &       \\
B3\,0822+39    &\xx1.2$^{*}$      &      8.6\xx     &    0.50            &    1.76  &$-1.3$  \\ 
B2\,0835+37    &\xx0.396$^{\ddag}$&     32.0\xx     &    1.62            &    1.26  &  0.3  \\
B2\,0927+35    &  $ \left\{ \matrix{ 0.100 \cr 1.000} \right \} $                    
               &  $ \left\{ \matrix{ 11.1  \cr 38.3}  \right \} $                     
                                                    &    1.70            &    1.15  &  0.4  \\
B3\,1049+38    &\xx1.018$^{\ddag}$&      8.5\xx     &    0.99            &    1.24  & $-0.2$\\
B3\,1101+38    &\xx0.030          &     41.4\xx     &    1.70            &    1.14  &  0.4  \\
B3\,1106+37    &\xx2.290$^{**}$   &      5.6\xx     &    1.70            &    1.29  &  0.3  \\
B2\,1129+37    &\xx1.060$^{*}$    &    162.7\xx     &    1.30            &    1.20  &  0.1  \\
B2\,1132+37    &\xx2.880$^{**}$   &      7.4\xx     &    1.10            &    1.52  & $-0.3$\\
B3\,1201+39    &\xx0.445$^{*\dag}$&      6.8\xx     &    1.90            &    1.11  &  0.5  \\
               &                  &                 &                    &          &       \\
\hline                                                                               
               &                  &                 &                    &          &       \\
B3\,1056+39    &\xx2.200$^{*}$    &     96.2\xx     &    1.70            &    0.82  &  0.7  \\
B2\,1204+34    &\xx0.11           &     105.0\xx    &    1.97            &    1.01  &  0.7  \\
B3\,1207+38    &\xx0.790$^{*\dag}$&    221.6\xx     &    1.90            &    0.94  &  0.7  \\
B2\,1225+36    &\xx1.975          &     0.6\xx      &    0.50            &    1.03  &  0.7  \\
B2\,1245+34    &\xx0.409$^{\ddag}$&     39.0\xx     &    1.80            &    1.06  &  0.5  \\   
               &                  &                 &
&          &       \\
\tableline
\tableline
\end{tabular}     
\end{center}
\end{table}
\newpage

\begin{table}[!t]
{\caption[junk] {\label{tab:fr23stats} The results of the statistical
analysis, run for three different sets of cosmological parameters, on
all members of the complete samples used in the analyses. }}
\begin{center}
\begin{tabular}{l}
\mc{1}{c}{FR\,IIs and DAs \ph{xxx} \cosone}\\[0.2cm]
\mc{1}{l}{
\begin{tabular}{cccc} 
\cline{2-4}
         & \mc{1}{|c}{$P$} & \mc{1}{|c}{$D$} & \mc{1}{|c|}{$z$} \\ 
\cline{2-4}
\hline
\mc{1}{|l|}{$\alpha$} & \mc{1}{r|}{$r_{P\alpha \vert zD} = 0.34$} & \mc{1}{r|}{$r_{D\alpha \vert Pz} = 0.48$} & \mc{1}{r|}{$r_{z\alpha \vert PD} = -0.32$}          \\ 
\mc{1}{|l|}{        } & \mc{1}{r|}{$\sigma = 5.83$} & \mc{1}{r|}{$\sigma = 8.59$          } & \mc{1}{r|}{$\sigma = -5.34$ }          \\ 
\cline{1-4} 
\mc{1}{|l|}{$z$}      & \mc{1}{r|}{$r_{Pz \vert \alpha D} = 0.75$} &  \mc{1}{r|}{$r_{Dz \vert P\alpha} = -0.06$}   &           \\ 
\mc{1}{|l|}{   }      & \mc{1}{r|}{$\sigma = 16.01$      } &  \mc{1}{r|}{$\sigma = -1.01$  }   &           \\ 
\cline{1-3} 
\mc{1}{|l|}{$D$}      & \mc{1}{r|}{$r_{PD \vert z\alpha} = -0.13$} & & \\ 
\mc{1}{|l|}{   }      & \mc{1}{r|}{$\sigma = -2.09$     } & &  \\ 
\cline{1-2} 
\end{tabular} 
} \\
\mc{1}{c}{}\\
\mc{1}{c}{FR\,IIs and DAs \ph{xxx} \costwo}\\[0.2cm]
\mc{1}{l}{
\begin{tabular}{cccc} 
\cline{2-4}
         & \mc{1}{|c}{$P$} & \mc{1}{|c}{$D$} & \mc{1}{|c|}{$z$} \\ 
\cline{2-4}
\hline
\mc{1}{|l|}{$\alpha$} & \mc{1}{r|}{$r_{P\alpha \vert zD} = 0.34$} & \mc{1}{r|}{$r_{D\alpha \vert Pz} = 0.48$} & \mc{1}{r|}{$r_{z\alpha \vert PD} = -0.24$}          \\ 
\mc{1}{|l|}{        } & \mc{1}{r|}{$\sigma = 5.75$} & \mc{1}{r|}{$\sigma = 8.58$          } & \mc{1}{r|}{$\sigma = -4.05$ }          \\ 
\cline{1-4} 
\mc{1}{|l|}{$z$}      & \mc{1}{r|}{$r_{Pz \vert \alpha D} = 0.65$} &  \mc{1}{r|}{$r_{Dz \vert P\alpha} = -0.19$}   &           \\ 
\mc{1}{|l|}{   }      & \mc{1}{r|}{$\sigma = 12.62$      } &  \mc{1}{r|}{$\sigma = -3.06$  }   &           \\ 
\cline{1-3} 
\mc{1}{|l|}{$D$}      & \mc{1}{r|}{$r_{PD \vert z\alpha} = -0.13$} & & \\ 
\mc{1}{|l|}{   }      & \mc{1}{r|}{$\sigma = -2.09$     } & &  \\ 
\cline{1-2} 
\end{tabular} 
} \\
\mc{1}{c}{}\\
\mc{1}{c}{FR\,IIs and DAs \ph{xxx} \costhree}\\[0.2cm]
\mc{1}{l}{
\begin{tabular}{cccc} 
\cline{2-4}
         & \mc{1}{|c}{$P$} & \mc{1}{|c}{$D$} & \mc{1}{|c|}{$z$} \\ 
\cline{2-4}
\hline
\mc{1}{|l|}{$\alpha$} & \mc{1}{r|}{$r_{P\alpha \vert zD} = 0.33$} & \mc{1}{r|}{$r_{D\alpha \vert Pz} = 0.48$} & \mc{1}{r|}{$r_{z\alpha \vert PD} = -0.32$}          \\ 
\mc{1}{|l|}{        } & \mc{1}{r|}{$\sigma = 5.65$} & \mc{1}{r|}{$\sigma = 8.46$          } & \mc{1}{r|}{$\sigma = -5.40$ }          \\ 
\cline{1-4} 
\mc{1}{|l|}{$z$}      & \mc{1}{r|}{$r_{Pz \vert \alpha D} = 0.76$} &  \mc{1}{r|}{$r_{Dz \vert P\alpha} = -0.05$}   &           \\ 
\mc{1}{|l|}{   }      & \mc{1}{r|}{$\sigma = 16.24$      } &  \mc{1}{r|}{$\sigma = -0.85$  }   &           \\ 
\cline{1-3} 
\mc{1}{|l|}{$D$}      & \mc{1}{r|}{$r_{PD \vert z\alpha} = -0.12$} & & \\ 
\mc{1}{|l|}{   }      & \mc{1}{r|}{$\sigma = -1.88$     } & &  \\ 
\cline{1-2} 
\end{tabular} 
} \\
\end{tabular}
\end{center}
\end{table}

\newpage

\begin{table}[!t]
{\caption[junk] {\label{tab:fr2stats} The results of the statistical
analysis, run for three different sets of cosmological parameters, on
just the sources with certain FR\,II structure from the complete
samples. }}
\begin{center}
\begin{tabular}{l}
\mc{1}{c}{FR\,IIs \ph{xxx} \cosone}\\[0.2cm]
\mc{1}{l}{
\begin{tabular}{cccc} 
\cline{2-4}
         & \mc{1}{|c}{$P$} & \mc{1}{|c}{$D$} & \mc{1}{|c|}{$z$} \\ 
\cline{2-4}
\hline
\mc{1}{|l|}{$\alpha$} & \mc{1}{r|}{$r_{P\alpha \vert zD} = 0.33$} & \mc{1}{r|}{$r_{D\alpha \vert Pz} = 0.37$} & \mc{1}{r|}{$r_{z\alpha \vert PD} = -0.31$}          \\ 
\mc{1}{|l|}{        } & \mc{1}{r|}{$\sigma = 4.91$} & \mc{1}{r|}{$\sigma = 5.58$          } & \mc{1}{r|}{$\sigma = -4.64$ }          \\ 
\cline{1-4} 
\mc{1}{|l|}{$z$}      & \mc{1}{r|}{$r_{Pz \vert \alpha D} = 0.74$} &  \mc{1}{r|}{$r_{Dz \vert P\alpha} = -0.03$}   &           \\ 
\mc{1}{|l|}{   }      & \mc{1}{r|}{$\sigma = 13.63$      } &  \mc{1}{r|}{$\sigma = -0.39$  }   &           \\ 
\cline{1-3} 
\mc{1}{|l|}{$D$}      & \mc{1}{r|}{$r_{PD \vert z\alpha} = -0.19$} & & \\ 
\mc{1}{|l|}{   }      & \mc{1}{r|}{$\sigma = -2.85$     } & &  \\ 
\cline{1-2} 
\end{tabular} 
} \\
\mc{1}{c}{}\\
\mc{1}{c}{FR\,IIs \ph{xxx} \costwo}\\[0.2cm]
\mc{1}{l}{
\begin{tabular}{cccc} 
\cline{2-4}
         & \mc{1}{|c}{$P$} & \mc{1}{|c}{$D$} & \mc{1}{|c|}{$z$} \\ 
\cline{2-4}
\hline
\mc{1}{|l|}{$\alpha$} & \mc{1}{r|}{$r_{P\alpha \vert zD} = 0.32$} & \mc{1}{r|}{$r_{D\alpha \vert Pz} = 0.37$} & \mc{1}{r|}{$r_{z\alpha \vert PD} = -0.24$}          \\ 
\mc{1}{|l|}{        } & \mc{1}{r|}{$\sigma = 4.86$} & \mc{1}{r|}{$\sigma = 5.57$          } & \mc{1}{r|}{$\sigma = -3.59$ }          \\ 
\cline{1-4} 
\mc{1}{|l|}{$z$}      & \mc{1}{r|}{$r_{Pz \vert \alpha D} = 0.61$} &  \mc{1}{r|}{$r_{Dz \vert P\alpha} = -0.18$}   &           \\ 
\mc{1}{|l|}{   }      & \mc{1}{r|}{$\sigma = 10.32$      } &  \mc{1}{r|}{$\sigma = -2.57$  }   &           \\ 
\cline{1-3} 
\mc{1}{|l|}{$D$}      & \mc{1}{r|}{$r_{PD \vert z\alpha} = -0.19$} & & \\ 
\mc{1}{|l|}{   }      & \mc{1}{r|}{$\sigma = -2.81$     } & &  \\ 
\cline{1-2} 
\end{tabular} 
} \\
\mc{1}{c}{}\\
\mc{1}{c}{FR\,IIs \ph{xxx} \costhree}\\[0.2cm]
\mc{1}{l}{
\begin{tabular}{cccc} 
\cline{2-4}
         & \mc{1}{|c}{$P$} & \mc{1}{|c}{$D$} & \mc{1}{|c|}{$z$} \\ 
\cline{2-4}
\hline
\mc{1}{|l|}{$\alpha$} & \mc{1}{r|}{$r_{P\alpha \vert zD} = 0.32$} & \mc{1}{r|}{$r_{D\alpha \vert Pz} = 0.36$} & \mc{1}{r|}{$r_{z\alpha \vert PD} = -0.31$}          \\ 
\mc{1}{|l|}{        } & \mc{1}{r|}{$\sigma = 4.75$} & \mc{1}{r|}{$\sigma = 5.44$          } & \mc{1}{r|}{$\sigma = -4.67$ }          \\ 
\cline{1-4} 
\mc{1}{|l|}{$z$}      & \mc{1}{r|}{$r_{Pz \vert \alpha D} = 0.74$} &  \mc{1}{r|}{$r_{Dz \vert P\alpha} = -0.02$}   &           \\ 
\mc{1}{|l|}{   }      & \mc{1}{r|}{$\sigma = 13.87$      } &  \mc{1}{r|}{$\sigma = -0.25$  }   &           \\ 
\cline{1-3} 
\mc{1}{|l|}{$D$}      & \mc{1}{r|}{$r_{PD \vert z\alpha} = -0.18$} & & \\ 
\mc{1}{|l|}{   }      & \mc{1}{r|}{$\sigma = -2.63$     } & &  \\ 
\cline{1-2} 
\end{tabular} 
} \\
\end{tabular}
\end{center}
\end{table}

\newpage

\begin{table}[!t]
{\caption[junk]{\label{tab:3cstats} The results of the statistical
analysis, run for three different sets of cosmological parameters, on
the 3C sample alone.}}
\begin{center}
\begin{tabular}{l}
\mc{1}{c}{\cosone}\\[0.2cm]
\mc{1}{l}{
\begin{tabular}{cccc} 
\cline{2-4}
         & \mc{1}{|c}{$P$} & \mc{1}{|c}{$D$} & \mc{1}{|c|}{$z$} \\ 
\cline{2-4}
\hline
\mc{1}{|l|}{$\alpha$} & \mc{1}{r|}{$r_{P\alpha \vert zD} = 0.02$} & \mc{1}{r|}{$r_{D\alpha \vert Pz} = 0.56$} & \mc{1}{r|}{$r_{z\alpha \vert PD} = 0.01$}          \\ 
\mc{1}{|l|}{        } & \mc{1}{r|}{$\sigma = 0.20$} & \mc{1}{r|}{$\sigma = 7.42$          } & \mc{1}{r|}{$\sigma = 0.12$ }          \\ 
\cline{1-4} 
\mc{1}{|l|}{$z$}      & \mc{1}{r|}{$r_{Pz \vert \alpha D} = 0.97$} &  \mc{1}{r|}{$r_{Dz \vert P\alpha} = -0.05$}   &           \\ 
\mc{1}{|l|}{   }      & \mc{1}{r|}{$\sigma = 25.24$      } &  \mc{1}{r|}{$\sigma = -0.63$  }   &           \\ 
\cline{1-3} 
\mc{1}{|l|}{$D$}      & \mc{1}{r|}{$r_{PD \vert z\alpha} = -0.03$} & & \\ 
\mc{1}{|l|}{   }      & \mc{1}{r|}{$\sigma = -0.38$     } & &  \\ 
\cline{1-2} 
\end{tabular} 
} \\
\mc{1}{c}{}\\
\mc{1}{c}{\costwo}\\[0.2cm]
\mc{1}{l}{
\begin{tabular}{cccc} 
\cline{2-4}
         & \mc{1}{|c}{$P$} & \mc{1}{|c}{$D$} & \mc{1}{|c|}{$z$} \\ 
\cline{2-4}
\hline
\mc{1}{|l|}{$\alpha$} & \mc{1}{r|}{$r_{P\alpha \vert zD} = 0.02$} & \mc{1}{r|}{$r_{D\alpha \vert Pz} = 0.56$} & \mc{1}{r|}{$r_{z\alpha \vert PD} = 0.03$}          \\ 
\mc{1}{|l|}{        } & \mc{1}{r|}{$\sigma = 0.28$} & \mc{1}{r|}{$\sigma = 7.42$          } & \mc{1}{r|}{$\sigma = 0.33$ }          \\ 
\cline{1-4} 
\mc{1}{|l|}{$z$}      & \mc{1}{r|}{$r_{Pz \vert \alpha D} = 0.96$} &  \mc{1}{r|}{$r_{Dz \vert P\alpha} = -0.12$}   &           \\ 
\mc{1}{|l|}{   }      & \mc{1}{r|}{$\sigma = 22.77$      } &  \mc{1}{r|}{$\sigma = -1.39$  }   &           \\ 
\cline{1-3} 
\mc{1}{|l|}{$D$}      & \mc{1}{r|}{$r_{PD \vert z\alpha} = -0.02$} & & \\ 
\mc{1}{|l|}{   }      & \mc{1}{r|}{$\sigma = -0.20$     } & &  \\ 
\cline{1-2} 
\end{tabular} 
} \\
\mc{1}{c}{}\\
\mc{1}{c}{\costhree}\\[0.2cm]
\mc{1}{l}{
\begin{tabular}{cccc} 
\cline{2-4}
         & \mc{1}{|c}{$P$} & \mc{1}{|c}{$D$} & \mc{1}{|c|}{$z$} \\ 
\cline{2-4}
\hline
\mc{1}{|l|}{$\alpha$} & \mc{1}{r|}{$r_{P\alpha \vert zD} = 0.03$} & \mc{1}{r|}{$r_{D\alpha \vert Pz} = 0.55$} & \mc{1}{r|}{$r_{z\alpha \vert PD} = -0.01$}          \\ 
\mc{1}{|l|}{        } & \mc{1}{r|}{$\sigma = 0.35$} & \mc{1}{r|}{$\sigma = 7.32$          } & \mc{1}{r|}{$\sigma = -0.13$ }          \\ 
\cline{1-4} 
\mc{1}{|l|}{$z$}      & \mc{1}{r|}{$r_{Pz \vert \alpha D} = 0.98$} &  \mc{1}{r|}{$r_{Dz \vert P\alpha} = -0.03$}   &           \\ 
\mc{1}{|l|}{   }      & \mc{1}{r|}{$\sigma = 26.20$      } &  \mc{1}{r|}{$\sigma = -0.33$  }   &           \\ 
\cline{1-3} 
\mc{1}{|l|}{$D$}      & \mc{1}{r|}{$r_{PD \vert z\alpha} = -0.04$} & & \\ 
\mc{1}{|l|}{   }      & \mc{1}{r|}{$\sigma = -0.47$     } & &  \\ 
\cline{1-2} 
\end{tabular} 
} \\
\end{tabular}
\end{center}
\end{table}

\newpage

\begin{table}[!t]
{\caption[junk] {\label{tab:exps} The parameters $n$ and $m$ from
equation~\ref{eq:nm} which have been fitted to the data for the three
different assumed cosmologies. }}
\begin{center} 
{\Large \cosone }
\end{center} 
\begin{center} 
{\Large FRIIs and DAs} 
\end{center} 
\begin{eqnarray}
n[r{{dz}\vert{pa}}]  &=&  0.40 \pm  0.30 \nonumber \\ 
n[r{{dz}\vert{p}}]   &=&  1.52 \pm  0.30 \nonumber \\ 
n[r{{dz}\vert{a}}]   &=&  0.86 \pm  0.20 \nonumber \\ 
m[r{{pd}\vert{za}}]  &=&  0.02 \pm  0.15 \nonumber \\ 
m[r{{pd}\vert{z}}]   &=&  0.01 \pm  0.20 \nonumber \\ 
m[r{{pd}\vert{a}}]   &=&  0.07 \pm  0.20 \nonumber      
\end{eqnarray}
\begin{center} 
{\Large \costwo }
\end{center} 
\begin{center} 
{\Large FRIIs and DAs} 
\end{center} 
\begin{eqnarray}
n[r{{dz\vert}{pa}}]  &=&  0.96 \pm  0.25 \nonumber \\ 
n[r{{dz\vert}{p}}]   &=&  1.87 \pm  0.30 \nonumber \\ 
n[r{{dz\vert}{a}}]   &=&  1.31 \pm  0.20 \nonumber \\ 
m[r{{pd\vert}{za}}]  &=&  0.01 \pm  0.15 \nonumber \\ 
m[r{{pd\vert}{z}}]   &=&  0.09 \pm  0.30 \nonumber \\ 
m[r{{pd\vert}{a}}]   &=&  0.26 \pm  0.20 \nonumber      
\end{eqnarray}
\begin{center} 
{\Large \costhree}
\end{center} 
\begin{center} 
{\Large FRIIs and DAs} 
\end{center} 
\begin{eqnarray}
n[r{{dz\vert}{pa}}]  &=&  0.35 \pm  0.30 \nonumber \\ 
n[r{{dz\vert}{p}}]   &=&  1.41 \pm  0.35 \nonumber \\ 
n[r{{dz\vert}{a}}]   &=&  0.76 \pm  0.20 \nonumber \\ 
m[r{{pd\vert}{za}}]  &=&  0.02 \pm  0.20 \nonumber \\ 
m[r{{pd\vert}{z}}]   &=&  0.07 \pm  0.30 \nonumber \\ 
m[r{{pd\vert}{a}}]   &=&  0.18 \pm  0.20 \nonumber      
\end{eqnarray}
\end{table}

\end{document}